%% file: asas5.tex
\documentclass[times,11pt]{article}
\input{include}

\newcommand{\FigurePs}[7]{\begin{figure}[htb]\vspace{#1}
\includegraphics{#4}
\FigCap{#2}\label{#3}
\end{figure}}

\begin{document}
\def\thefootnote{\fnsymbol{footnote}}
\begin{Titlepage}
\Title{The All Sky Automated Survey. 
Variable Stars in the 0$^{\rm h}$ - 6$^{\rm h}$ 
Quarter of the Southern Hemisphere.
\footnote{Based on observations obtained
at the Las Campanas Observatory of the Carnegie Institution of Washington.}}
\Author{G.~~P~o~j~m~a~\'n~s~k~i}{Warsaw University Observatory,
Al~Ujazdowskie~4, 00-478~Warsaw, Poland\\
e-mail: gp@sirius.astrouw.edu.pl}

\end{Titlepage}

\vspace*{-12pt}
\Abstract{
This paper describes the first part of the photometric data from the 
$9\arcd \times 9\arcd$
ASAS cameras monitoring the whole southern hemisphere in $V$-band. 
Data acquisition and reduction pipeline is described and preliminary 
list of variable stars is presented. Over 1,300,000 stars brighter than
$V$=15 on 40,000 frames were analyzed and  3126 were found to be variable 
(1046 eclipsing,
778 regular pulsating, 132 Mira and 1170 other, mostly SR, IR and LPV stars).
Periodic light curves have been classified using the fully automated
algorithm, which is described in  detail.
Basic photometric properties are presented in the tables and thumbnail
light curves are printed for reference.
All photometric data is  available over the INTERNET at
http://www.astrouw.edu.pl/$\sim$gp/asas/asas.html or http://archive.princeton.edu/$\sim$asas.
}
{Catalogs -Stars:variables:general-Surveys}

\vspace*{-6pt} 
\section{Introduction}
The All Sky Automated Survey (Pojma{\'n}ski 1997, 1998, 2000)
has finally achieved its important goal - photometric 
monitoring of the large part of the sky (Paczy{{\'n}}ski 1997). 
After installing two wide-field 
($9\arcd \times 9\arcd$) cameras in October 2000 (Pojma{\'n}ski 2001)  ASAS gained 
capability to measure brightness of the observable stars on the nightly basis.

The prototype ASAS system using small commercial CCD camera, 135~mm
f/1.8 telephoto lens and $I$-band (Schott RG-9, 3mm) filter
was used in the years 1997-2000 to  
monitor 0.7\% of the sky to the limiting magnitude of $I\sim13$.
During 3 years of operation it has collected over $50 \times 10^6$ measurements
of over 150.000 stars and detected almost 4000 variable stars.

The instrument, data acquisition and reduction pipeline,
the ASAS Catalog and summary of the variability search were described 
by Pojma{\'n}ski (1997, 1998, 2000),

In summer 2000 we have installed upgraded ASAS-3 system in the dome of the
10" astrograph in the Las Campanas Observatory (operated by the Carnegie Institution 
of Washington). New hardware consisted of the three independent instruments
each equipped with the automated paralactic mount,
imaging optics with standard filter, 2K $\times $ 2K (14 um pixels)
CCD camera and dedicated computer. 

Two wide-field systems are equipped with the Minolta 200/2.8 APO-G telephoto
lenses giving superb sharpness (FWHM $< 2$ pixels) but
also strong vignetting (40-50 \% in the corners). Field of view 
is $8.8\arcd \times 8.8\arcd$. The two systems are equipped with standard $I$ and $V$ filters.

Narrow field instrument is $D=250$~mm, $F=750$~mm Cassegrain system with 
three element, Wyne-type field corrector. It gives sharp images 
(FWHM $< 2.2$ pixels) in the field of $2\arcd$ diameter. 
With $2048 \times 2048$
CCD field of view is $2.2\arcd \times 2.2\arcd$. This system has $I$ 
filter in its optical path.

Three AP-10 (Apogee) commercial cameras were purchased for the project.
High readout noise ($>10 e^-$) and 14 bit ADC were accepted as a trade-off 
for 5 sec readout time. Unfortunately two of three systems did not 
meet factory specifications (increased noise resulting in substantially
reduced sensitivity, failed  thermo-cooler) and had to be sent for repair
after first months of operation. ASAS-3 was left with only wide-field
$V$-band camera running.

\section{Observations}

\FigurePs{7cm}
{Standard deviation $\sigma_V$ {\em vs.} $V$\.-band
magnitudes measured with the wide-field $V$ camera and 3-minute 
exposures for the field centered on SMC. Larger dots
denote detected variables. Small dots lying above 95 centile 
represent mostly "long-term" variables not recognized in this work.
}{mag_disp}{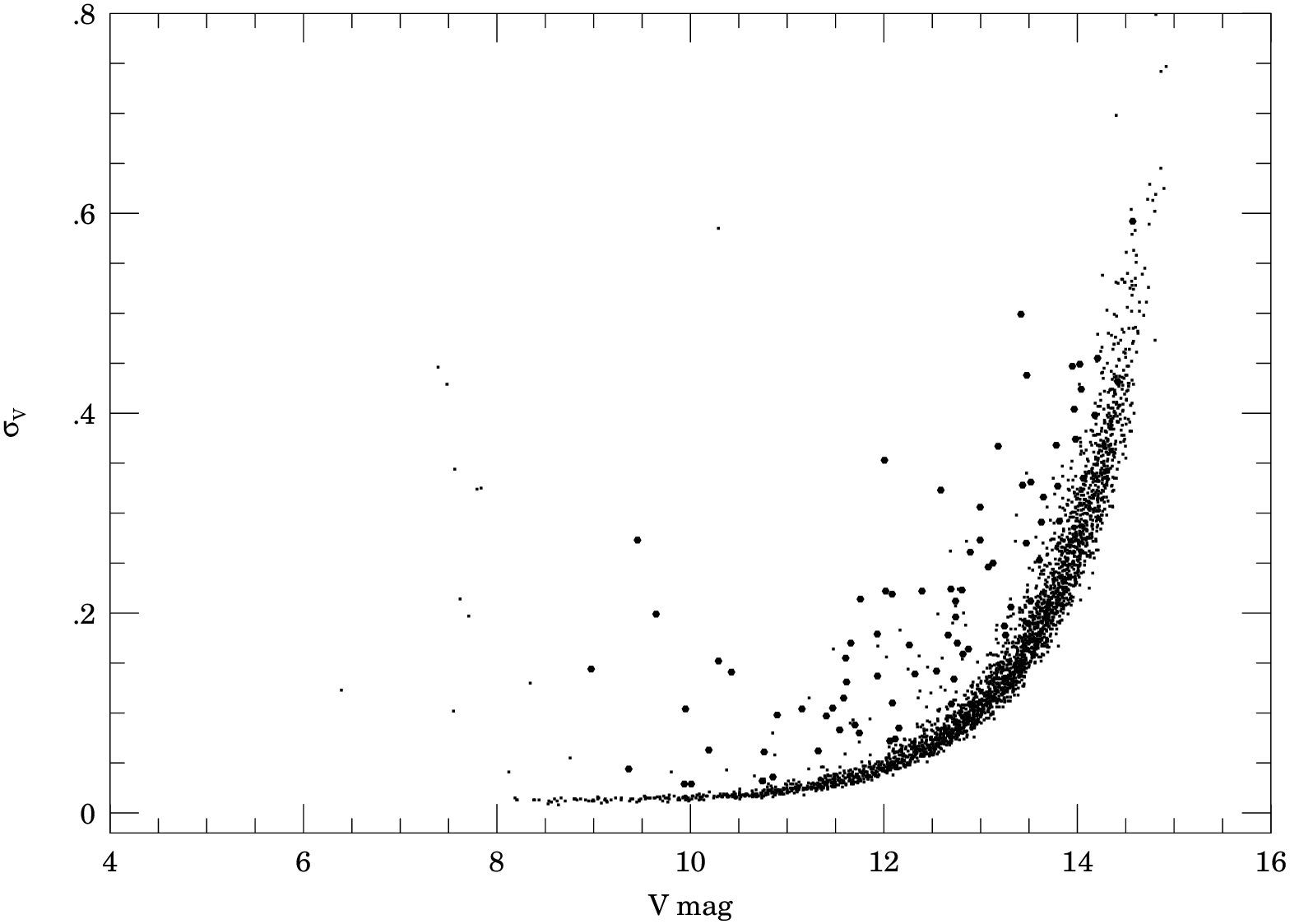}{40}{45}{-20} 

Although mechanical quality of the mount allows for unguided exposures as 
long as 5-10 minutes for wide-field systems, only 3 minute exposures
are taken in $V$-band giving the limiting magnitude of 
almost 15.0 and causing saturation of the stars brighter than $V=8.5$.

\FigurePs{7cm}{Coverage of the sky with $V$-band observations
after one year of ASAS-3 operation.
White color corresponds to the area observed at least 250 times, dark 
gray - at least 30. Only fields centered within $0^{\rm h} < \alpha < 6^{\rm h}$
are analyzed in this paper.
}{mapa}{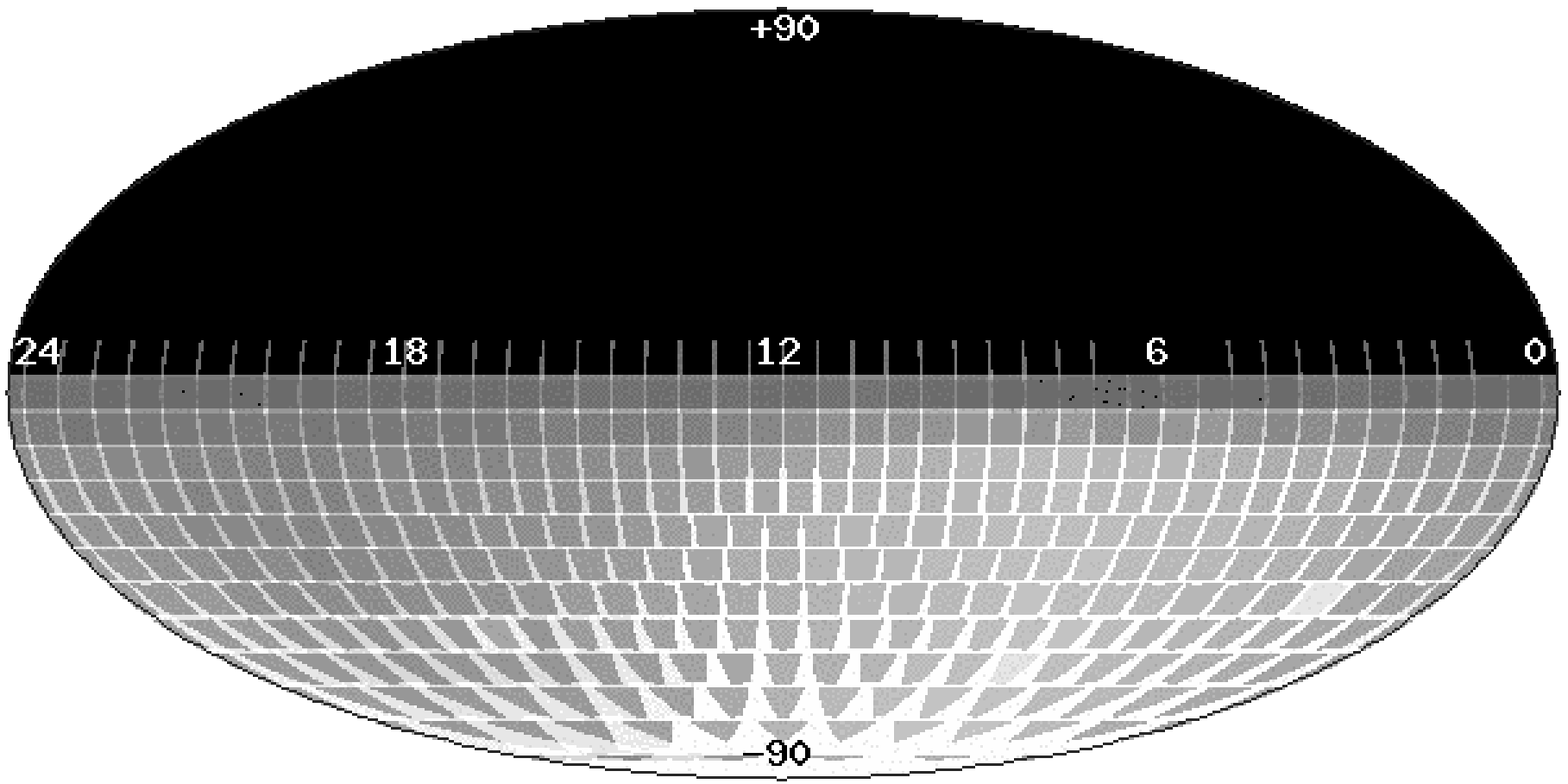}{70}{20}{10}

The whole sky was divided into 709 $8\arcd \times 8\arcd$ 
fields of which 422 (60\%)
with $\delta < +20\arcd$ can be observed from Las Campanas (up to 300 on a 
single
summer night). Due to the obscuration by the dome
only fields with $\delta < +2\arcd$  were observed by the $V$ instrument.

The routine observing schedule for each wide-field camera consists of 
cycling through the list of fields and picking up the one that is the
most suitable for observation at the moment. Selection algorithm was
not uniform enough, so equatorial fields were observed two times less often 
than the circumpolar ones.

Although the sky-flats are not best suited for the wide-field
instruments we are using them both as the first order correction for vignetting
and pixel-to-pixel sensitivity variation. Each night a series 
of sky-flat exposures is taken after the sunset, followed by the dark
exposures.

Between 160 and 200 frames per night in $V$-band are taken each night  
(depending on the season), enabling camera to cover the whole 
available sky in less than two days. The raw data stream of one ASAS-3 
instrument is about 1.5 GB per night. The loss-less 
compression reduces this stream to 0.7 GB per night (2-3 DAT-3 tapes per
month), so we are considering some lossy compression in future.

\section{Data Reduction}

The data reduction pipe-line used for ASAS 1-2 observations
was described by Pojma{\'n}ski (1998). ASAS-3 incorporates only some minor
changes to that schematics.

\FigurePs{6cm}{Difference between Hipparcos and ASAS-3 magnitudes
for all available stars.}{hip}{fig3.ps}{100}{-50}{-70}

Current optics gives sharp images and produces less scattered light than the
previous one. In fact images can be so sharp (FWHM $\sim$ 1.4-1.6 
pixels) that only aperture photometry can properly deal with them.
Therefore we are now making simultaneous photometry through five
apertures varying in size from 2 to 6 pixels in diameter.
Each aperture data is process separately, so one can use data obtained
with the smallest one for the faint objects and with the largest one
for the brightest objects in the catalog. 

Reducing ASAS-2 data we partly managed to correct photometry for saturation 
and bleeding and have included overexposed stars in the ASAS Catalog.
Unfortunately current CCD cameras do not allow for such correction
and current photometry of the saturated images is practically useless.

Less scattered light in the optics (partly due to the smaller opening: 
f/2.8 {\em vs.} f/1.8) have substantially improved the quality of the
sky-flats allowing us to obtain good first-order correction of the
vignetting. 

The astrometry is now based on the ACT catalog. Using 3-d order
polynomials in X and Y we usually obtain positional accuracy
better than 0.2 pixels ( $< 3$ arcsec).

The zero-point offset of our photometry is based on the Hipparcos 
(Perryman {\em et al.} 1997) data.
A few hundred Hipparcos stars are usually located in each 9x9 deg field,
We use them for precise offset calibration.  The rms
scatter between ASAS and Hipparcos calculated for all stars lying close to 
the frame center
is about 0.015 mag (Fig. \ref{hip}a) Close to the edges it 
is significantly larger 
(Fig. \ref{hip}b) due to systematic effects caused by non-perfect 
flat-fielding. Erros are most prominent (Fig. \ref{hip}c) when we compare 
brightness of the stars located
in the overlapping areas of the neighboring fields (Fig. \ref{hip}c).

The final catalog is divided in two parts now. The first, flexible one is
relatively slow, but is capable to store information about all sources
detected on the frame. This catalog is regularly analyzed and 
significant measurements are moved into the second,
fast-access part of the catalog. The remaining data (usually single
entries caused by the image flaws) are archived separately.

All data processing is normally done in real time on the 
instrument's computer, but due to the necessity of fine-tuning of 
many elements of the pipe-line we are reprocessing the data
from the tapes at the Warsaw head-quarters.

\section{Variability Search}

Data analyzed in this paper cover the period of one year (2001)
of unattended operation.  During that time
all southern sky has been covered with observations at least 50-150 times.
Over 50.000 frames have been collected in $V$-band. More than 10.000 
for $0^h - 6^h$ quarter of the Southern Hemisphere alone.
1.3 million stars have been measured in this area more than 40 times.
These stars were subject to variability analysis similar to that
performed for the ASAS-2 data (Pojma{\'n}ski 2000).

First, light curves for each star in the field were extracted and
median magnitude and dispersion  were calculated (in each aperture
separately). These were plotted on the magnitude-dispersion diagram  
(Fig. \ref{mag_disp}) and stars
lying above the 95\% centile were selected for the further 
analysis. 

For each suspected star AOV period analysis (Schwarzenberg-Czerny 1989) was
performed. Stars with statistics value larger than 10. were accepted.
The other stars were subject to long-term variability tests: variance
analysis in variable-length bins and trend analysis (average number of
consecutive observations showing the same direction of brightness
change). Proper thresholds for these observables were selected after
tests. Finally all stars showing dispersion above 99.9\% centile were
added to the set.

As many as 18.000 stars have passed selection criteria and had to be
further inspected. About 3.200 have passed visual inspection.
The large number of rejected stars was due to several reasons:
First - small number of observations, which usually forces AOV algorithm
to produce spurious (or at least hard to verify) frequencies.
Second - trends (slopes) in brightness changes of some stars which were partly 
produced by the increasing defocus of the optics in the course
of the last two months. We decided not to include any light variation
of this kind, although we are convinced some of these are real.
Third - saturation, which has disappeared for some stars after they've 
got defocused. 

The relative number of the variable stars detected so far (0.2\% of the total)
is 10 times lower than that obtained by ASAS 1-2 in the selected fields.
This has been caused by several factors. 

First, we have investigated
only stars above 95\% centile in the magnitude-dispersion diagram. while
previously we have inspected almost 75\% of the stars,
discovering many low-amplitude variables. This will be done also for
current data, but not earlier than more efficient algorithms for data 
selection are worked out.

Second, data span is one year now {\em vs.} almost three years previously.
Since we have also omitted most of the slowly varying objects much
lower number of long-term variables has been discoverd. 
Previously we have discoverd about 400 periodic stars  (0.26\%) and 
3400 long-term ones (2.2\%), while now about 2200 (0.16\%) periodic and
1000 (0.08\%) long-term.

Third, number of data points for each star is on average several times 
smaller than in ASAS-2 database.

Fourth, data sampling was different - a few times per night previously
{\em vs.} once per night now. This resulted in significant aliasing.
We will solve this problem in future
observing each field for some time in the "high frequency" mode - at
least several times per night. As for now we can only present our best
choice and try to improve lists available on the INTERNET as soon as new
data are available.

506 variable stars observed  by ASAS-2 are located in the currently analyzed  
region.  466 of them were observed by the ASAS-3 system. Most of the
others were faint in $I$ band, so probably were to faint for the  $V$
instrument. One was bright (053820-6937.4 $I\sim9.69$), but since it
is irregular variable it might have faded below our $V$ detection limit.
Only about 100 of 466 stars were detected to be variable in $V$-band by the 
present analysis.

62 other objects were selected by the detection algorithm but
were rejected after visual inspection. Only 3 of these were
of "periodic" type and in all cases primary rejection reason was small
number of points and low power in spectrum. 
For example ASAS~053936-7958.6 with an amplitude of
0.1 and only 40 data points was not found to be variable by the
present analysis, while ASAS-2 has easily detected its variability
with almost 4000 measurements. 

Most of the other light curves that did not
pass visual inspection were of the "long-term" type  and were rejected 
because their slow varaition or sparse data were not 
convincing enough.

The other 300 (65\%) stars were not even preselected by the selection
algorithm, again due to small number of data points or minuit 
brightness changes.
In most cases however, plotting their light-curves folded 
with the known period clearly reveals periodic variation.

\section{Variability Classification}

  Only 2.000 of the 18.000 stars in the initial set of the 
candidate variable stars were selected by  a quite robust long-term
selection algorithm. Almost 1.500 of them have passed visual inspection.

The other 16.000 stars were selected by the AOV algorithm, thus their 
folded light 
curves had to be inspected visually one by one. Unfortunately, the relative 
number of true variables dramatically decreases as the AOV statistics comes
closer to threshold, making verification process highly non-efficient.

Survey catalogs do not necessary need to provide object classification, but
their usefulness increases if they do. However, since volume of the data 
is growing rapidly, such step requires some automation
providing necessary speed, repeatability and consistency.

Several attempts have been made so far to create classification machines
based on neural networks and machine learning 
(e.g. Eyer and Blake 2002, Wozniak et al. 2002b) or on direct parametric 
analysis (e.g. Udalski \etal 1999a, for cepheids, Ruci{\'n}ski 1993, 1997,
Szyma{\'n}ski, Kubiak, Udalski 2001 for contact binaries).

Here we propose simplified approach that mimics nonlinear neural network 
behavior. Its parameters ("weights") can be easily defined in terms of
the light curve parameters. "Teaching" in this case is done by
defining the multidimensional solids describing selected classes.
Classification could then be done by selecting the closest solid in the
multidimensional parametric space. To simplify our task we use
carefully selected two-dimensional cross-section of that space.

Ruci{\'n}ski (1993, 1997) has shown that light curves of the contact 
configurations 
of W UMa systems can be easily described using only two coefficients, $a_2$
and $a_4$, of the cosine series $\sum a_i \cos (2\pi i \theta)$. We have
extended this approach tested behavior of the semi-detached and
detached configurations in the $a_2$-$a_4$ plane. 

Several thousands theoretical light curves were calculated using the
Wilson-Devinney (1971) code for a wide range of binary parameters: masses
$M_1, M_2$ temperatures $T_1, T_2$ and radii $R_1, R_2$ 
characteristic for the main sequence O...M stars, periods $P$
varying from the contact configuration to 100 days and inclinations $i$ in the
range of $60\arcd - 90\arcd$. For each combination of $M_1, M_2$ and $P$ we 
have varied $R_1$ and $R_2$ between the main sequence and the Roche lobe sizes.

\FigurePs{13cm}{Distribution of light curves in the $a_2$ - $a_4$ plane.
a) Data obtained from the Wilson-Devinney simulation are plotted.
Grey area encloses contact models calculated by Ruci{\'n}ski (1993). 
Two continuous lines
delineate contact and detached configurations, while
semi-detached systems are located between dashed lines. b) Eclipsing variables
from the OGLE Bulge database (Wozniak 2002a). c)  ASAS eclipsing binaries (this
paper). d) OGLE Bulge pulsating variables.}{a2a4}{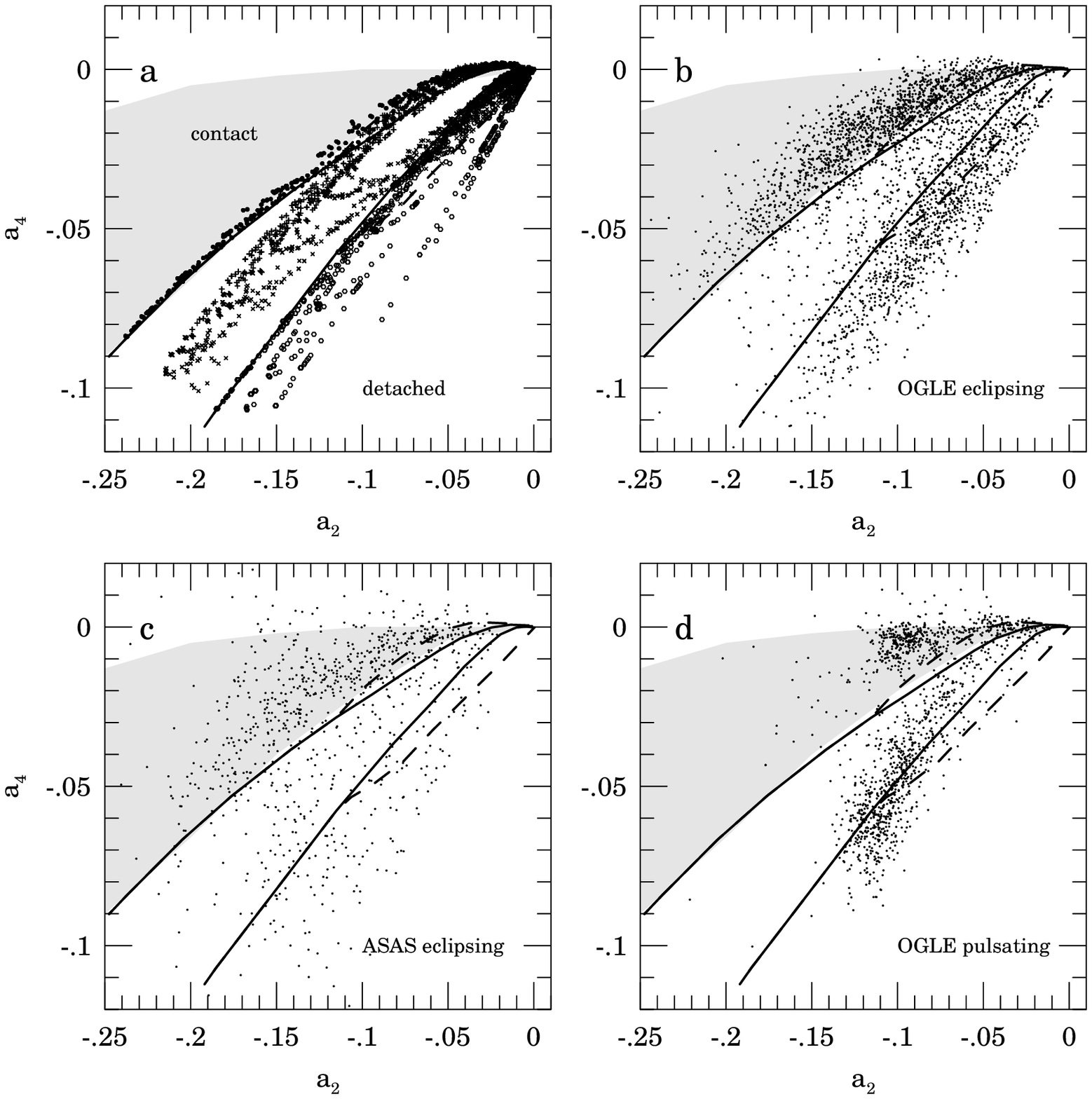}{70}{-10}{0}

\FigurePs{7cm}{Separation of eclipsing and pulsating variables in 
a) $b_2$ - $b_4$ and b) $a_4$ - $b_4 $ planes. Data plotted are
for the OGLE Bulge variables.}{b2b4}{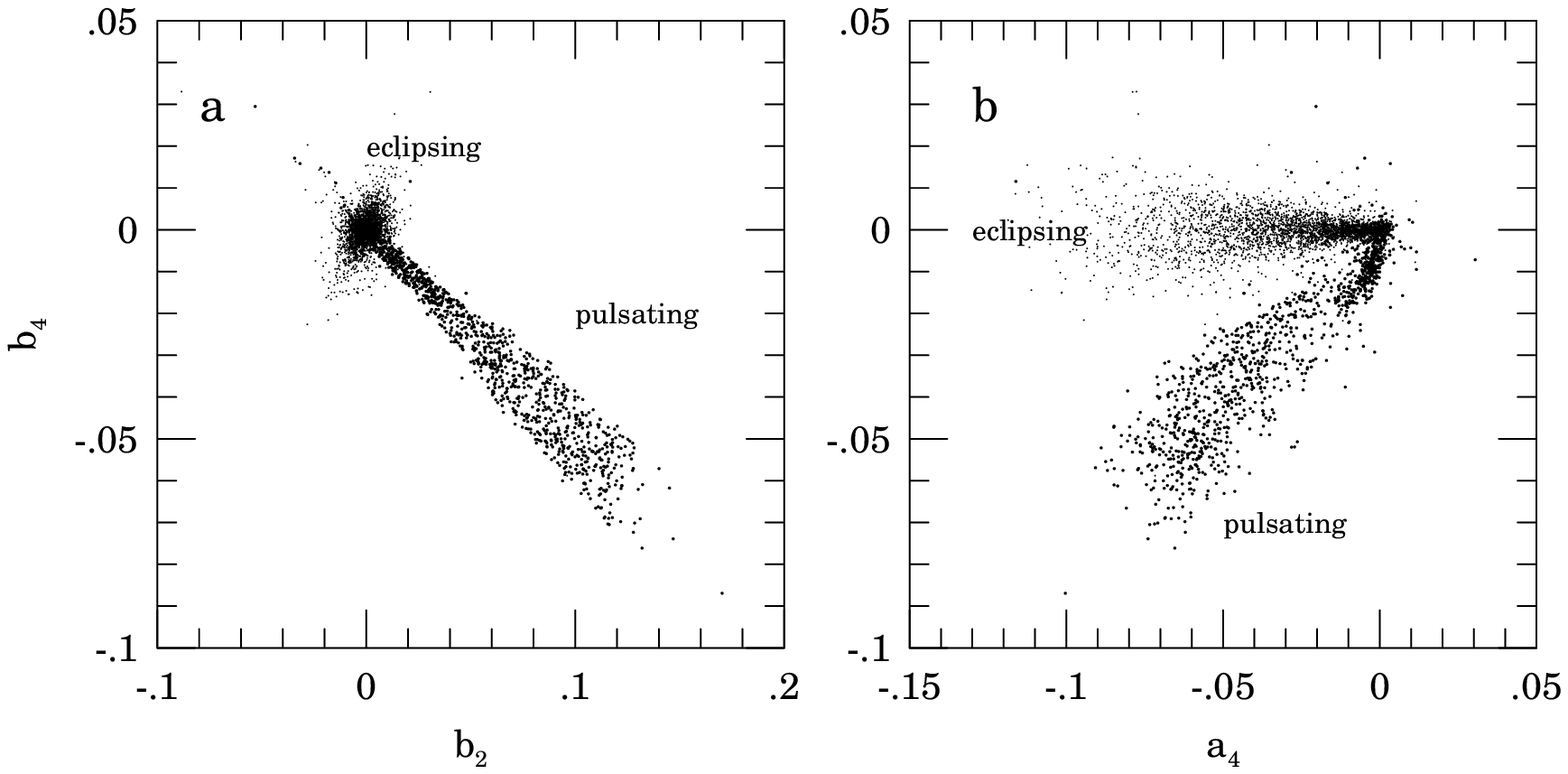}{70}{-10}{0}

Harmonic series of the form:
\beq
\sum_{i=1}^4 ( a_i\cos (2\pi i \varphi) + b_i\sin (2\pi i \varphi))
\eeq
was then fitted to all light curves.

Fig. \ref{a2a4}a shows coefficients $a_2$ and $a_4$. 
Filled and open dots represent
contact and detached configurations respectively,  while pluses and crosses
denote semi-detached configurations with primary and secondary filling its
Roche lobe, respectively. Grey area covers contact configurations 
calculated by Ruci{\'n}ski (1993, Fig. 6).  

Fig. \ref{a2a4}a clearly indicates that in most cases contact, semi-detached and
detached configurations can be unambiguously distinguished in the $a_2$-$a_4$
plane. However, looking at individual light curves one can note some
confusion introduced by widely used $EW$, $EB$, $EA$ classification of the light
curves:  contact or almost contact configurations do not necessary have 
minima of the same depth; there are frequent semi-detached configurations 
with equal minima; detached binaries do not necessary have flat maxima. 
Thus, although often used in such context, $EW$, $EB$ and $EA$ classes do not 
correspond entirely to the contact/detached discrimination. Since we feel
that the later is physically more interesting, we would prefer to 
recognize ``c'', ``sd'', and ``d'' configurations. Thus we will classify as
$EC$ contact configurations lying above the upper line in
Fig. \ref{a2a4}a, as $ED$ detached binaries located  below the lower line,
and as $ESD$ remaining semi-detached systems located between the dashed lines.

In case of real data some additional steps have to be taken before applying
Fourier decomposition. First the ``base'' frequency, $f_0$,  that
equals to the frequency of the eclipsing system and is half of the pulsation 
frequency for the pulsating stars has to be located in the power
spectrum. AOV spectrum puts most of the power either into $2f_0$,  
$f_0$ or even $2/3f_0$, depending on the light curve shape and distribution of
observations - complicating automation of the process. 

Fortunately a slightly
slower ORTPER code (AOV Multiharmonic Periodogram for Uneven Sampling, 
Schwarzenberg-Czerny, 1996) usually  properly selects $2f_0$
for the pulsating stars and $f_0$ or $2f_0$ for eclipsing light curves. 
The final test on  $f_0$ is done performing 
Fourier decomposition using $f_0$. In properly phased light curve  we
expect the $a_2$ term to be dominant, so if we obtain 
$a_4/a_2 > 1.$ we have to halve the initial value of $f_0$.

The second step necessary for direct comparison of the Fourier
coefficients is consistent determination of the zero phase. We do that by 
locating minimum on the reconstructed light curve.

In the present work we use only the first 4 harmonics. This reduces our
ability to recognize tiny features on the light curves in return
for smaller coefficient scatter. 

Large data set of uniform light curves is necessary to efficiently divide 
parametric volume into class solids. We have used 3 major sources of data:
our own (ASAS) light curves that were provisionally classified into $DSCT$, 
$RRAB$, $RRC$, $DCEP$, $EW$, $EB$ and $EA$ classes; 
OGLE SMC and LMC cepheids (Udalski \etal 1999a, 1999b) 
carefully divided into $DCEP_{FU}$ and $DCEP_{FO}$ (fundamental and first
overtone) classes; and subset of the
OGLE Bulge variables (Wozniak 2002a), that were provisionally divided into
pulsating and eclipsing groups.

In Fig. \ref{a2a4}b the OGLE Bulge eclipsing variables
(Wozniak 2002a) are plotted, while in  Fig. \ref{a2a4}c ASAS data from this 
paper. It is interesting that in case of OGLE data significant deficiency
of the semi-detached systems could be spotted.

In Fig. \ref{a2a4}d the OGLE Bulge pulsating stars are
plotted to show that $a_2$-$a_4$ plane cannot be used for separating
pulsating and eclipsing systems. Instead $b_2$-$b_4$ or $a_4$-$b_4$ coefficient
pairs (Fig. \ref{b2b4}ab) provide necessary segregation.

\FigurePs{13cm}{Distribution of light curve parameters in the 
$a_{42}$ - $\varphi_{42}$  and $log P$ - $a_{42}$  planes for OGLE (a,c) 
and ASAS (b,d) variables. Tiny dots in the upper panel are eclipsing
binaries, while larger ones - pulsating.}{af42}{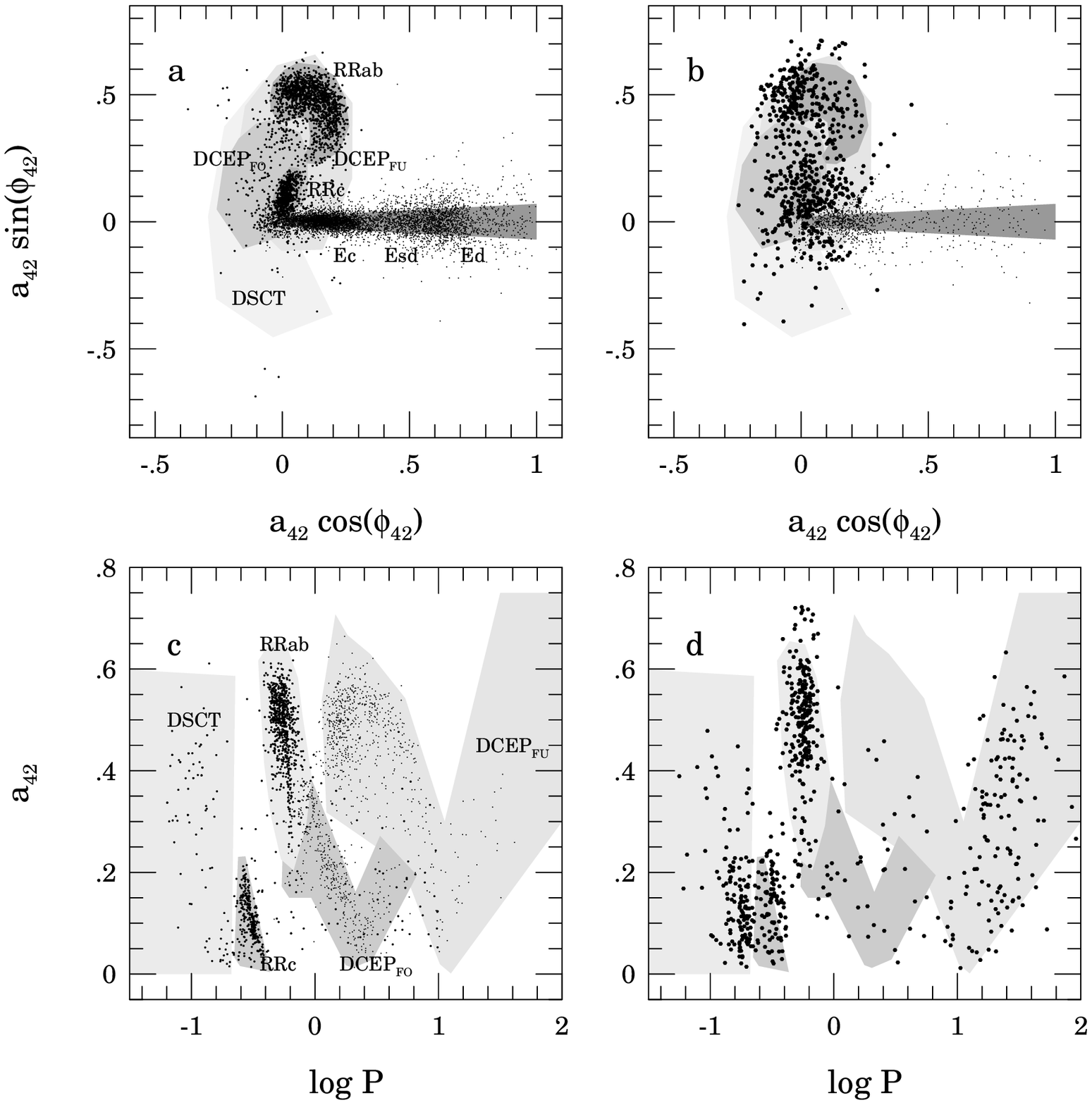}{70}{-10}{0}

Finally we need to recognize different types of pulsating stars. We have
done this using  $a_{42}=\frac{a_4}{a_2}$ {\em vs.} 
$\varphi_{42}=\varphi_4-\varphi_2$ 
(radial coordinates, Fig. \ref{af42}a,b)  and $\log P$ {\em vs.}
$a_{42}$ (Fig. \ref{af42}c,d) diagrams, on which categorized ASAS data and 
OGLE Bulge pulsating stars  were plotted. The later one is an equivalent of the
well known $R_{21}$ {\em vs.} $\log P$ diagram used for cepheid classification. 
Although ASAS data were scattered, they provided necessary clues to define 
sharp polygons enclosing different classes of OGLE Bulge variables in both
planes.

Besides the period $P$  three other non-Fourier parameters describing light
curves were needed to reduce the number of wrong classifications:
two describing difference in minima and maxima levels and the third
describing vertical asymmetry of the light curve (fraction of the
measurements brighter than the average). They were
necessary, because the fourth order harmonics could not properly recognize
narrow, unequal minima or sharp maxima.

Process of automated classification is now straightforward: 
each class is described by a set of polygons in several cross-sections 
(planes) of the
parametric space. In each plane score function is calculated: it equals
1.0 inside the polygon  and  decreases exponentially outside the polygon. 
The rate of decrease depends on the accuracy with which we determine the 
observable, so even large distances from the polygons could be accepted
for noisy data. All scores are multiplied and the class with the highest 
result is selected as a winner. If other classes have similar
scores (at least 75\% of the maximum) multiple classification is accepted.
If the highest score is below 0.50 uncertainty flag (:) is added, while
for very low scores (less then 0.2) general class of $PULSE$ is applied.
So far we have defined three eclipsing classes: $EC$, $ESD$, $ED$ and
six pulsating: $DSCT$, $RRAB$, $RRC$, $DCEP_{FU}$, $DCEP_{FO}$, $M$.

Such algorithm has many advantages. It is fast,  easy to verify and very
flexible. Adding new observables (like location in color-magnitude or
period-luminosity diagrams, spectral type  etc.) is instant and needs 
only defining a new polygon in the parametric space.
The obvious disadvantage is that only predefined classes could be
selected.

In this paper, using only one color photometric data, we are in fact
classifying data according to the  the light curve shape and period only.
Therefore caution must be taken not to over-interpret results:
Many $RRC$ objects might be in fact $EC$ or $DSCT$. The
algorithm will usually recognize and mark such situations.
$DCEP$ classes are based on SMC/LMC data and will not entirely correspond
to Galactic objects. $DCEP_{FU}$ are usually easily recognized by their
$RRAB$-like shapes, but around $\log P = 1.$ many sinusoidal cases
are encountered and probability of wrong classification increases.
$M$ (Mira) class was defined for objects with $\Delta V > 2.$ and $P >
70^{\rm d}$, so some SR, IR and even CV objects might have been also classified as $M$.

We have discovered many ``M'' shaped variables,
both in ASAS and OGLE data, with periods of tens of days. The algorithm
tends to classify them as $EC$ or $ESD$, implying they are
contact or semi-detached giant stars. In fact in case of OGLE variables
most of them are located close to the Red Clump (Udalski, 2002), so there is no
inconsistency. That is why we did not restrict $EC$ class to the short 
periods only. 

Another problem with automated classification is caused by semiregular
and irregular variables, which often show strong signal in the power
spectrum. Unless detected by other means these stars have a good chance to
be falsely classified. We have tried to define $SR$ and $IR$ classes
in terms of long-term variations in the light curve, but results are not
satisfactory yet. For the present analysis we decided to relay on the initial
visual inspection of light curves, during which we have selected about 1100
irregular ones that were marked as $MISC$ and were not subject to automated 
classification. These are mostly semiregular ($SR$), irregular ($IR$), slow 
irregular ($L$), $LBV$ and other less regularly changing stars with time scales
of variation between 10 and 200 days.

\section{The catalog}

Current list of candidate variable stars in the first quarter of the
Southern Hemisphere contains  3126 stars.

For each star the following data are provided: ASAS identification $ID$
(coded from the star's $\alpha_{2000}$ and $\delta_{2000}$ in the 
form: $hhmmss-ddmm.m$), period $P$ in days (or characteristic time
scale of variation for irregular objects), $T_0$ -  epoch of minimum 
brightness
(for eclipsingi stars) or maximum brightness (for pulsating stars), $V_{max}$ -
brightness at maximum , $\Delta V$ - amplitude of variation, $Type$ - one
of the predefined classes: $EC$, $ESD$, $ED$, $DSCT$, $RRC$, $RRAB$, 
$DCEP_{FU}$, $DCEP_{FO}$, $M$, $PULS$ and $MISC$.

67 stars marked with ``:'' were inspected visually, and in about 20 cases 
original classification was changed (in most cases for highly distorted
$ED$ curves, that were classified as pulsating ones). 
680 objects have multiple classification. For 330 cases this is
exclusively due to $EC/ESD$ or $ED/ESD$ confusion, but for 180 other this is a more
serious $EC/RRC$ or $EC/DSCT$ double classification. Quite often visual
inspection may help to remove such degeneracy.

$ELL$ and $SR$ suffixes were added in several (54 and 45) cases for 
almost purely sinusoidal light curves and for evident long term changes
respectively. 

Search for detected variables in the SIMBAD database revealed only
800 matches with stars known to be variable
(almost 200  of them are in the the SMC and LMC fields), so
over 75\% of stars are probably new detections.

Table \ref{vartab} contains a compact version of the
catalog. Only four columns are 
listed  for each star: identification $ID$, period $P$, $V$, and $\Delta V$. 
Column $ID$ also contains some 
flags - ":" if classification was uncertain, "?" if multiple classes
were assigned, "v" if SIMBAD lists a star to be variable.
Table \ref{tabvar} summarizes classification results.

Figs. 7 through 39 in the Appendix show thumbnail pictures of all
light curves. Only $ID$ is given for each. For periodic variables
phase in the range (-0.1 - 2.1) is plotted along the $x$-axis, while for 
Mira's and miscellaneous stars -  HJD in the range (2451800-2452300).
On the $y$-axis magnitudes are given. Larger ticks always mark
1 magnitude intervals and vertical span is never smaller than 1 mag.

The full catalog of variable stars observed by the ASAS system
containing more classification details and complete photometric data
is available over the INTERNET:\\
\centerline{ http://www.astrouw.edu.pl/$\sim$gp/asas/asas.html}
 or 
\centerline{http://archive.pinceton.edu/$\sim$asas}.

\tabcolsep 3pt
\input{vartab}
\tabcolsep 5pt
\MakeTable{|l|r|l|r|}{8cm}{\label{tabvar}
Number of various types of variable stars detected on 5000 sq. deg by ASAS-3 $V$ camera.}{
\hline
\multicolumn{1}{|c|}{Type} & \multicolumn{1}{c|}{Count} & \multicolumn{1}{c|}{Type} & \multicolumn{1}{c|}{Count}\\
\hline
$DCEP_{FU}$	& 192	& $EC$	& 655 \\
$DCEP_{FO}$	& 69	& $ESD$	& 250 \\
$DSCT$		& 148	& $ED$	& 141 \\
$RRAB$		& 247	& $M$	& 132 \\	
$RRC$		& 113 	& $MISC$& 1170 \\
$PULS$		&  9	&&\\
\hline
}

\section{Conclusions}

We have presented preliminary results of the search for variable stars
in the southern hemisphere. Only a quarter of available data covering
right ascension between $0^{\rm h}$ - $6^{\rm h}$ has been presented 
in this paper. Over 3000 variable stars were found so far
among 1,300,000 stars brighter than $V\sim 15$ mag.
Comparison to the ASAS-2 data suggests, that number of variable stars
should increase 3-10 times, as more observations are collected.

Most of these stars belong to the "periodic" class, as we have
not included many candidate objects with the "long-term" variation.
There is a large subset of bright ($V<12$ mag) detached binaries -
useful for distance scale calibration - perfect targets for small size 
spectroscopic instruments.

Fully automated classification algorithm that we have described in this paper,
efficiently separates stars into predefined classes. For a good quality data
its efficiency reaches 100\% in those regions of the parametric 
space, where the classes do not overlap. For poor data or overlapping regions
alternative classes are always proposed.
Currently our algorithm works properly only if the highest peak in the power 
spectrum corresponds to the true frequency or one of its harmonics, 
which is not necessary true for noisy, poorly sampled data.
What is still missing is a robust algorithm that will allow to separate
light curves into strictly periodic and irregular ones.

However, at this time we feel it is better to publish imperfect data and
to allow a broad range of astronomers to see the diversity of raw results
and to work on variety of scientific topics.  Note a recent paper by
Pietrukowicz (2001), who used the past ASAS data to determine period 
changes in cepheids in the LMC.  

Processing of the remaining ASAS data is in progress, so we plan to release 
next parts of this catalog soon. Also, four ASAS instruments are 
taking data, covering the whole southern sky each night.
Over 120,000 images have been collected so far in $V$, $I$  and
$R$ bands, covering the whole hemisphere more than 300 times. Soon
real time photometry will be achieved, finally allowing us to setup the Early
Warning System. 

\section{Acknowledgments}
This project was made possible by a generous gift from Mr. William
Golden to Dr. Bohdan Paczy{\'n}ski, and funds from Princeton University.  It is a
great pleasure to thank Dr. B. Paczy{\'n}ski for his initiative, interest,
valuable discussions, and the funding of this project.

I am indebted to the OGLE collaboration for the use of facilities of the
Warsaw telescope at LCO, for their permanent support and maintenance of the
ASAS instrumentation, and to The Observatories of the 
Carnegie Institution of Washington for providing
the excellent site for the observations.
I am especially indebted to Marcin Kubiak for providing selected light curves
from the  OGLE Bulge database.

This research has made use of the SIMBAD database,
operated at CDS, Strasbourg, France.

This work was partly supported by the KBN 2P03D01416 grant.

\vspace{1.5cm}
\begin{center}
References
\end{center}

\noindent
\begin{itemize}
\leftmargin 0pt
\itemsep -5pt
\parsep -5pt
\refitem{Eyer, L. and Blake, C.}{2002}{Radial and Nonradial Pulsations as Probes of Stellar Physics, ASP Conference Series}{Vol. 259}{160}
\refitem{Paczy{\'n}ski, B.}{1997}{~}{~}{``The Future of Massive Variability
  Searches'', in {\it Proceedings of 12th IAP Colloquium}: 
  ``Variable Stars and the Astrophysical Returns of Microlensing Searches'', 
  Paris (Ed. R. Ferlet), p.~357}
\refitem{Perryman, M.A.C. et al.}{ 1997}{ Astron. Astroph}{ 323}{ L49}
\refitem{Pojma{\'n}ski, G.}{1997}{Acta Astr.}{47}{467}
\refitem{Pojma{\'n}ski, G.}{1998}{Acta Astr.}{48}{35}
\refitem{Pojma{\'n}ski, G.}{2000}{Acta Astr.}{50}{177}
\refitem{Pojma{\'n}ski, G.}{2001}{in: {\it Small Telescope Astronomy on 
  Global Scales, ASP Conference Series Vol. 246,
  IAU Colloquium}}{183}{53}
\refitem{Pietrukowicz, P.}{2001}{Acta Astr.}{51}{247}
\refitem{Ruci{\'n}ski, S.}{1993}{PASP}{105}{1433}
\refitem{Ruci{\'n}ski, S.}{1997}{Astron. J.}{113}{1112}
\refitem{Schwarzenberg-Czerny, A.}{1989}{MNRAS}{241}{153}
\refitem{Schwarzenberg-Czerny, A.}{1996}{Astrophys. J.}{460}{L107}
\refitem{Szyma{\'n}ski, M., Kubiak, M., Udalski, A.}{2001}{Acta Astr.}{51}{259}
\refitem{Udalski, A., Soszy{\'n}ski, I., Szyma{\'n}ski, M., Kubiak, M.,
Pietrzy{\'n}ski, G., Wozniak, P., {\.Z}ebru{\'n}, K.}{1999a}{Acta Asr.}{49}{223}
\refitem{Udalski, A., Soszy{\'n}ski, I., Szyma{\'n}ski, M., Kubiak, M.,
Pietrzy{\'n}ski, G., Wozniak, P., {\.Z}ebru{\'n}, K.}{1999b}{Acta Astr.}{49}{437}
\refitem{Udalski, A.}{2002}{private communication.}{}{}
\refitem{Wilson, R. E. and Devinney, E. J.}{1971}{Astrophys. J.}{166}{605}
\refitem{Wozniak, P., Udalski, A., Szyma{\'n}ski, M., Kubiak, M., 
Pietrzy{\'n}ski, G., Soszy{\'n}ski, I., {\.Z}ebru{\'n}, K.}{2002a}{Acta Astr.}{52}{129}
\refitem{Wozniak, P. {\em et al.}}{2002b}{American Astronomical Society Meeting}{199}{130.04}
\end{itemize}
\clearpage
\input{appendix_short}
\end{document}

%% file: include
\newcommand{\etal}{{\it et al.\ }}

\newcommand{\beq}{\begin{equation}
  \renewcommand{\int}{\intop\limits}
  \renewcommand{\oint}{\ointop\limits}}
\newcommand{\eeq}{\end{equation}}
\newcommand{\beqarr}{\par\begin{minipage}{11cm} \begin{eqnarray*}}
\newcommand{\eeqarr}{\end{eqnarray*} \end{minipage} \hfill 
   \stepcounter{equation}{\rm (\theequation)}\vspace{3mm}\linebreak}
\newcommand{\bdm}{\begin{displaymath}
  \renewcommand{\int}{\intop\limits}
  \renewcommand{\oint}{\ointop\limits}}
\newcommand{\edm}{\end{displaymath}}

\newcommand{\up}[1]{\ifmmode^{\rm #1}\else$^{\rm #1}$\fi}

\newcommand{\arcd}{\ifmmode^{\circ}\else$^{\circ}$\fi}
\newcommand{\arcm}{\ifmmode{'}\else$'$\fi}
\newcommand{\arcs}{\ifmmode{''}\else$''$\fi}

\newcounter{pagefrom}
\newcounter{pageto}
\newcounter{volume}
\newcounter{year}

\newenvironment{Titlepage}{
\vspace*{2cm}
  \begin{center}
}{
  \end{center}\par\vspace{3mm}
}

\newcommand{\Title}[1]{{\large\bf\boldmath #1 \\[3mm] {\footnotesize by} 
\\[3mm]}}

\newcommand{\Author}[2]{{\large\spaceskip 2pt plus 1pt minus 1pt #1}\\[3mm]
   {\small #2}\\[6mm]}

\newcommand{\Received}[1]{}

\newcommand{\Abstract}[2]{{\footnotesize\begin{center}ABSTRACT\end{center}
\vspace{1mm}\par#1\par
\noindent
{\bf Key words:~~}{\it #2}}}

\newcommand{\FigCap}[1]{\footnotesize\par\noindent Fig.\  % 
  \refstepcounter{figure}\thefigure. #1\par}

\newcommand{\TabCap}[2]{\begin{center}\parbox[t]{#1}{\begin{center}
  \small {\spaceskip 2pt plus 1pt minus 1pt T a b l e}
  \refstepcounter{table}\thetable \\[2mm]
  \footnotesize #2 \end{center}}\end{center}}

\newcommand{\TableFont}{\footnotesize}

\newcommand{\MakeTable}[4]{\begin{table}[htb]\TabCap{#2}{#3}
  \begin{center} \TableFont \begin{tabular}{#1} #4 
  \end{tabular}\end{center}\end{table}}

\newcommand{\MakeTableSep}[4]{\begin{table}[p]\TabCap{#2}{#3}
  \begin{center} \TableFont \begin{tabular}{#1} #4 
  \end{tabular}\end{center}\end{table}}

{
\footnotesize \frenchspacing

\renewcommand{\and}{{\rm and }}
\section{{\rm REFERENCES}}
\sloppy \hyphenpenalty10000
\begin{list}{}{\leftmargin1cm\listparindent-1cm
\itemindent\listparindent\parsep0pt\itemsep0pt}}%
{\end{list}\vspace{2mm}}

\def\TYLDA{~}
\newlength{\DW}
\newcommand{\refitem}[5]{\item[]{#1} #2%
\def\REFARG{#3}\ifx\REFARG\TYLDA\else, {\it#3}\fi
\def\REFARG{#4}\ifx\REFARG\TYLDA\else, {\bf#4}\fi
\def\REFARG{#5}\ifx\REFARG\TYLDA\else, {#5}\fi.}

%% file: vartab
\renewcommand{\TableFont}{\tiny}
\MakeTableSep{|l|r|r|r||l|r|r|r||l|r|r|r|}{10cm}{\label{vartab}
ASAS Catalog of Variable Stars. 0$^{\rm h}$ - 6$^{\rm h}$ Quarter of the Southern Hemisphere.}{
\hline
\multicolumn{1}{|c|}{ID} & \multicolumn{1}{c|}{$P$} & \multicolumn{1}{c|}{$V$} & \multicolumn{1}{c|}{$\Delta~V$} & \multicolumn{1}{|c|}{ID} & \multicolumn{1}{c|}{$P$} & \multicolumn{1}{c|}{$V$} & \multicolumn{1}{c|}{$\Delta~V$} & \multicolumn{1}{|c|}{ID} & \multicolumn{1}{c|}{$P$} & \multicolumn{1}{c|}{$V$} & \multicolumn{1}{c|}{$\Delta~V$}\\
\hline
\multicolumn{12}{|c|}{\em  Stars classified as EC.}\\
000108-3330.1~ & 0.46658 & 11.51 & 0.43 & 000202-6653.3~ & 0.326576 & 12.16 & 0.60 & 000309-3456.3~ & 0.265418 & 12.40 & 0.47\\
000336-1452.4~ & 1.05465 & 10.93 & 0.32 & 000424-7437.9~? & 0.47113 & 12.42 & 0.22 & 000425-5346.4~ & 0.288260 & 13.82 & 0.97\\
000504-3727.0~? & 0.40251 & 12.43 & 0.32 & 000514-0732.6~? & 0.38617 & 12.57 & 0.97 & 000622-7621.8~? & 0.35196 & 13.65 & 0.66\\
000650-3537.5~ & 0.38318 & 12.19 & 0.35 & 000815-2414.7~? & 0.262427 & 13.25 & 0.51 & 000909-6650.8~ & 0.329324 & 12.39 & 0.44\\
000932-6905.4~ & 0.36725 & 11.68 & 0.31 & 001215-0718.6~? & 0.76778 & 11.77 & 0.36 & 001229-1052.5~? & 0.76042 & 11.43 & 0.39\\
001313-5258.7~ & 0.299992 & 12.04 & 0.36 & 001356-3820.4~? & 0.87903 & 12.26 & 0.31 & 001447-3914.6~v & 0.36436 & 11.81 & 0.84\\
001508-5335.4~ & 0.239485 & 13.44 & 0.69 & 001547-3105.7~v & 0.320823 & 11.72 & 0.68 & 001721-7155.0~v & 0.59483 & 10.06 & 0.51\\
001750-5223.0~ & 0.38384 & 13.36 & 0.55 & 001846-3940.6~v & 0.41361 & 9.19 & 0.28 & 001913-5608.9~ & 0.289702 & 13.17 & 0.64\\
001913-6302.2~? & 0.262686 & 13.43 & 1.03 & 002021-2519.3~? & 0.41273 & 11.32 & 0.24 & 002102-7547.6~ & 0.239260 & 12.10 & 0.79\\
002112-6251.6~? & 0.42100 & 12.23 & 0.49 & 002123-3925.1~ & 0.326428 & 12.89 & 0.76 & 002322-4226.8~? & 0.62833 & 11.14 & 0.20\\
002328-2041.8~ & 0.41469 & 9.86 & 0.47 & 002329-6831.5~? & 0.67250 & 12.23 & 0.24 & 002345-2223.9~? & 0.286220 & 13.13 & 0.32\\
002449-2744.3~ & 0.313670 & 12.90 & 0.73 & 002539-4229.1~ & 0.44929 & 10.16 & 0.20 & 002806-1648.0~ & 0.46808 & 12.71 & 0.53\\
002811-7026.8~ & 1.01637 & 10.80 & 0.41 & 002821-1453.3~ & 0.40266 & 11.54 & 0.44 & 002821-2904.1~ & 0.269896 & 12.00 & 0.55\\
002857-0812.7~? & 0.267421 & 12.59 & 0.45 & 003009-3114.8~? & 0.343184 & 12.60 & 0.54 & 003225-2452.6~v & 0.56628 & 10.02 & 0.28\\
003336-4349.3~v & 0.42240 & 11.06 & 0.69 & 003406-5103.0~ & 174 & 9.58 & 0.17 & 003514-4143.2~? & 0.74857 & 9.47 & 0.29\\
003628-2540.4~v & 0.51155 & 10.31 & 0.68 & 003647-1958.0~? & 0.336706 & 13.83 & 0.73 & 004009-2002.9~? & 0.46068 & 12.70 & 0.31\\
004111-4344.9~? & 0.36523 & 13.38 & 0.45 & 004114-1800.2~? & 0.207484 & 13.72 & 0.70 & 004240-2956.7~ & 0.301682 & 11.05 & 0.42\\
004312-5008.7~? & 0.304819 & 13.56 & 0.58 & 004618-3108.6~ & 0.54671 & 11.53 & 0.50 & 004625-1602.9~? & 0.89908 & 12.73 & 0.41\\
004634-3331.0~ & 0.299138 & 12.59 & 0.83 & 004647-6002.5~ & 0.46625 & 12.84 & 0.49 & 004717-1941.6~ & 0.48881 & 11.31 & 0.40\\
004801-4240.0~ & 0.42970 & 12.40 & 0.62 & 004843-6143.7~?v & 1.05386 & 9.39 & 0.20 & 004857-3718.6~ & 0.37503 & 11.35 & 0.48\\
005008-4130.5~ & 0.335422 & 11.74 & 0.34 & 005016-6318.5~? & 0.273454 & 13.25 & 0.58 & 005521-4427.7~ & 0.340220 & 12.56 & 0.43\\
005530-1106.6~ & 0.54326 & 10.13 & 0.52 & 005543-0205.6~v & 0.52242 & 10.43 & 0.56 & 005702-7925.7~? & 0.55230 & 12.13 & 0.32\\
005735-5246.0~ & 0.291417 & 12.16 & 0.33 & 005830-3659.9~ & 14.55 & 11.13 & 0.28 & 005848-3027.1~ & 0.41897 & 13.42 & 0.82\\
005853-0642.9~? & 0.318030 & 11.76 & 0.28 & 005856-7941.3~ & 0.292275 & 13.24 & 0.52 & 010007-0744.5~ & 0.70443 & 11.60 & 0.41\\
010055-7511.9~ & 0.55933 & 12.70 & 0.63 & 010303-4531.3~ & 0.350845 & 13.15 & 0.61 & 010327-1210.4~ & 0.58662 & 10.21 & 0.24\\
011027-1749.2~ & 0.34232 & 13.23 & 0.70 & 011031-1804.1~ & 0.44542 & 11.93 & 0.44 & 011054-1857.6~ & 0.46669 & 12.07 & 0.41\\
011113-7910.6~v & 0.284032 & 12.28 & 0.87 & 011116-1425.3~ & 0.60605 & 10.99 & 0.18 & 011213-1956.6~? & 0.39336 & 11.26 & 0.27\\
011236-6429.5~ & 0.37226 & 11.71 & 0.48 & 011243-8051.1~ & 42.7 & 10.44 & 0.23 & 011315-6411.6~? & 9.550 & 10.33 & 0.19\\
011509-6155.2~? & 0.45601 & 12.69 & 0.31 & 011520-5539.2~? & 0.37359 & 12.65 & 0.38 & 011612-0659.6~ & 0.324740 & 12.66 & 0.63\\
011638-3942.5~v & 0.37992 & 10.32 & 0.52 & 011816-4853.2~ & 0.40504 & 12.96 & 0.46 & 011833-4217.5~ & 0.293548 & 11.49 & 0.20\\
011910-2746.2~ & 0.39666 & 12.30 & 0.48 & 012040-5702.5~? & 0.39063 & 11.70 & 0.30 & 012133-2907.8~v & 0.42290 & 10.44 & 0.45\\
012244-7328.6~? & 0.75718 & 12.52 & 0.23 & 012336-2515.9~? & 0.291046 & 12.00 & 0.27 & 012350-6558.5~? & 0.38472 & 12.83 & 0.50\\
012450-3241.4~ & 0.308971 & 11.45 & 0.58 & 012513-1144.3~ & 0.35288 & 13.29 & 0.80 & 012755-2301.7~v & 0.58621 & 11.06 & 0.63\\
012841-5037.0~? & 0.93595 & 12.05 & 0.26 & 012917-7243.4~ & 181 & 11.59 & 0.74 & 013000-3040.5~ & 0.284678 & 12.90 & 0.53\\
013003-1257.4~ & 0.309407 & 13.16 & 0.77 & 013107-4359.6~v & 0.71987 & 12.79 & 0.79 & 013416-0724.7~ & 0.73471 & 11.85 & 0.43\\
013549-7546.0~? & 0.95085 & 11.60 & 0.10 & 013624-6500.6~ & 0.41796 & 11.99 & 0.42 & 013711-3459.3~v & 0.46426 & 11.02 & 0.96\\
013733-0208.4~ & 0.83909 & 11.76 & 0.74 & 014226-4557.0~v & 0.234724 & 12.65 & 0.67 & 014255-2007.5~ & 0.36595 & 11.11 & 0.37\\
014434-7401.7~? & 0.256202 & 13.64 & 0.68 & 014447-7500.1~? & 0.320917 & 12.30 & 0.23 & 014450-7536.8~ & 0.39699 & 13.59 & 0.74\\
014605-4934.1~? & 0.351035 & 10.32 & 0.22 & 014613-4656.9~? & 10.889 & 9.32 & 0.09 & 014656-0945.1~?v & 0.48596 & 10.92 & 0.73\\
014725-1704.0~ & 0.282927 & 11.86 & 0.35 & 014838-5718.6~? & 52.7 & 11.42 & 0.16 & 014854-2053.6~v & 0.316849 & 10.40 & 0.74\\
014957-0811.8~? & 0.44858 & 12.25 & 0.62 & 015017-7733.4~? & 0.268746 & 13.25 & 0.57 & 015024-4658.4~v & 0.306617 & 12.23 & 0.55\\
015053-2515.3~ & 0.43173 & 12.46 & 0.53 & 015216-2026.9~ & 0.290812 & 13.24 & 0.69 & 015311-2105.7~? & 1.4777 & 11.45 & 0.14\\
015438-0409.2~ & 0.38764 & 11.54 & 0.45 & 015531-2633.1~ & 0.47923 & 11.77 & 0.37 & 015536-2432.8~ & 0.42424 & 12.17 & 0.35\\
015606-0044.3~v & 0.74078 & 10.96 & 0.50 & 015705-5051.0~ & 191 & 11.01 & 0.39 & 015749-2740.4~ & 0.341711 & 13.07 & 0.56\\
015911-1743.7~? & 0.44780 & 13.65 & 0.64 & 015915-7059.5~ & 0.292020 & 11.58 & 0.44 & 015931-1130.8~? & 0.47145 & 12.19 & 0.31\\
015937-0331.0~ & 0.63152 & 9.35 & 0.33 & 020012-1812.5~v & 0.79044 & 10.58 & 1.03 & 020119-3704.9~v & 0.308128 & 9.86 & 0.46\\
020156-4218.7~? & 1.8660 & 10.96 & 0.13 & 020332-3005.1~ & 0.36814 & 10.68 & 0.35 & 020340-2500.8~? & 0.74346 & 13.44 & 0.70\\
020800-2858.7~? & 0.42204 & 11.98 & 0.32 & 020933-5945.1~ & 0.36842 & 12.13 & 0.26 & 021147-7139.0~ & 0.39085 & 11.30 & 0.24\\
021229-6911.5~ & 0.36128 & 10.93 & 0.31 & 021306-3247.0~? & 0.311751 & 11.14 & 0.14 & 021503-6549.1~? & 0.297448 & 13.16 & 0.55\\
021544-2024.2~? & 0.41046 & 13.76 & 0.74 & 021552-3643.2~ & 0.346348 & 12.27 & 0.73 & 021625-4312.2~? & 0.41656 & 12.71 & 0.34\\
021649-4145.2~ & 0.37148 & 10.25 & 0.33 & 021900-2305.5~? & 0.61105 & 12.75 & 0.31 & 021940-6138.2~ & 0.50904 & 9.64 & 0.45\\
022014-0252.0~ & 0.64346 & 10.61 & 0.40 & 022133-4216.2~ & 0.251578 & 12.86 & 0.44 & 022153-1740.7~v & 0.338960 & 11.59 & 0.69\\
022211-6343.9~? & 0.297511 & 12.20 & 0.61 & 022358-1247.7~? & 0.80069 & 10.51 & 0.19 & 022406-5125.2~ & 0.39060 & 12.56 & 0.46\\
022508-4239.5~ & 0.40408 & 11.41 & 0.52 & 022517-2354.2~ & 0.36267 & 12.47 & 0.68 & 022556-4921.9~ & 0.35990 & 12.01 & 0.80\\
022611-4313.7~ & 0.260948 & 13.26 & 0.64 & 022810-0659.4~ & 0.36984 & 11.50 & 0.33 & 022830-2848.2~? & 0.54228 & 12.98 & 0.28\\
022846-0229.3~ & 0.58467 & 11.56 & 0.72 & 022902-3025.8~ & 0.38652 & 10.04 & 0.27 & 022917-7745.3~? & 0.60214 & 11.49 & 0.23\\
022938-7149.0~? & 0.289622 & 12.51 & 0.41 & 023056-1644.7~ & 1.6531 & 11.66 & 0.40 & 023130-1252.4~ & 0.258565 & 13.75 & 0.78\\
023135-1620.6~? & 0.292432 & 13.62 & 0.63 & 023152-3837.4~ & 0.58873 & 9.91 & 0.38 & 023215-5002.6~? & 0.327647 & 11.91 & 0.29\\
023304-3627.9~? & 0.74267 & 12.55 & 0.30 & 023309-1717.6~ & 0.290404 & 13.09 & 0.63 & 023432-3515.2~ & 0.35629 & 12.60 & 0.62\\
023525-1518.7~ & 2.8709 & 14.12 & 1.08 & 023830-4648.0~? & 0.66836 & 9.64 & 0.36 & 023833-1417.9~v & 0.44079 & 9.47 & 0.58\\
023848-1859.1~? & 9.959 & 12.15 & 0.26 & 023930-7456.6~ & 0.42418 & 12.06 & 0.27 & 023949-2618.1~ & 0.49485 & 10.65 & 0.34\\
024010-4555.0~ & 0.88986 & 10.67 & 0.41 & 024055-6644.1~? & 0.50557 & 12.64 & 0.28 & 024144-2601.8~v & 0.61790 & 13.13 & 0.63\\
024145-6734.4~? & 0.35583 & 12.00 & 0.30 & 024207-5940.0~ & 0.42135 & 11.79 & 0.19 & 024429-7636.8~?: & 0.45929 & 11.36 & 0.57\\
024507-1807.9~ & 0.36493 & 10.94 & 0.47 & 024508-2821.1~ & 0.86285 & 13.35 & 0.60 & 024526-5941.9~ & 0.42882 & 12.48 & 0.52\\
024536-2932.1~? & 0.325294 & 12.92 & 0.31 & 024557-2918.8~? & 0.38751 & 12.55 & 0.63 & 024612-4017.9~ & 0.53651 & 12.44 & 0.34\\
024633-7010.1~? & 0.41533 & 13.02 & 0.36 & 024746-5105.1~ & 0.77569 & 10.12 & 0.36 & 024750-7539.6~? & 0.61363 & 12.29 & 0.41\\
024811-1518.1~ & 0.36658 & 12.76 & 0.71 & 024828-2732.7~ & 0.75220 & 9.96 & 0.27 & 024840-3309.2~ & 0.35205 & 12.82 & 0.77\\
024959-1634.7~ & 0.38264 & 12.62 & 0.47 & 025016-4649.2~? & 0.271753 & 12.48 & 0.53 & 025018-3241.8~ & 0.322037 & 12.16 & 0.79\\
025115-2525.4~ & 0.55927 & 10.14 & 0.23 & 025211-6925.7~ & 0.42122 & 10.70 & 0.45 & 025616-2336.5~ & 0.35219 & 12.49 & 0.47\\
025618-2705.5~? & 1.2576 & 14.60 & 1.02 & 025619-7431.1~? & 0.336582 & 12.80 & 0.47 & 025643-5210.4~ & 0.37723 & 12.88 & 0.59\\
025853-4308.9~ & 0.40598 & 12.65 & 0.49 & 025931-3238.9~v & 0.72604 & 10.32 & 0.63 & 025958-1627.4~ & 0.84475 & 10.13 & 0.32\\
030005-1108.5~ & 0.77337 & 12.21 & 0.29 & 030031-7707.7~? & 0.56410 & 13.36 & 0.42 & 030050-4412.2~ & 0.81368 & 11.57 & 0.31\\
030218-0238.0~? & 0.92650 & 11.54 & 0.27 & 030235-5625.1~ & 0.42804 & 12.71 & 0.55 & 030254-1554.1~ & 1.01565 & 10.07 & 0.31\\
030313-2036.9~v & 0.334978 & 12.56 & 0.57 & 030315-2311.2~ & 0.45660 & 11.72 & 0.47 & 030459-4214.0~? & 0.44585 & 11.54 & 0.22\\
030521-0619.9~? & 0.41130 & 11.26 & 0.34 & 030615-6521.6~? & 0.81988 & 10.22 & 0.16 & 030617-6812.5~ & 0.41612 & 9.51 & 0.46\\
030701-5608.1~v & 0.62511 & 10.96 & 0.78 & 030703-2602.5~ & 0.52669 & 10.03 & 0.28 & 030707-7935.2~ & 0.36216 & 13.52 & 0.54\\
030735-5721.9~? & 0.37646 & 11.85 & 0.41 & 030827-1609.1~ & 0.47637 & 12.99 & 0.52 & 030953-0653.6~v & 0.44528 & 11.15 & 0.56\\
031133-3257.8~ & 0.45131 & 13.46 & 0.47 & 031141-0043.8~ & 0.79359 & 12.22 & 0.56 & 031235-1931.9~v & 0.62856 & 12.47 & 0.44\\
031246-1704.0~ & 0.38469 & 13.11 & 0.47 & 031252-0744.3~ & 0.36011 & 12.11 & 0.72 & 031341-8045.4~ & 0.292302 & 12.77 & 0.44\\
031506-1707.9~? & 0.94768 & 13.26 & 0.29 & 031509-5144.2~? & 21.46 & 9.75 & 0.13 & 031529-1146.2~ & 0.42979 & 10.82 & 0.28\\
031619-6718.9~? & 0.43043 & 13.59 & 0.56 & 031649-1956.5~? & 12.97 & 10.00 & 0.19 & 031740-5318.6~? & 0.39285 & 12.48 & 0.31\\
031751-3615.5~ & 0.39508 & 12.05 & 0.67 & 031808-8231.4~? & 78.8 & 11.12 & 0.23 & 031909-3507.0~? & 2.2613 & 11.22 & 0.14\\
031916-3842.1~ & 0.49110 & 10.18 & 0.45 & 031928-1944.0~ & 0.37306 & 13.10 & 0.59 & 032027-1650.5~? & 0.307546 & 11.86 & 0.33\\
032038-5902.4~v & 0.35496 & 13.64 & 0.68 & 032053-1703.5~? & 0.76784 & 11.73 & 0.24 & 032202-5112.8~ & 8.746 & 12.16 & 0.23\\
\hline
}
\clearpage

\addtocounter{table}{-1}
\MakeTableSep{|l|r|r|r||l|r|r|r||l|r|r|r|}{10cm}{Continued}{
\hline
\multicolumn{1}{|c|}{ID} & \multicolumn{1}{c|}{$P$} & \multicolumn{1}{c|}{$V$} & \multicolumn{1}{c|}{$\Delta~V$} & \multicolumn{1}{|c|}{ID} & \multicolumn{1}{c|}{$P$} & \multicolumn{1}{c|}{$V$} & \multicolumn{1}{c|}{$\Delta~V$} & \multicolumn{1}{|c|}{ID} & \multicolumn{1}{c|}{$P$} & \multicolumn{1}{c|}{$V$} & \multicolumn{1}{c|}{$\Delta~V$}\\
\hline
\multicolumn{12}{|c|}{\em  Stars classified as EC.}\\
032212-4001.0~ & 0.335329 & 12.66 & 0.64 & 032351-4024.3~? & 0.55273 & 11.78 & 0.23 & 032439-0209.5~ & 0.82571 & 12.31 & 0.53\\
032455-0559.5~? & 35.3 & 12.98 & 0.34 & 032513-4002.7~ & 0.40473 & 12.64 & 0.59 & 032521-2037.6~? & 0.69259 & 14.18 & 0.80\\
032629-3101.3~ & 0.63867 & 10.81 & 0.43 & 032635-5229.5~? & 0.36929 & 13.11 & 0.42 & 032709-5333.1~ & 0.96640 & 11.26 & 0.17\\
032736-7250.9~ & 0.309847 & 10.54 & 0.71 & 032754-5152.7~? & 0.52510 & 13.88 & 0.60 & 032812-2503.5~ & 0.315501 & 11.37 & 0.65\\
033142-5927.4~ & 0.37799 & 12.17 & 0.44 & 033148-2631.7~ & 0.300162 & 13.89 & 0.73 & 033205-7259.0~? & 0.47139 & 12.18 & 0.28\\
033527-7445.0~? & 0.40170 & 12.98 & 0.56 & 033543-2021.1~ & 0.327086 & 11.60 & 0.70 & 033623-0755.6~ & 0.56809 & 11.69 & 0.52\\
033702-4131.7~v & 0.292340 & 9.59 & 0.30 & 033710-7053.9~? & 0.41435 & 11.18 & 0.18 & 033725-0126.7~ & 0.47504 & 11.09 & 0.38\\
033809-1913.4~ & 0.341545 & 12.52 & 0.67 & 033856-3633.0~? & 0.38980 & 12.34 & 0.28 & 034135-1143.0~? & 0.40725 & 13.52 & 0.67\\
034202-5145.2~ & 0.40921 & 10.82 & 0.51 & 034215-2008.8~ & 0.282038 & 12.26 & 0.74 & 034327-3937.4~?v & 80.0 & 9.54 & 0.11\\
034340-4505.2~? & 0.35408 & 13.51 & 0.55 & 034347-7727.6~ & 0.48606 & 10.61 & 0.35 & 034444-6105.8~ & 0.284815 & 9.49 & 0.37\\
034703-0704.2~ & 0.37772 & 11.04 & 0.42 & 034809-5839.8~ & 0.42924 & 10.48 & 0.34 & 034901-1656.5~ & 0.42369 & 12.76 & 0.44\\
034911-4236.9~v & 0.82762 & 11.06 & 0.64 & 034927-0609.7~? & 0.53867 & 13.39 & 0.74 & 034931-0431.2~ & 0.48018 & 9.03 & 0.40\\
034936-2441.2~ & 0.49377 & 12.63 & 0.45 & 034949-0552.5~? & 0.36264 & 12.36 & 0.32 & 035014-1739.1~ & 0.35215 & 12.19 & 0.41\\
035020-8017.4~ & 0.62241 & 11.94 & 0.45 & 035036-6056.2~? & 0.85882 & 11.96 & 0.79 & 035120-0331.4~ & 0.49090 & 12.38 & 0.46\\
035153-1031.8~v & 0.50764 & 8.30 & 0.51 & 035200-2155.8~ & 0.335164 & 10.69 & 0.43 & 035232-3111.7~ & 0.36753 & 12.35 & 0.58\\
035248-2926.2~ & 0.314441 & 9.92 & 0.35 & 035307-1701.2~ & 0.46187 & 10.97 & 0.34 & 035320-7948.6~?v & 0.41244 & 9.07 & 0.18\\
035324-2902.4~ & 0.33933 & 13.74 & 0.51 & 035422-3605.6~ & 0.290447 & 13.52 & 0.53 & 035444-1456.1~?v & 0.63220 & 9.37 & 0.78\\
035454-1330.7~? & 0.86815 & 13.05 & 0.45 & 035511-4617.3~ & 0.316715 & 12.40 & 0.29 & 035559-3513.7~ & 0.46435 & 11.48 & 0.48\\
035643-1927.3~ & 0.316366 & 13.56 & 0.43 & 035658-1451.2~? & 0.68161 & 12.83 & 0.46 & 035710-7414.6~ & 1.0943 & 12.89 & 0.32\\
035734-5421.5~v & 0.49010 & 8.29 & 0.69 & 035756-2623.1~? & 0.41021 & 12.14 & 0.38 & 035851-5110.6~ & 0.310653 & 9.45 & 0.32\\
035943-4112.0~? & 2.0989 & 11.61 & 0.17 & 035951-4944.6~? & 0.42445 & 12.59 & 0.25 & 040053-4434.3~ & 0.71782 & 12.36 & 0.43\\
040300-5750.7~? & 0.339716 & 11.87 & 0.42 & 040444-1624.8~ & 0.49260 & 10.69 & 0.45 & 040455-3351.8~? & 0.46068 & 10.16 & 0.23\\
040508-0939.7~? & 0.34841 & 12.94 & 0.56 & 040528-6536.2~ & 0.294770 & 11.91 & 0.40 & 040531-4341.7~ & 0.340220 & 12.56 & 0.43\\
040550-5402.6~ & 0.36012 & 11.51 & 0.36 & 040633-4729.4~ & 0.40637 & 11.38 & 0.37 & 040637-2740.1~v & 0.63848 & 10.16 & 0.59\\
040756-3848.9~ & 0.40244 & 12.13 & 0.41 & 040822-0804.5~? & 0.42002 & 11.10 & 0.17 & 040855-6620.6~ & 0.38658 & 12.88 & 0.47\\
040908-6007.5~? & 0.97527 & 11.75 & 0.35 & 040911-4906.9~ & 0.285999 & 12.84 & 0.51 & 041025-2747.8~? & 0.38639 & 13.03 & 0.78\\
041037-3855.7~ & 0.42693 & 10.55 & 0.33 & 041049-4949.8~ & 0.335015 & 13.06 & 0.27 & 041138-4438.0~v & 0.89456 & 9.42 & 0.21\\
041148-1147.5~v & 0.41692 & 11.51 & 0.77 & 041208-1335.4~? & 0.45942 & 11.96 & 0.23 & 041209-1028.2~v & 0.321499 & 8.37 & 0.67\\
041233-5843.1~? & 37.0 & 9.85 & 0.16 & 041321-3613.1~ & 0.37046 & 11.87 & 0.33 & 041339-2433.5~? & 0.45701 & 12.74 & 0.25\\
041344-2804.1~ & 0.63348 & 11.50 & 0.32 & 041419-4626.9~v & 0.311635 & 11.48 & 0.86 & 041436-5441.9~ & 0.45771 & 12.13 & 0.37\\
041450-0340.1~? & 0.41925 & 12.46 & 0.39 & 041451-6814.2~ & 0.297281 & 12.47 & 0.39 & 041635-2506.3~ & 0.38747 & 11.36 & 0.41\\
041731-6649.6~? & 0.53059 & 12.79 & 0.30 & 041805-7547.1~? & 0.47112 & 13.03 & 0.35 & 041813-0615.6~ & 0.327760 & 12.43 & 0.39\\
041826-2157.9~? & 0.41593 & 12.27 & 0.41 & 041938-3004.7~ & 0.346748 & 11.45 & 0.51 & 041946-0518.0~ & 0.35383 & 12.66 & 0.53\\
042031-2813.9~ & 0.40040 & 12.06 & 0.43 & 042041-2905.1~? & 0.292692 & 13.86 & 0.66 & 042049-0451.3~ & 0.49090 & 11.35 & 0.34\\
042052-3945.7~? & 0.311885 & 11.98 & 0.20 & 042103-2629.5~ & 0.39586 & 9.99 & 0.36 & 042134-3825.8~? & 0.37736 & 11.73 & 0.36\\
042149-4317.5~? & 4.728 & 10.21 & 0.08 & 042158-1756.5~? & 0.40819 & 12.01 & 0.21 & 042215-1004.6~? & 0.295518 & 10.26 & 0.20\\
042317-1302.7~ & 0.67951 & 10.75 & 0.34 & 042401-5006.4~? & 0.310360 & 11.52 & 0.30 & 042509-1937.1~ & 0.57386 & 12.96 & 0.48\\
042518-2347.4~ & 0.57131 & 13.57 & 0.65 & 042534-4152.1~ & 0.42528 & 11.19 & 0.28 & 042539-3803.6~v & 0.94376 & 9.93 & 0.19\\
042549-6030.7~? & 0.56389 & 12.39 & 0.34 & 042559-2129.0~v & 0.331925 & 10.71 & 0.63 & 042609-3420.9~ & 0.48694 & 13.47 & 0.53\\
042708-1238.8~ & 0.35763 & 12.46 & 0.55 & 042716-2135.0~? & 0.60109 & 10.54 & 0.26 & 042745-5812.0~ & 0.90899 & 12.57 & 0.31\\
042844-4258.2~? & 0.61788 & 13.72 & 0.59 & 042851-4035.3~? & 0.312817 & 10.93 & 0.15 & 042925-3334.6~v & 0.63419 & 9.98 & 0.47\\
043046-4813.9~ & 0.35714 & 12.69 & 0.64 & 043128-0043.9~v & 0.46561 & 12.46 & 0.57 & 043204-2401.2~ & 0.50540 & 12.37 & 0.85\\
043249-7345.5~ & 0.74382 & 12.01 & 0.37 & 043325-2356.3~v & 0.62358 & 8.62 & 0.30 & 043345-7429.2~v & 0.43503 & 13.13 & 0.53\\
043348-0754.7~ & 0.85136 & 12.08 & 0.23 & 043423-3634.2~ & 0.83006 & 11.04 & 0.27 & 043433-0919.2~ & 0.46476 & 11.37 & 0.62\\
043437-3152.3~? & 0.84037 & 12.54 & 0.28 & 043450-1905.7~? & 0.55988 & 10.72 & 0.24 & 043559-0450.1~? & 0.58349 & 11.24 & 0.35\\
043639-0923.2~? & 0.35378 & 12.78 & 0.64 & 043842-0249.9~ & 0.67095 & 11.65 & 0.48 & 043852-0931.9~ & 0.37875 & 12.02 & 0.47\\
043933-5627.6~ & 0.53523 & 12.72 & 0.38 & 043933-7146.6~ & 1.6319 & 11.18 & 0.35 & 043940-3402.1~ & 0.54515 & 11.67 & 0.50\\
043953-2900.3~ & 0.53056 & 12.12 & 0.27 & 043953-3604.0~ & 0.76315 & 10.52 & 0.33 & 043954-5634.2~? & 0.35473 & 13.16 & 0.57\\
044005-3027.5~? & 0.330880 & 13.27 & 0.69 & 044007-6131.5~ & 0.47276 & 11.89 & 0.45 & 044017-2410.7~ & 0.41632 & 13.75 & 0.73\\
044026-6617.9~? & 0.35466 & 12.80 & 0.38 & 044040-7149.6~? & 0.65980 & 10.44 & 0.20 & 044053-4326.3~ & 0.91595 & 13.51 & 0.66\\
044106-0121.5~? & 0.221495 & 10.51 & 0.27 & 044120-2400.9~ & 0.40537 & 13.28 & 0.97 & 044216-2549.5~ & 0.254893 & 12.65 & 0.59\\
044308-1631.5~ & 0.294096 & 13.51 & 0.63 & 044336-8448.6~v & 1.1494 & 9.16 & 0.53 & 044439-2423.1~? & 0.41627 & 12.96 & 0.45\\
044447-5427.0~ & 0.309680 & 10.92 & 0.35 & 044516-7060.0~? & 0.49562 & 9.80 & 0.22 & 044536-2353.0~ & 0.33105 & 13.10 & 0.50\\
044624-4936.3~ & 0.284723 & 13.16 & 0.62 & 044625-6620.8~? & 0.258164 & 10.93 & 0.23 & 044639-4737.6~? & 0.308405 & 13.11 & 0.34\\
044655-5903.6~ & 0.49273 & 13.26 & 0.62 & 044659-1437.4~ & 0.52725 & 10.91 & 0.60 & 044813-5348.1~v & 0.49692 & 12.19 & 0.78\\
044853-0911.9~ & 0.48963 & 13.71 & 0.89 & 044907-4953.9~? & 1.1197 & 10.05 & 0.16 & 044908-4908.1~ & 0.40361 & 12.19 & 0.36\\
044925-1259.3~? & 0.47345 & 13.11 & 0.47 & 045022-1815.0~ & 0.323727 & 11.78 & 0.45 & 045101-1214.4~? & 1.6922 & 9.52 & 0.10\\
045151-2934.3~ & 0.38150 & 12.56 & 0.52 & 045156-1629.8~ & 0.345823 & 10.90 & 0.32 & 045211-2511.7~ & 0.57889 & 11.41 & 0.31\\
045218-0345.4~? & 0.48120 & 12.59 & 0.41 & 045319-4313.5~ & 0.39325 & 12.02 & 0.36 & 045322-0322.9~v & 0.61402 & 11.43 & 0.43\\
045343-4951.5~? & 0.42383 & 12.25 & 0.18 & 045417-5924.7~? & 0.294544 & 13.17 & 0.36 & 045450-1135.6~? & 0.43104 & 12.98 & 0.52\\
045540-1930.3~? & 0.80573 & 11.82 & 0.36 & 045555-2619.5~? & 0.34877 & 13.70 & 0.81 & 045657-3308.0~ & 0.48253 & 10.60 & 0.63\\
045706-5516.7~ & 0.41321 & 12.25 & 0.61 & 045707-7207.9~? & 0.41839 & 12.08 & 0.45 & 045713-3535.5~? & 0.65228 & 12.90 & 0.40\\
045720-3708.6~ & 0.34152 & 12.88 & 0.76 & 045722-8022.9~ & 0.36693 & 12.34 & 1.06 & 045724-1927.1~ & 0.42337 & 13.33 & 0.55\\
045822-4930.2~ & 0.47132 & 12.81 & 0.34 & 045841-8301.8~? & 0.95872 & 13.24 & 0.46 & 045858-2238.7~? & 0.46070 & 9.45 & 0.20\\
045920-1758.9~? & 0.315347 & 13.07 & 0.78 & 045948-6508.4~? & 0.343027 & 11.20 & 0.21 & 050048-7029.9~v & 0.38735 & 11.35 & 0.43\\
050056-3007.9~? & 1.4572 & 12.98 & 0.40 & 050135-0838.6~ & 0.313364 & 11.31 & 0.99 & 050204-0248.1~ & 0.321983 & 11.05 & 0.72\\
050233-2359.0~ & 0.320560 & 13.41 & 0.63 & 050308-0622.0~ & 0.58479 & 11.52 & 0.39 & 050334-2521.9~ & 0.41406 & 11.09 & 0.31\\
050417-1922.8~ & 0.286919 & 11.70 & 0.46 & 050425-2051.3~ & 1.11780 & 12.65 & 0.37 & 050437-2432.3~ & 0.42866 & 10.41 & 0.32\\
050505-0733.7~v & 0.92650 & 11.12 & 0.59 & 050512-5812.2~? & 0.323526 & 12.78 & 0.56 & 050530-6059.0~? & 38.8 & 9.98 & 0.16\\
050537-5755.6~ & 0.55781 & 11.48 & 0.25 & 050611-0546.3~ & 0.85811 & 11.28 & 0.35 & 050617-2007.8~v & 0.44864 & 9.45 & 0.33\\
050617-2619.3~ & 0.36020 & 13.05 & 0.53 & 050618-2007.9~v & 0.44864 & 9.45 & 0.33 & 050631-3832.6~ & 0.45560 & 11.30 & 0.46\\
050631-4833.3~? & 0.302175 & 13.41 & 0.48 & 050653-2611.2~? & 0.98917 & 11.30 & 0.13 & 050656-2936.0~ & 0.39837 & 9.89 & 0.62\\
050734-5421.3~ & 0.47623 & 12.96 & 0.31 & 050750-2319.2~? & 1.07747 & 11.63 & 0.15 & 050904-2727.2~? & 0.340474 & 13.52 & 0.44\\
050905-0741.7~? & 0.298088 & 13.17 & 0.75 & 050909-7707.9~ & 0.36932 & 12.80 & 0.78 & 050922-1932.5~ & 0.270842 & 12.30 & 0.41\\
050937-5618.5~ & 1.03555 & 10.53 & 0.41 & 050957-2305.4~ & 0.42874 & 11.45 & 0.39 & 051006-1146.0~ & 0.60336 & 12.73 & 0.58\\
051009-5650.3~ & 0.66997 & 11.11 & 0.27 & 051016-6133.1~ & 0.41446 & 12.14 & 0.40 & 051101-6709.9~?v & 1.2358 & 12.96 & 0.35\\
051114-0833.4~v & 0.42340 & 9.32 & 0.61 & 051130-5035.0~? & 0.54940 & 14.36 & 0.54 & 051134-1034.4~ & 0.59444 & 11.25 & 0.45\\
051135-3120.4~? & 21.37 & 9.86 & 0.09 & 051207-2919.9~ & 0.41311 & 12.33 & 0.65 & 051228-3741.5~? & 0.346387 & 12.42 & 0.25\\
051247-2308.1~ & 0.43369 & 13.36 & 0.67 & 051301-2131.7~ & 0.35608 & 11.92 & 0.46 & 051305-6624.7~? & 0.280563 & 13.19 & 0.38\\
051332-8719.7~ & 0.38408 & 12.42 & 0.51 & 051339-3301.3~ & 0.76465 & 9.35 & 0.19 & 051348-4027.2~ & 0.67572 & 10.45 & 0.30\\
051353-1701.2~ & 0.341836 & 11.66 & 0.55 & 051405-2802.8~ & 0.309576 & 13.20 & 0.56 & 051452-2741.4~? & 0.51350 & 12.85 & 0.28\\
051459-7356.3~ & 0.34572 & 11.78 & 0.39 & 051521-5509.2~ & 0.54527 & 13.49 & 0.69 & 051527-5555.0~? & 0.43582 & 13.14 & 0.45\\
051531-2845.0~v & 0.67234 & 9.56 & 0.39 & 051543-5624.7~ & 0.42900 & 12.69 & 0.35 & 051649-2524.7~? & 0.40425 & 13.35 & 0.64\\
051700-5555.4~ & 0.79037 & 10.32 & 0.51 & 051733-5447.2~ & 1.2018 & 11.12 & 0.30 & 051738-3639.7~ & 0.44432 & 11.66 & 0.41\\
\hline
}
\clearpage

\addtocounter{table}{-1}
\MakeTableSep{|l|r|r|r||l|r|r|r||l|r|r|r|}{10cm}{Continued}{
\hline
\multicolumn{1}{|c|}{ID} & \multicolumn{1}{c|}{$P$} & \multicolumn{1}{c|}{$V$} & \multicolumn{1}{c|}{$\Delta~V$} & \multicolumn{1}{|c|}{ID} & \multicolumn{1}{c|}{$P$} & \multicolumn{1}{c|}{$V$} & \multicolumn{1}{c|}{$\Delta~V$} & \multicolumn{1}{|c|}{ID} & \multicolumn{1}{c|}{$P$} & \multicolumn{1}{c|}{$V$} & \multicolumn{1}{c|}{$\Delta~V$}\\
\hline
\multicolumn{12}{|c|}{\em  Stars classified as EC.}\\
051757-1146.1~?v & 9.085 & 11.57 & 0.14 & 051758-3714.8~ & 0.36139 & 13.37 & 0.50 & 051800-1730.5~ & 0.58491 & 11.45 & 0.33\\
051801-6041.4~? & 0.248267 & 10.36 & 0.12 & 051832-6813.6~v & 0.285461 & 11.00 & 0.71 & 051901-4556.6~? & 0.45727 & 11.89 & 0.23\\
051954-3554.1~v & 0.56519 & 10.97 & 0.57 & 052025-0630.2~ & 0.39026 & 13.47 & 0.49 & 052042-1215.1~? & 0.307820 & 13.00 & 0.36\\
052046-3232.5~? & 0.52963 & 12.82 & 0.35 & 052109-4449.3~ & 0.55460 & 11.10 & 0.40 & 052117-0943.5~? & 0.46695 & 10.24 & 0.23\\
052121-3540.6~? & 0.37074 & 11.71 & 0.31 & 052126-2959.6~? & 0.97938 & 10.37 & 0.10 & 052134-6335.2~ & 0.47379 & 12.15 & 0.38\\
052201-6852.3~?v & 49.8 & 13.17 & 0.48 & 052214-7156.3~v & 0.77662 & 10.69 & 0.61 & 052234-4635.3~ & 0.44342 & 11.33 & 0.47\\
052313-0907.7~ & 0.40198 & 12.18 & 0.41 & 052446-1301.2~? & 0.306843 & 13.22 & 0.59 & 052452-2809.2~ & 0.275778 & 10.61 & 0.37\\
052615-0642.4~? & 1.07166 & 12.59 & 0.47 & 052630-1950.2~ & 0.319652 & 13.52 & 0.65 & 052642-4337.1~ & 0.245713 & 12.96 & 0.81\\
052650-8135.2~v & 0.46166 & 8.17 & 0.49 & 052651-6121.3~ & 0.55720 & 12.87 & 0.68 & 052746-2357.9~? & 0.61936 & 11.86 & 0.39\\
052754-3125.9~ & 0.312883 & 12.88 & 0.40 & 052819-1931.3~ & 0.66466 & 9.65 & 0.38 & 052851-3010.2~ & 0.302101 & 11.31 & 0.41\\
052941-1742.8~? & 0.229750 & 13.29 & 0.52 & 053030-1844.5~? & 0.51502 & 11.64 & 0.17 & 053042-4815.1~? & 10.710 & 9.82 & 0.12\\
053056-3844.3~? & 0.92980 & 10.25 & 0.19 & 053107-0757.7~? & 0.44333 & 11.88 & 0.43 & 053113-3610.1~ & 0.330847 & 12.72 & 0.63\\
053115-5222.0~ & 0.38654 & 11.37 & 0.47 & 053121-0723.7~? & 0.51746 & 11.95 & 0.29 & 053122-1540.1~ & 0.36159 & 12.96 & 0.73\\
053123-5812.2~ & 0.344282 & 11.81 & 0.66 & 053126-0058.6~? & 0.85222 & 10.39 & 0.23 & 053211-2302.0~ & 0.47408 & 12.73 & 0.41\\
053259-1354.0~ & 0.308853 & 12.72 & 0.58 & 053355-1204.4~? & 0.40681 & 11.13 & 0.20 & 053420-2606.6~ & 0.298081 & 14.24 & 0.96\\
053458-3825.4~? & 0.261736 & 12.86 & 0.79 & 053558-4208.7~? & 0.74261 & 12.91 & 0.51 & 053602-2926.9~ & 0.36670 & 12.27 & 0.47\\
053610-1827.2~? & 0.47277 & 12.97 & 0.33 & 053654-5304.9~ & 0.51747 & 11.53 & 0.26 & 053717-2339.7~ & 0.41111 & 11.85 & 0.39\\
053722-2618.0~ & 0.35757 & 12.46 & 0.50 & 053728-7218.9~?v & 0.52777 & 12.04 & 0.88 & 053750-3756.2~? & 0.43477 & 12.73 & 0.37\\
053756-2358.0~? & 0.317048 & 13.69 & 0.97 & 053758-1602.9~ & 0.323175 & 12.12 & 0.55 & 053808-3314.9~ & 0.36942 & 12.97 & 0.50\\
053817-2731.2~ & 0.319101 & 11.83 & 0.39 & 053845-0820.0~ & 0.344270 & 11.92 & 0.73 & 053854-2315.8~ & 0.50440 & 12.47 & 0.65\\
053905-1458.0~? & 0.97490 & 11.58 & 0.21 & 053920-4733.5~? & 0.33565 & 13.35 & 0.78 & 053930-0808.9~? & 0.44399 & 10.14 & 0.33\\
053959-5343.7~? & 10.954 & 11.90 & 0.32 & 054000-6828.7~v & 0.36222 & 11.38 & 0.57 & 054010-4214.5~ & 0.76226 & 11.80 & 0.40\\
054050-1126.0~ & 0.65397 & 12.29 & 0.30 & 054134-6044.1~ & 0.34702 & 12.15 & 0.68 & 054156-2440.1~ & 0.76047 & 13.11 & 0.68\\
054159-1129.1~ & 0.307690 & 12.87 & 0.44 & 054203-2730.2~ & 0.45606 & 12.56 & 0.38 & 054220-5929.0~? & 0.38661 & 13.72 & 0.66\\
054236-2215.9~ & 0.53189 & 13.43 & 0.59 & 054241-5316.5~ & 0.232730 & 12.87 & 0.84 & 054311-5845.6~ & 0.43708 & 10.12 & 0.45\\
054354-0243.6~ & 0.43810 & 11.57 & 0.43 & 054546-1249.9~? & 0.70491 & 12.18 & 0.39 & 054651-2615.5~? & 0.47535 & 13.39 & 0.39\\
054829-2807.5~ & 1.1694 & 11.95 & 0.29 & 054905-2554.4~ & 0.313061 & 11.82 & 0.50 & 054911-1902.6~ & 0.314238 & 13.74 & 0.84\\
054926-2352.5~? & 59.7 & 10.80 & 0.19 & 055000-0725.9~ & 0.322599 & 12.22 & 0.84 & 055019-5520.3~ & 0.41350 & 13.34 & 0.65\\
055209-2742.8~ & 0.90342 & 12.28 & 0.27 & 055303-2322.5~ & 0.33670 & 12.51 & 0.45 & 055317-5629.1~ & 0.39317 & 11.43 & 0.30\\
055321-1703.7~ & 0.33784 & 13.55 & 0.55 & 055426-1233.5~ & 1.08746 & 11.12 & 0.37 & 055501-7241.6~v & 0.343841 & 10.29 & 0.82\\
055558-0840.6~ & 0.78805 & 12.96 & 0.52 & 055624-5919.5~ & 0.61917 & 11.88 & 0.45 & 055708-2738.5~ & 0.53583 & 12.74 & 0.78\\
055827-1739.8~ & 0.41459 & 10.15 & 0.39 & 060011-1549.8~ & 0.41690 & 12.02 & 0.40 & 060239-1108.5~? & 0.312823 & 10.88 & 0.19\\
060244-1927.2~ & 0.40953 & 13.18 & 0.68 & 060501-1028.3~? & 0.42417 & 13.18 & 0.29 & 060646-2522.3~? & 0.32537 & 13.25 & 0.55\\
060706-1712.5~ & 0.91461 & 12.99 & 0.48 & 060935-2119.9~? & 0.90967 & 11.09 & 0.30 & 061019-2333.4~ & 0.77329 & 11.63 & 0.64\\
061059-2012.7~v & 1.0703 & 9.57 & 0.79 & 061627-7426.8~ & 0.63045 & 10.76 & 0.39 & 062252-7501.9~ & 0.257701 & 11.46 & 0.46\\
064047-8815.4~ & 0.43863 & 11.82 & 0.45 & 073106-8625.2~ & 1.4893 & 12.20 & 0.77 & 084617-8833.7~ & 0.267129 & 12.59 & 0.65\\
100123-8813.5~? & 0.65226 & 11.97 & 0.28 & 115933-8546.0~?v & 0.61128 & 11.61 & 0.27 & 123244-8726.4~v & 0.338519 & 11.41 & 0.74\\
131308-8528.6~?v & 0.55245 & 11.51 & 0.27 & 213300-8325.2~ & 9.266 & 10.94 & 0.23 & 221220-8453.8~ & 0.52799 & 11.99 & 0.26\\
223203-7852.9~ & 0.279247 & 11.73 & 0.54 & 223709-8728.8~ & 0.84838 & 11.67 & 0.60 & 230401-8045.7~? & 37.6 & 10.72 & 0.17\\
231745-8239.4~ & 0.45619 & 12.42 & 0.39 & 232348-7454.0~ & 0.334050 & 11.38 & 0.38 & 232416-6222.1~v & 0.35773 & 9.79 & 0.50\\
232658-6337.0~? & 0.75639 & 13.50 & 0.46 & 233214-8250.8~ & 0.325393 & 12.91 & 0.57 & 233356-6911.2~ & 0.95370 & 11.87 & 0.28\\
233607-5632.1~ & 0.322861 & 13.05 & 0.53 & 233735-6512.9~? & 0.35841 & 13.31 & 0.53 & 233819-7742.4~? & 0.279352 & 13.59 & 0.81\\
233830-4512.4~ & 0.40390 & 11.97 & 0.39 & 233858-3830.3~? & 0.48675 & 11.83 & 0.30 & 234239-6055.7~ & 0.268751 & 13.20 & 0.63\\
234307-1852.5~ & 0.48626 & 11.09 & 0.26 & 234336-3206.7~v & 0.42409 & 11.98 & 0.50 & 234558-2113.7~ & 0.303694 & 11.78 & 0.48\\
234716-7842.3~ & 0.78917 & 13.85 & 0.58 & 234718-0805.2~v & 0.48141 & 10.38 & 0.35 & 234750-6604.3~ & 0.38054 & 13.05 & 0.54\\
234823-4054.7~ & 0.34720 & 12.85 & 0.37 & 234933-1851.7~ & 0.37430 & 13.57 & 0.63 & 235006-8444.3~ & 0.62847 & 12.90 & 0.59\\
235103-0055.8~ & 0.52159 & 12.80 & 0.52 & 235103-1904.5~ & 0.47078 & 11.90 & 0.42 & 235151-4111.3~ & 0.327418 & 13.16 & 0.47\\
235239-2135.2~? & 1.01014 & 10.96 & 0.24 & 235354-1923.1~? & 0.63267 & 13.67 & 0.79 & 235424-5756.5~ & 0.58420 & 11.47 & 0.47\\
235451-2806.6~ & 0.41360 & 11.47 & 0.22 & 235620-6411.5~ & 0.38893 & 12.60 & 0.35 & 235620-6535.6~? & 0.49013 & 11.97 & 0.34\\
235621-7058.8~ & 0.326685 & 12.36 & 0.51 & 235749-5027.0~ & 0.43088 & 13.63 & 0.49 & 235818-3356.8~ & 0.276060 & 12.59 & 0.88\\
235900-0931.6~ & 0.256220 & 13.64 & 0.51 &  &  &  &  &  &  &  & \\
\multicolumn{12}{|c|}{\em  Stars classified as ESD.}\\
000053-1717.5~? & 0.297986 & 12.66 & 0.59 & 000147-5714.5~? & 0.47035 & 11.03 & 0.20 & 000221-2929.6~? & 6.294 & 12.32 & 0.40\\
000229-5653.9~? & 26.55 & 10.15 & 0.11 & 000955-2808.6~? & 0.220859 & 11.70 & 0.15 & 001109-5150.3~? & 0.63829 & 13.41 & 0.77\\
002156-2240.4~? & 0.54045 & 11.95 & 0.12 & 002221-2936.8~ & 0.64658 & 12.11 & 0.39 & 002257-2039.4~v & 0.76730 & 10.35 & 0.32\\
002525-5741.0~? & 0.36059 & 12.18 & 0.20 & 002541-3449.7~? & 0.268644 & 9.39 & 0.11 & 002655-3952.9~?v & 0.75534 & 8.94 & 0.48\\
002657-5844.8~? & 0.352222 & 11.38 & 0.27 & 002807-7922.0~v & 0.70110 & 13.08 & 0.78 & 002811-5919.4~ & 1.9463 & 9.67 & 0.34\\
002845-5732.9~ & 25.40 & 10.49 & 0.19 & 002904-1713.0~v & 0.62162 & 9.87 & 0.17 & 003355-1608.7~? & 0.84474 & 12.22 & 0.38\\
003513-3256.0~ & 0.50537 & 11.60 & 1.03 & 003537-4041.2~? & 0.78426 & 12.90 & 0.45 & 003544-2514.0~? & 1.3004 & 9.76 & 0.10\\
003831-1914.6~? & 1.1381 & 13.58 & 0.50 & 004004-7442.6~ & 0.339826 & 10.73 & 0.10 & 004041-4258.5~ & 0.83623 & 12.34 & 0.26\\
004430-3606.5~ & 0.246539 & 9.62 & 0.15 & 004619-7251.0~?v & 0.37664 & 13.48 & 0.71 & 004642-5437.5~ & 0.273367 & 10.89 & 0.34\\
005056-1747.0~? & 0.73511 & 10.85 & 0.45 & 005133-7559.2~? & 2.7054 & 13.18 & 0.66 & 005149-7159.9~?v & 0.87086 & 10.50 & 1.33\\
005235-4521.2~? & 0.61950 & 10.20 & 0.11 & 005633-7603.0~? & 0.267084 & 11.65 & 0.19 & 010129-7256.5~? & 5.393 & 13.28 & 0.68\\
010720-7524.1~ & 0.63474 & 12.05 & 0.39 & 010946-2013.0~?v & 0.256486 & 9.29 & 0.19 & 011322-6139.8~ & 9.365 & 13.14 & 0.37\\
011406-5349.9~ & 0.78065 & 11.66 & 0.37 & 011839-1849.6~ & 1.6884 & 12.31 & 0.30 & 011959-5816.5~? & 0.48990 & 11.27 & 0.16\\
012048-4742.8~? & 0.269707 & 12.16 & 0.26 & 012842-5356.2~? & 1.02464 & 10.95 & 0.35 & 013227-4954.3~? & 0.45495 & 11.36 & 0.22\\
013556-5228.7~? & 0.246048 & 11.45 & 0.15 & 013904-7840.7~? & 1.8990 & 12.07 & 0.28 & 014025-4138.0~?v & 0.33867 & 13.48 & 0.75\\
014253-5351.7~? & 0.36407 & 11.13 & 0.11 & 014831-8126.2~? & 0.316200 & 14.14 & 0.84 & 014933-1937.6~?v & 0.340812 & 11.20 & 0.70\\
015010-7412.9~? & 0.282760 & 13.78 & 0.68 & 015136-4310.6~?v & 0.90356 & 11.03 & 0.57 & 015200-2800.4~ & 0.44384 & 9.84 & 0.33\\
015337-3116.8~? & 0.36877 & 12.57 & 0.26 & 015423-5344.9~? & 0.65621 & 12.33 & 0.38 & 015713-6923.7~? & 0.340337 & 12.60 & 0.26\\
015902-2901.9~ & 0.294960 & 13.26 & 0.49 & 020230-8202.0~? & 0.47298 & 10.12 & 0.14 & 020345-1953.4~? & 0.303122 & 9.84 & 0.20\\
020352-4738.4~? & 0.35988 & 11.35 & 0.13 & 020511-0732.7~? & 0.44838 & 12.73 & 0.45 & 020643-0914.3~? & 0.50925 & 12.27 & 0.83\\
020850-1802.9~ & 0.41471 & 13.45 & 0.60 & 020917-4718.1~? & 0.34632 & 12.27 & 0.30 & 021341-5942.0~? & 0.44999 & 10.78 & 0.22\\
021742-0816.7~? & 1.4634 & 11.16 & 0.33 & 021948-4702.7~? & 0.53997 & 11.40 & 0.20 & 022544-5013.0~ & 0.270447 & 11.02 & 0.15\\
022721-1257.0~? & 0.67684 & 13.14 & 0.79 & 022837-5021.1~? & 0.87987 & 11.48 & 0.22 & 023018-3335.7~ & 0.43177 & 12.73 & 0.35\\
023458+0005.9~? & 0.40647 & 11.50 & 0.21 & 023916-2349.7~? & 0.51378 & 11.55 & 0.17 & 024012-2645.0~? & 0.64670 & 12.44 & 0.33\\
024051-7326.2~? & 0.351287 & 9.83 & 0.14 & 024101-4836.8~? & 0.42280 & 13.10 & 0.53 & 024138-0307.7~? & 1.1158 & 12.95 & 0.51\\
024206-7326.2~? & 0.276255 & 12.36 & 0.57 & 024233-5739.6~? & 14.74 & 11.02 & 0.13 & 024425-4224.9~? & 0.37718 & 12.30 & 0.21\\
024658-3215.2~? & 0.63648 & 11.10 & 0.13 & 024951-7151.1~? & 0.44955 & 12.43 & 0.29 & 025030-0305.0~? & 1.4756 & 10.84 & 0.41\\
025237-6213.7~? & 0.55904 & 12.60 & 0.54 & 025305-2031.7~? & 0.45886 & 12.52 & 0.35 & 025414-4706.8~? & 0.266771 & 13.38 & 0.52\\
025535-0219.9~ & 0.79274 & 11.59 & 0.54 & 025859-3523.8~? & 0.40742 & 12.00 & 0.34 & 030241-7332.9~? & 28.27 & 9.89 & 0.18\\
030503-3822.5~ & 0.40277 & 12.33 & 0.61 & 030728-4155.5~? & 0.41976 & 12.59 & 0.48 & 031000-1206.3~? & 3.4177 & 12.55 & 1.24\\
031000-7303.1~ & 0.283816 & 13.08 & 0.35 & 031013-5742.7~? & 0.266125 & 12.40 & 0.43 & 031029-5721.5~?: & 2.5041 & 11.46 & 0.19\\
031159-3759.2~? & 1.4225 & 12.57 & 0.25 & 031208-5248.9~? & 0.59487 & 12.51 & 0.33 & 031420-7127.6~ & 40.8 & 10.48 & 0.12\\
\hline
}
\clearpage

\addtocounter{table}{-1}
\MakeTableSep{|l|r|r|r||l|r|r|r||l|r|r|r|}{10cm}{Continued}{
\hline
\multicolumn{1}{|c|}{ID} & \multicolumn{1}{c|}{$P$} & \multicolumn{1}{c|}{$V$} & \multicolumn{1}{c|}{$\Delta~V$} & \multicolumn{1}{|c|}{ID} & \multicolumn{1}{c|}{$P$} & \multicolumn{1}{c|}{$V$} & \multicolumn{1}{c|}{$\Delta~V$} & \multicolumn{1}{|c|}{ID} & \multicolumn{1}{c|}{$P$} & \multicolumn{1}{c|}{$V$} & \multicolumn{1}{c|}{$\Delta~V$}\\
\hline
\multicolumn{12}{|c|}{\em  Stars classified as ESD.}\\
031612-2205.5~v & 0.77195 & 12.55 & 0.75 & 031718-4709.0~ & 0.304966 & 12.81 & 0.65 & 032048-3142.0~ & 1.03480 & 10.03 & 0.23\\
032212-5725.1~? & 0.35588 & 12.87 & 0.48 & 032246-2105.4~? & 38.0 & 9.52 & 0.16 & 032428-1033.7~? & 0.46110 & 12.35 & 0.32\\
032736-7730.8~? & 14.88 & 11.62 & 0.12 & 032743-3854.3~? & 0.44747 & 12.27 & 0.82 & 032906-1949.5~? & 31.46 & 10.96 & 0.12\\
033010-8108.4~?v & 0.36951 & 13.19 & 0.97 & 033131-5000.4~? & 0.35984 & 13.14 & 0.40 & 033307-2417.0~ & 0.54367 & 12.71 & 0.78\\
033437-0907.8~? & 0.54183 & 12.19 & 0.31 & 033520-8015.6~ & 0.48877 & 12.01 & 0.29 & 033536-7540.3~? & 0.344403 & 12.20 & 0.29\\
033642-7908.2~? & 0.61399 & 10.23 & 0.12 & 033716-1938.4~? & 0.78522 & 12.00 & 0.23 & 033921-6425.5~ & 0.57088 & 12.18 & 0.47\\
034019-3649.5~? & 0.76948 & 11.44 & 0.19 & 034242-5823.2~ & 0.41940 & 11.31 & 0.15 & 034331-4515.5~ & 0.70299 & 10.47 & 0.36\\
034400-5141.5~? & 1.4633 & 11.85 & 0.13 & 034512-3236.3~ & 0.47655 & 12.08 & 0.51 & 034921-2052.5~?v & 0.84305 & 9.40 & 0.42\\
035046-1339.9~? & 0.286065 & 12.81 & 0.28 & 035210-3433.1~? & 0.46360 & 13.42 & 0.58 & 040004-2902.4~? & 1.6368 & 9.67 & 0.09\\
040013-3347.7~? & 1.0949 & 14.34 & 1.12 & 040014-0746.0~ & 0.64048 & 10.79 & 0.48 & 040054-4052.1~? & 0.306390 & 13.45 & 0.60\\
040225-3200.4~? & 0.332681 & 12.47 & 0.85 & 040410-0758.4~? & 0.312262 & 13.17 & 0.35 & 040637-5634.5~ & 0.77997 & 8.73 & 0.31\\
040938-3915.8~? & 26.29 & 12.69 & 0.28 & 041204-7228.9~? & 0.41580 & 13.72 & 0.79 & 041210-5906.0~? & 0.64190 & 11.25 & 0.46\\
041353-0113.0~? & 0.35175 & 11.22 & 0.26 & 041454-5415.4~? & 0.36325 & 12.30 & 0.45 & 041501-6932.2~v & 0.74360 & 10.61 & 0.12\\
041652-0453.8~? & 0.286827 & 12.83 & 0.42 & 041707-4552.7~? & 0.81712 & 12.36 & 0.27 & 041723-2231.4~ & 0.68401 & 13.05 & 0.58\\
041946-3235.2~? & 0.76435 & 11.93 & 0.26 & 042047-4515.4~v & 0.51385 & 9.31 & 0.61 & 042143-5715.9~? & 0.84571 & 12.37 & 0.19\\
042401-4949.0~? & 40.1 & 10.10 & 0.17 & 042519-4430.0~? & 0.54308 & 13.27 & 0.50 & 042547-3506.9~? & 0.40511 & 12.80 & 0.43\\
042554-3426.4~ & 13.68 & 10.55 & 0.20 & 042618-3349.6~ & 0.86285 & 12.81 & 0.44 & 042701-0933.4~? & 0.46827 & 13.04 & 0.55\\
042702-3255.7~? & 3.0642 & 12.14 & 0.28 & 043052-6649.9~? & 0.319732 & 12.15 & 0.17 & 043544-2854.9~ & 0.90760 & 11.64 & 0.45\\
043837-2034.9~? & 1.4186 & 13.45 & 1.01 & 044405-1945.7~? & 0.51803 & 12.81 & 0.45 & 044455-4137.2~? & 0.67735 & 12.04 & 0.40\\
044537-0547.6~? & 0.45854 & 13.03 & 0.57 & 044645-5201.5~v & 1.5431 & 9.82 & 0.39 & 044703-2237.0~ & 2.1529 & 10.00 & 0.12\\
044837-7610.5~?v & 0.75721 & 13.24 & 0.45 & 045249-1955.0~? & 1.10473 & 9.98 & 0.15 & 045509-6440.4~? & 0.37641 & 12.60 & 0.32\\
045641-3911.5~? & 0.58972 & 10.85 & 0.36 & 045707-1927.7~? & 0.36612 & 12.87 & 0.49 & 045716-6633.9~?v & 4.342 & 13.65 & 0.65\\
050032-0914.3~ & 1.1193 & 12.31 & 0.55 & 050049-3212.7~ & 0.333740 & 12.38 & 0.64 & 050136-3354.8~? & 0.309241 & 12.40 & 0.35\\
050208-2204.9~? & 6.880 & 8.68 & 0.17 & 050520-0806.8~? & 0.90323 & 12.55 & 0.40 & 050527-6743.3~v & 4.049 & 11.01 & 0.61\\
050533-2610.2~? & 0.48011 & 11.24 & 0.20 & 050554-6810.8~?v & 0.93546 & 11.76 & 0.46 & 050628-3400.4~ & 2.9382 & 12.43 & 0.77\\
050645-1534.1~ & 0.40373 & 14.80 & 0.91 & 050726-1347.1~? & 0.49005 & 13.27 & 0.55 & 050728-0524.4~?v & 2.2549 & 10.42 & 0.17\\
050800-0550.6~v & 0.82195 & 10.94 & 0.34 & 050805-4242.9~ & 1.7161 & 11.30 & 0.28 & 050825-2326.7~? & 1.3994 & 13.62 & 0.99\\
050836-5709.6~? & 0.34554 & 13.19 & 0.50 & 050936-1019.2~? & 0.35724 & 13.65 & 0.73 & 050940-1247.7~? & 0.41426 & 11.54 & 0.22\\
051049-0832.5~? & 1.5556 & 10.80 & 0.14 & 051126-2636.1~? & 0.40699 & 12.96 & 0.40 & 051126-4413.1~ & 0.73337 & 12.35 & 0.66\\
051153-0428.6~ & 0.73746 & 10.91 & 0.44 & 051206-2928.8~? & 0.338587 & 12.57 & 0.31 & 051211-1312.0~v & 0.91544 & 9.99 & 0.57\\
051338-0510.5~? & 1.1825 & 10.62 & 0.22 & 051414-2142.1~v & 0.52944 & 11.48 & 0.89 & 051424-5321.8~? & 0.57897 & 12.85 & 0.49\\
051542-4219.8~? & 0.52459 & 10.87 & 0.21 & 051546-0644.6~? & 0.90317 & 10.86 & 0.23 & 051700-4852.2~ & 0.57833 & 11.85 & 0.82\\
051947-4259.0~ & 1.2309 & 12.25 & 0.51 & 052018-5605.8~? & 0.52235 & 12.69 & 0.25 & 052149-6519.0~? & 1.2670 & 12.91 & 0.34\\
052202-3430.3~? & 0.63198 & 10.41 & 0.16 & 052246-3917.1~?: & 5.062 & 10.40 & 0.62 & 052419-4614.1~ & 1.00763 & 11.00 & 0.30\\
052526-2256.2~: & 6.762 & 10.31 & 0.89 & 052625-3354.5~? & 0.42686 & 13.04 & 0.46 & 052727-6711.9~v & 4.830 & 12.84 & 0.84\\
052818-1856.5~? & 0.65808 & 13.78 & 0.81 & 052833-6836.2~? & 0.57685 & 11.36 & 0.18 & 053253-2031.0~? & 0.72814 & 11.60 & 0.12\\
053423-0952.9~ & 0.72431 & 12.71 & 0.94 & 053451-3025.2~v & 0.51282 & 11.44 & 0.62 & 053516-1210.8~? & 0.67143 & 11.64 & 0.14\\
053647-3155.3~? & 0.32917 & 13.19 & 0.47 & 053948-2631.8~ & 1.6040 & 12.54 & 1.08 & 053955-1240.2~? & 8.311 & 12.10 & 0.60\\
054029-4454.6~? & 0.64221 & 11.03 & 0.17 & 054234-3415.7~? & 2.6922 & 10.57 & 0.13 & 054250-3318.8~? & 0.34185 & 13.72 & 0.82\\
054632-1221.1~? & 2.0559 & 11.18 & 0.14 & 054654-2528.9~ & 57.0 & 11.96 & 0.18 & 054852-1724.4~? & 0.87048 & 12.31 & 0.27\\
054921-0938.4~? & 0.35936 & 12.00 & 0.25 & 055015-0330.4~ & 1.3684 & 11.73 & 0.49 & 055017-1100.7~ & 0.90198 & 11.83 & 0.62\\
055252-1103.4~ & 0.80630 & 10.97 & 0.54 & 055809-1112.1~? & 0.51565 & 13.07 & 0.60 & 055828-6056.9~ & 13.23 & 11.11 & 0.22\\
055901-5421.6~? & 0.70736 & 13.05 & 0.36 & 060240-1646.0~? & 26.17 & 12.75 & 1.12 & 060247-2227.7~? & 0.342557 & 13.64 & 0.81\\
060249-2926.2~? & 0.67896 & 11.60 & 0.24 & 060424-1636.3~ & 0.66361 & 11.07 & 0.45 & 060522-6947.0~? & 0.59303 & 12.15 & 0.28\\
060606-2539.0~? & 3.950 & 9.79 & 0.18 & 060900-2906.4~? & 0.48902 & 12.49 & 0.43 & 074401-8907.6~ & 0.79802 & 12.82 & 0.53\\
221846-8438.4~? & 2.1349 & 11.66 & 0.18 & 221940-8359.1~? & 0.93756 & 13.00 & 0.68 & 225436-8327.9~ & 0.52988 & 12.98 & 0.31\\
231336-7556.0~ & 0.53091 & 11.96 & 0.68 & 232545-7257.8~ & 0.67009 & 12.23 & 0.50 & 232748-8613.3~? & 1.3958 & 9.36 & 0.12\\
232906-6659.4~? & 0.35723 & 12.34 & 0.46 & 233041-6953.5~ & 0.63147 & 11.32 & 0.23 & 233652-5742.8~? & 12.45 & 11.67 & 0.18\\
233803-5926.5~?v & 1.2658 & 11.30 & 1.67 & 233842-6047.8~? & 17.99 & 10.91 & 0.13 & 234008-6726.3~? & 1.09055 & 11.61 & 0.43\\
234425-0546.6~? & 0.328746 & 9.82 & 0.30 & 234520-3100.4~? & 0.88348 & 10.94 & 0.27 & 235117-8517.4~? & 0.46073 & 12.30 & 0.33\\
235157-5725.4~ & 0.39258 & 11.06 & 0.23 & 235210-3200.7~? & 0.35007 & 12.46 & 0.41 & 235428-1236.6~? & 1.9837 & 12.24 & 0.57\\
235757-0109.8~? & 12.34 & 10.59 & 0.42 &  &  &  &  &  &  &  & \\
\multicolumn{12}{|c|}{\em  Stars classified as ED.}\\
000030-3937.5~v & 2.5545 & 10.79 & 1.22 & 001438-1948.7~ & 1.4698 & 11.28 & 0.25 & 001855-7954.9~v & 0.90310 & 11.28 & 0.77\\
003016-4628.0~v & 5.414 & 12.46 & 1.11 & 003245-6535.1~ & 0.92846 & 11.44 & 0.37 & 003449-5124.4~ & 1.05861 & 12.98 & 1.07\\
003620-2120.8~? & 0.40949 & 13.47 & 0.53 & 003640-5557.8~ & 0.91962 & 11.78 & 0.45 & 003655-0552.5~ & 6.975 & 9.87 & 0.45\\
003713-6349.2~v & 0.86588 & 9.88 & 0.36 & 004550-7849.3~v & 3.5358 & 9.88 & 2.40 & 004817-2805.3~ & 0.84664 & 10.18 & 0.19\\
005143-6457.3~ & 2.3616 & 11.54 & 0.67 & 010319-3745.2~ & 1.12264 & 10.78 & 0.75 & 010538-8003.7~ & 8.070 & 10.10 & 0.44\\
010921-2408.3~ & 2.4106 & 11.18 & 0.49 & 011328-3821.1~ & 0.44559 & 11.78 & 0.33 & 011513-5149.1~ & 1.7809 & 13.60 & 0.66\\
011748-8119.8~ & 1.9251 & 12.28 & 0.67 & 011750-3753.4~ & 0.97032 & 11.93 & 0.73 & 012424-2753.4~ & 2.0654 & 10.62 & 0.18\\
012534-4148.6~v & 6.883 & 10.70 & 0.76 & 012546-3956.2~?v & 1.2608 & 11.51 & 0.50 & 012940-8309.2~ & 0.68986 & 12.87 & 0.45\\
013020-8332.0~v & 4.489 & 11.42 & 1.10 & 013112-7556.9~ & 1.06334 & 9.93 & 0.33 & 014101-0643.6~? & 2.3064 & 10.39 & 0.58\\
014122-5520.0~ & 4.635 & 11.54 & 0.47 & 014613-7951.9~ & 1.3170 & 10.67 & 0.38 & 015650-2111.7~v & 2.2702 & 10.56 & 0.50\\
015838-5331.7~v & 1.2381 & 9.63 & 0.98 & 020150-6447.1~ & 2.3838 & 12.66 & 0.74 & 020902-3720.9~ & 4.124 & 12.07 & 0.42\\
020931-8518.2~ & 1.3303 & 9.90 & 0.64 & 021258-8124.0~ & 7.458 & 10.51 & 0.41 & 022053-4856.7~? & 1.1936 & 10.44 & 0.56\\
022136-3712.8~v & 2.4348 & 10.33 & 1.48 & 022915-7827.9~ & 2.3250 & 10.07 & 0.58 & 023434-3438.3~?v & 4.278 & 11.30 & 0.84\\
023455-7315.5~ & 1.05502 & 11.80 & 0.35 & 023526-1605.2~? & 1.6031 & 12.57 & 0.54 & 023539-4504.2~v & 5.784 & 9.11 & 0.91\\
024458-1746.8~v & 2.2323 & 9.30 & 0.29 & 024519-5407.4~ & 2.4057 & 12.01 & 0.50 & 024644+0107.9~ & 3.691 & 10.41 & 0.57\\
024946-3825.6~ & 0.46322 & 11.70 & 0.64 & 025136-6734.3~?v & 0.68586 & 13.09 & 0.85 & 025701-0252.3~? & 1.2853 & 11.67 & 0.49\\
030518-1838.2~v & 0.80613 & 12.47 & 1.56 & 030524-6641.1~v & 6.682 & 11.45 & 0.81 & 030545-8545.0~v & 0.90709 & 12.21 & 1.42\\
030807-2445.6~v & 0.91820 & 10.16 & 0.55 & 030847-1927.6~v & 1.2985 & 12.92 & 0.63 & 030956-4459.5~v & 1.8721 & 9.26 & 0.39\\
031952-2436.8~v & 2.4301 & 10.14 & 0.43 & 032123-1017.1~v & 1.5575 & 11.68 & 0.88 & 032704-4552.9~v & 2.6078 & 11.32 & 0.58\\
033113-4706.3~? & 1.6231 & 12.60 & 0.36 & 033240-7553.4~? & 1.05588 & 12.26 & 0.39 & 033431-3924.7~v & 4.224 & 9.56 & 0.77\\
033615-6114.1~ & 3.4927 & 11.94 & 0.57 & 033856-7224.5~v & 0.81707 & 12.08 & 0.73 & 034351-8400.1~v & 2.5996 & 11.44 & 1.22\\
034413-4116.8~v & 0.72240 & 8.94 & 0.60 & 034512-1624.6~v & 2.9597 & 8.86 & 0.49 & 034746-0836.7~ & 2.8764 & 9.41 & 0.88\\
034944-8349.9~ & 1.7927 & 11.40 & 0.34 & 035418-4306.6~v & 7.259 & 11.91 & 0.82 & 035427-5834.7~ & 1.8263 & 10.21 & 0.27\\
035858-1713.8~v & 4.977 & 10.09 & 0.71 & 040057-7753.8~v & 2.2756 & 12.23 & 0.80 & 040249-0926.0~ & 3.4144 & 11.58 & 0.73\\
040310-7632.9~ & 2.0358 & 9.97 & 0.33 & 040403-7427.2~v & 2.8783 & 10.91 & 0.37 & 040506-3110.2~v & 0.90140 & 8.52 & 0.50\\
040907-4828.6~v & 2.3832 & 8.52 & 0.24 & 041947-3210.1~? & 0.64318 & 11.80 & 0.44 & 042347-2426.2~ & 2.3038 & 10.25 & 0.44\\
042535-1848.0~v & 2.6306 & 11.98 & 1.06 & 042535-6045.4~v & 2.0846 & 8.71 & 0.45 & 042631-5214.5~ & 0.58756 & 14.23 & 1.07\\
042641-2457.0~? & 1.2873 & 11.50 & 0.58 & 042735-2513.9~?v & 1.4267 & 10.40 & 0.41 & 043041-1913.7~ & 1.7974 & 11.98 & 0.50\\
043201-1744.8~v & 9.298 & 10.84 & 0.78 & 043650-3622.1~ & 1.10473 & 11.58 & 0.26 & 043653-4456.2~ & 1.9717 & 9.81 & 0.23\\
044300-3245.7~ & 2.0054 & 11.19 & 0.36 & 044420-3141.3~? & 4.335 & 11.71 & 0.16 & 044733-8322.1~ & 2.5695 & 11.81 & 1.68\\
044748-4759.3~ & 1.07770 & 10.77 & 0.44 & 045021-4546.8~v & 13.96 & 9.61 & 0.88 & 045304-0700.4~ & 1.6224 & 11.35 & 0.29\\
\hline
}
\clearpage

\addtocounter{table}{-1}
\MakeTableSep{|l|r|r|r||l|r|r|r||l|r|r|r|}{10cm}{Continued}{
\hline
\multicolumn{1}{|c|}{ID} & \multicolumn{1}{c|}{$P$} & \multicolumn{1}{c|}{$V$} & \multicolumn{1}{c|}{$\Delta~V$} & \multicolumn{1}{|c|}{ID} & \multicolumn{1}{c|}{$P$} & \multicolumn{1}{c|}{$V$} & \multicolumn{1}{c|}{$\Delta~V$} & \multicolumn{1}{|c|}{ID} & \multicolumn{1}{c|}{$P$} & \multicolumn{1}{c|}{$V$} & \multicolumn{1}{c|}{$\Delta~V$}\\
\hline
\multicolumn{12}{|c|}{\em  Stars classified as ED.}\\
045605-4800.7~? & 2.2259 & 10.23 & 0.21 & 045907-2302.9~ & 5.259 & 11.37 & 0.40 & 045927-7732.8~? & 1.03698 & 12.20 & 0.28\\
050040-1802.8~ & 1.09577 & 11.68 & 0.41 & 050205-2842.8~ & 3.3023 & 10.98 & 0.63 & 050322-4337.0~ & 1.06362 & 13.09 & 0.44\\
050456-5308.4~?v & 0.86185 & 10.51 & 0.86 & 050513-2312.8~ & 1.7545 & 12.82 & 0.69 & 050604-3509.1~? & 0.74615 & 11.75 & 0.29\\
050712-2054.8~ & 1.2161 & 12.79 & 0.37 & 050717-1909.2~ & 2.5258 & 10.93 & 0.59 & 051010-1452.3~v & 773 & 11.63 & 1.20\\
051023-6846.4~? & 74.5 & 10.42 & 0.49 & 051433-2519.7~ & 0.89307 & 12.58 & 0.50 & 051815-4534.8~v & 1.8381 & 10.07 & 0.55\\
051837-5803.2~: & 5.492 & 11.26 & 0.26 & 051950-3155.0~? & 1.4668 & 10.72 & 0.17 & 052041-5619.1~ & 1.3445 & 10.84 & 0.43\\
052543-2935.2~: & 4.271 & 12.02 & 0.74 & 052800-3334.6~ & 6.186 & 11.28 & 0.73 & 052919-1617.3~ & 0.66085 & 11.17 & 0.42\\
053104-1102.2~ & 1.09203 & 11.11 & 0.56 & 053200-3320.5~ & 0.73285 & 12.44 & 0.42 & 053503-6843.7~ & 1.0864 & 12.20 & 0.41\\
053638-1247.0~ & 3.0274 & 11.72 & 0.67 & 053854-3337.9~v & 2.7520 & 11.84 & 0.47 & 054004-0119.7~v & 4.633 & 10.58 & 0.81\\
054020-2901.4~ & 2.4789 & 11.49 & 1.32 & 054136-1339.6~ & 0.93301 & 11.96 & 0.87 & 054447-2250.9~v & 3.723 & 9.45 & 0.49\\
054531-1746.6~ & 7.979 & 10.64 & 0.28 & 054937-2114.4~ & 2.0164 & 9.95 & 0.29 & 055010-5648.4~ & 1.1685 & 8.91 & 0.27\\
055035-1022.0~ & 2.6096 & 12.00 & 0.41 & 055655-7320.5~? & 2.9278 & 10.24 & 0.23 & 055708-0728.2~ & 5.804 & 12.15 & 0.48\\
055918-2013.4~v & 1.2885 & 9.83 & 1.50 & 060325-5528.4~ & 3.3228 & 11.37 & 0.82 & 060352-2454.1~v & 4.460 & 10.84 & 0.91\\
060450-1314.3~ & 0.79702 & 12.40 & 0.71 & 060649-2137.6~ & 4.288 & 11.33 & 1.93 & 060927-1501.7~ & 0.87708 & 9.37 & 0.36\\
220923-8615.0~ & 0.92688 & 11.67 & 0.62 & 230727-7818.3~ & 1.9080 & 9.28 & 0.33 & 232326-6948.6~? & 1.03956 & 12.05 & 0.38\\
233023-5825.6~?v & 5.463 & 9.77 & 1.18 & 234621-5441.7~ & 3.3178 & 12.38 & 0.64 & 235052-2316.7~ & 1.4023 & 9.42 & 0.29\\
\multicolumn{12}{|c|}{\em  Stars classified as DSCT.}\\
000116-6037.0~v & 0.122072 & 10.03 & 0.35 & 000316-7342.2~? & 0.201026 & 13.72 & 0.54 & 000410-5252.9~? & 0.131583 & 11.66 & 0.25\\
000412-2055.1~ & 0.178994 & 11.57 & 0.17 & 001026-3739.9~ & 0.153047 & 12.06 & 0.33 & 001123-1732.5~? & 0.177967 & 12.78 & 0.43\\
001513-7016.7~? & 0.209465 & 11.89 & 0.19 & 002134-1345.3~?: & 0.164569 & 12.43 & 0.28 & 002424-7102.9~? & 0.169003 & 10.99 & 0.32\\
002450-5235.8~? & 0.171659 & 13.19 & 0.49 & 002524-4655.5~?v & 0.218481 & 10.43 & 0.17 & 003323-4849.7~? & 0.165699 & 10.84 & 0.27\\
003344-1532.6~ & 0.198584 & 10.43 & 0.15 & 004100-2541.9~? & 0.180747 & 12.07 & 0.23 & 004445-1503.6~?: & 0.137691 & 12.96 & 0.50\\
004753-3245.4~: & 0.202992 & 10.61 & 0.08 & 004852-3625.0~? & 0.172294 & 12.07 & 0.18 & 004912-4322.5~ & 0.130639 & 13.65 & 0.57\\
005115-5904.0~v & 0.146458 & 13.09 & 0.63 & 005221-1907.4~? & 0.205720 & 13.50 & 0.50 & 010111-6843.1~ & 0.091612 & 12.09 & 0.44\\
010435-1916.7~? & 0.177791 & 12.68 & 0.48 & 010844-0340.1~ & 0.194607 & 11.44 & 0.83 & 011224-8541.0~? & 0.173886 & 12.74 & 0.37\\
012222-2906.9~? & 0.205090 & 10.61 & 0.08 & 012349-3643.7~ & 0.127768 & 9.99 & 0.24 & 012555-3558.8~? & 0.154133 & 13.54 & 0.60\\
013437-3928.2~ & 0.145186 & 12.09 & 0.23 & 014001-1024.5~ & 0.173991 & 12.49 & 0.29 & 014435-7745.3~v & 0.205717 & 14.35 & 1.38\\
014554-4730.2~ & 0.150850 & 11.23 & 0.19 & 015245-7445.7~ & 0.136873 & 10.04 & 0.08 & 015307-5056.5~ & 0.091002 & 9.71 & 0.29\\
015307-6757.3~? & 0.140889 & 10.27 & 0.15 & 015455-2626.9~ & 0.206455 & 12.48 & 0.21 & 015619-4040.3~ & 0.191174 & 11.88 & 0.20\\
015724-2626.2~? & 0.211946 & 11.67 & 0.20 & 015736-3112.6~ & 0.180976 & 11.76 & 0.21 & 020846-7020.8~? & 0.191314 & 13.10 & 0.47\\
021313-4054.8~ & 0.115507 & 12.38 & 0.41 & 021923-8613.4~ & 0.177578 & 13.08 & 0.55 & 021947-7037.8~ & 0.146672 & 12.64 & 0.37\\
022531-1859.7~? & 0.186926 & 12.84 & 0.56 & 022818-4856.0~? & 0.174036 & 12.58 & 0.33 & 023134-3549.1~ & 0.210595 & 11.83 & 0.31\\
023226-2848.2~? & 0.178662 & 12.07 & 0.18 & 023401-6536.6~v & 0.199460 & 9.48 & 0.17 & 023546-5603.7~? & 0.172731 & 13.43 & 0.43\\
023836-5600.7~? & 0.186163 & 11.97 & 0.26 & 023903-1637.1~ & 0.200669 & 14.31 & 0.66 & 024506-4128.3~? & 0.101698 & 12.17 & 0.23\\
024727-0936.2~? & 0.187181 & 11.01 & 0.22 & 024738-2702.4~ & 0.134958 & 13.48 & 0.49 & 024754-4546.7~ & 0.134506 & 12.03 & 0.15\\
025237-4927.1~? & 0.181748 & 10.49 & 0.41 & 025252-3754.5~? & 0.198962 & 11.68 & 0.14 & 025347-1307.7~ & 0.185929 & 12.04 & 0.30\\
025508-4052.1~? & 0.206136 & 11.53 & 0.33 & 025743-3351.6~ & 0.064873 & 11.36 & 0.20 & 025845-3833.3~ & 0.121284 & 11.48 & 0.25\\
030503-5642.6~? & 0.196234 & 13.25 & 0.51 & 030737-1301.7~ & 0.190600 & 11.33 & 0.27 & 031319-3100.4~? & 0.173388 & 12.90 & 0.39\\
031406-2035.5~? & 0.197313 & 10.34 & 0.14 & 031504-6429.2~ & 0.220207 & 12.73 & 0.30 & 031636-4957.3~ & 0.208692 & 11.19 & 0.23\\
031826-0950.4~ & 0.144948 & 12.10 & 0.18 & 032246-7237.8~ & 0.129425 & 12.07 & 0.33 & 032350-1439.8~ & 0.145393 & 12.70 & 0.47\\
032917-2138.0~ & 0.179375 & 13.33 & 0.43 & 033109-3742.7~? & 0.183400 & 11.50 & 0.21 & 033219-3539.3~ & 0.060689 & 10.70 & 0.23\\
033627-4109.9~ & 0.177565 & 12.48 & 0.36 & 033715-2832.6~ & 0.056138 & 13.40 & 0.75 & 034018-2651.4~: & 0.143222 & 13.82 & 0.98\\
034534-1123.2~? & 0.132551 & 13.13 & 0.79 & 040137-5342.1~ & 0.094855 & 12.26 & 0.42 & 040320-8342.0~ & 0.085660 & 11.88 & 0.30\\
040558-8344.2~? & 0.230377 & 12.41 & 0.33 & 041345-5256.8~? & 0.178287 & 13.21 & 0.36 & 042232-1325.8~ & 0.201619 & 12.37 & 0.18\\
042337-6032.2~? & 0.173824 & 14.10 & 0.84 & 042750-1846.3~ & 0.113415 & 11.01 & 0.27 & 043401-3334.0~ & 0.200877 & 12.29 & 0.25\\
043616-1904.4~? & 0.192372 & 12.41 & 0.30 & 043623-4811.6~? & 0.192181 & 11.31 & 0.41 & 043720-2928.1~? & 0.169624 & 12.66 & 0.86\\
043748-6516.6~? & 0.138084 & 12.47 & 0.42 & 043907-4649.0~ & 0.201566 & 12.80 & 0.40 & 044114-5505.9~? & 0.188373 & 12.10 & 0.35\\
044225-2617.5~? & 0.227757 & 12.32 & 0.21 & 044325-1028.9~ & 0.188280 & 11.90 & 0.20 & 044526-4906.8~? & 0.201264 & 13.04 & 0.39\\
044916-1503.5~? & 0.167064 & 13.11 & 0.41 & 045338-2906.6~? & 0.192781 & 12.17 & 0.45 & 045425-6804.7~ & 0.153416 & 12.20 & 0.28\\
045636-2740.8~ & 0.078465 & 12.62 & 0.57 & 045800-4555.6~ & 0.103096 & 12.89 & 0.90 & 045801-5011.2~ & 0.093966 & 12.70 & 0.41\\
045956-5829.9~ & 0.134604 & 8.75 & 0.22 & 050148-7614.3~? & 0.172780 & 12.55 & 0.29 & 050226-1525.2~? & 0.156772 & 13.20 & 0.47\\
050318-3449.7~ & 0.144399 & 12.88 & 0.44 & 050402-2116.9~? & 0.153324 & 12.80 & 0.30 & 050545-1754.0~? & 0.172556 & 11.74 & 0.13\\
050645-5903.1~?v & 0.213593 & 9.34 & 0.22 & 050702-0047.6~?v & 0.215970 & 10.51 & 0.21 & 051144-6242.9~? & 0.195234 & 13.23 & 0.56\\
051512-3523.6~: & 0.245344 & 12.99 & 1.04 & 051614-2955.1~? & 0.156993 & 12.40 & 0.50 & 051752-2552.3~? & 0.150510 & 13.69 & 0.58\\
052135-2448.0~ & 0.169032 & 12.80 & 0.61 & 052301-1208.1~? & 0.164740 & 11.96 & 0.27 & 052707-8334.7~ & 0.139924 & 11.98 & 0.28\\
053123-4731.9~? & 0.131549 & 13.26 & 0.48 & 053438-1756.0~? & 0.147025 & 12.99 & 0.33 & 053512-5801.2~v & 0.104758 & 12.14 & 0.15\\
053607-1538.4~? & 0.175075 & 11.95 & 0.21 & 053652-0824.8~?v & 0.173241 & 12.54 & 0.28 & 053714-5158.2~ & 0.143875 & 13.17 & 0.46\\
053940-5934.1~ & 0.158203 & 12.95 & 0.46 & 054049-5527.8~? & 0.149118 & 12.46 & 0.40 & 054057-1810.6~ & 0.160241 & 12.09 & 0.18\\
054215-1124.7~? & 0.155481 & 13.94 & 1.24 & 054252-1725.2~? & 0.111756 & 13.42 & 0.67 & 054514-2204.6~? & 0.178080 & 12.84 & 0.34\\
054727-1323.2~? & 0.172267 & 12.84 & 0.33 & 055933-1835.5~? & 0.197731 & 12.30 & 0.25 & 055941-3039.9~v & 0.144938 & 9.54 & 0.16\\
060226-2526.9~? & 0.155634 & 12.95 & 0.66 & 060228-0807.1~: & 0.200464 & 10.93 & 0.23 & 060521-5800.6~? & 0.181880 & 13.43 & 0.57\\
060726-7655.6~ & 0.200137 & 9.57 & 0.23 & 154315-8648.1~ & 0.142939 & 11.42 & 0.69 & 230125-8514.1~ & 0.207031 & 12.75 & 0.39\\
233302-6722.2~? & 0.192630 & 11.08 & 0.19 & 233439-6341.1~? & 0.117861 & 12.60 & 0.26 & 233610-5006.9~? & 0.147404 & 12.71 & 0.37\\
233816-4129.5~ & 0.197581 & 12.72 & 0.29 & 234003-4738.6~ & 0.111420 & 11.37 & 0.47 & 234112-5341.7~ & 0.219595 & 12.63 & 0.29\\
234203-4051.0~ & 0.173522 & 13.38 & 0.69 & 234846-0808.7~v & 0.197822 & 9.17 & 0.44 & 234849-6946.9~ & 0.196619 & 12.11 & 0.27\\
235026-5337.8~? & 0.198453 & 12.05 & 0.23 & 235144-6728.8~? & 0.182003 & 10.99 & 0.18 & 235849-0853.3~ & 0.177163 & 13.33 & 0.43\\
235951-3343.5~ & 0.167164 & 11.22 & 0.64 &  &  &  &  &  &  &  & \\
\multicolumn{12}{|c|}{\em  Stars classified as RRC.}\\
000507-1204.6~? & 0.35013 & 13.64 & 0.52 & 001231-1402.1~ & 0.326337 & 13.33 & 0.58 & 003514-0415.0~? & 0.34463 & 12.92 & 0.81\\
003629-2011.7~? & 0.245848 & 11.50 & 0.22 & 004627-5750.5~? & 0.36789 & 13.61 & 0.54 & 004729-6522.3~ & 0.40994 & 13.77 & 0.70\\
004929-2723.2~?v & 0.35471 & 13.33 & 0.52 & 005231-6110.5~ & 0.287839 & 12.56 & 0.36 & 005409-5708.5~? & 0.297671 & 13.10 & 0.40\\
005618-4122.0~?: & 0.335374 & 12.68 & 0.29 & 005924-1556.6~ & 0.289576 & 12.47 & 0.48 & 010029-4404.2~? & 0.325583 & 12.29 & 0.24\\
010117-4556.6~ & 0.37377 & 11.82 & 0.30 & 011148-6355.3~? & 0.39644 & 13.35 & 0.57 & 011643-3759.6~? & 0.308660 & 13.19 & 0.58\\
011800-5511.4~? & 0.262641 & 12.08 & 0.33 & 011831-6755.1~v & 0.40578 & 11.56 & 0.44 & 012138-7350.2~? & 0.263843 & 13.45 & 0.34\\
012301-7234.4~ & 0.316314 & 13.06 & 0.54 & 013251-4715.8~ & 0.317882 & 9.55 & 0.11 & 013344-5217.7~ & 0.286989 & 12.90 & 0.57\\
013412-5026.5~ & 0.336951 & 12.36 & 0.46 & 013450-3909.0~? & 0.288137 & 12.49 & 0.40 & 013931-3322.3~ & 0.316309 & 12.40 & 0.45\\
014226-3027.6~ & 0.273712 & 12.12 & 0.53 & 014312-2342.5~ & 0.307141 & 12.36 & 0.55 & 014500-3003.6~v & 0.37738 & 11.14 & 0.53\\
014525-1846.6~ & 0.34542 & 13.18 & 0.51 & 015649-0650.0~? & 0.221601 & 11.65 & 0.23 & 015752-0532.1~ & 0.301467 & 11.89 & 0.42\\
020010-2433.1~ & 0.324301 & 12.39 & 0.40 & 020013-1728.7~ & 0.258105 & 12.09 & 0.21 & 020728-5752.2~v & 0.37503 & 10.90 & 0.43\\
020740-7748.8~v & 0.310666 & 12.55 & 0.56 & 021012-2856.7~ & 0.320204 & 12.19 & 0.53 & 022011-2325.4~ & 0.329154 & 12.86 & 0.46\\
022358-4538.8~? & 0.319599 & 12.25 & 0.21 & 022637-4119.7~: & 0.293921 & 10.14 & 0.11 & 023319-7336.7~v & 0.287146 & 11.81 & 0.57\\
023706-4257.8~v & 0.311326 & 8.84 & 0.52 & 024636-0002.4~: & 0.40366 & 10.91 & 0.15 & 024728-3209.1~ & 0.33697 & 13.36 & 0.62\\
025134-4748.1~v & 0.311356 & 12.13 & 0.45 & 025344-2652.9~?v & 0.280430 & 13.62 & 0.62 & 025705-2826.9~ & 0.39315 & 12.69 & 0.52\\
025928-3105.3~? & 0.324935 & 12.90 & 0.39 & 030015-0459.7~: & 0.315903 & 12.66 & 0.50 & 030114-1059.7~ & 0.36311 & 13.44 & 0.64\\
\hline
}
\clearpage

\addtocounter{table}{-1}
\MakeTableSep{|l|r|r|r||l|r|r|r||l|r|r|r|}{10cm}{Continued}{
\hline
\multicolumn{1}{|c|}{ID} & \multicolumn{1}{c|}{$P$} & \multicolumn{1}{c|}{$V$} & \multicolumn{1}{c|}{$\Delta~V$} & \multicolumn{1}{|c|}{ID} & \multicolumn{1}{c|}{$P$} & \multicolumn{1}{c|}{$V$} & \multicolumn{1}{c|}{$\Delta~V$} & \multicolumn{1}{|c|}{ID} & \multicolumn{1}{c|}{$P$} & \multicolumn{1}{c|}{$V$} & \multicolumn{1}{c|}{$\Delta~V$}\\
\hline
\multicolumn{12}{|c|}{\em  Stars classified as RRC.}\\
030405-3952.1~? & 0.300939 & 11.42 & 0.24 & 030433-4157.7~ & 0.38991 & 12.41 & 0.55 & 030528-3058.7~ & 0.304610 & 12.69 & 0.45\\
030615-2612.8~?v & 0.40614 & 13.43 & 0.57 & 031124-3719.5~ & 0.322599 & 13.75 & 0.74 & 031238-7448.3~ & 0.312059 & 13.63 & 0.62\\
031315-3008.2~ & 0.283661 & 12.18 & 0.55 & 031320-1726.8~ & 0.262896 & 13.61 & 0.50 & 031408-3446.4~ & 0.312434 & 11.54 & 0.53\\
031535-1647.0~: & 0.35132 & 14.52 & 0.91 & 031626-4249.0~? & 0.34517 & 13.60 & 0.85 & 032820-6458.7~v & 0.35480 & 12.63 & 0.45\\
033021-5517.9~ & 0.280492 & 13.72 & 0.42 & 033250-4830.5~? & 0.218309 & 12.30 & 0.28 & 033623-2129.1~ & 0.316991 & 13.72 & 0.71\\
033634-0711.6~ & 0.307785 & 12.10 & 0.24 & 033822-6259.1~? & 0.342568 & 9.67 & 0.10 & 034033-0856.9~? & 0.34561 & 12.05 & 0.45\\
034349-1328.3~? & 0.309994 & 13.61 & 0.53 & 035428-4930.6~? & 0.315495 & 11.62 & 0.22 & 035658-5446.4~ & 0.264197 & 12.09 & 0.54\\
035813-6629.6~?v & 0.36058 & 11.43 & 0.53 & 040527-4637.5~? & 0.39313 & 13.65 & 0.54 & 040625-2339.6~v & 0.327081 & 12.21 & 0.42\\
041056-6258.7~ & 0.300575 & 13.08 & 0.56 & 041950-1559.1~? & 0.346849 & 11.25 & 0.16 & 043036-3608.3~ & 0.244161 & 13.18 & 0.45\\
043741-3904.1~? & 0.36013 & 12.80 & 0.57 & 043753-7604.8~ & 0.316099 & 13.20 & 0.56 & 043917-3251.8~ & 0.286085 & 11.18 & 0.39\\
043930-3233.1~? & 0.35798 & 10.45 & 0.17 & 044146-1809.7~? & 0.293995 & 11.07 & 0.14 & 044850-3728.4~: & 0.252144 & 13.95 & 1.03\\
045213-5620.8~ & 0.35036 & 13.61 & 0.71 & 045815-2244.5~ & 0.274048 & 12.11 & 0.41 & 045914-6935.8~v & 0.328945 & 11.34 & 0.43\\
050319-0427.0~: & 0.247440 & 11.39 & 0.14 & 051337-3342.7~? & 0.37181 & 12.96 & 0.60 & 051618-2936.3~? & 0.241792 & 10.70 & 0.10\\
051714-1142.0~ & 0.236170 & 11.07 & 0.17 & 052545-1007.4~? & 0.38351 & 12.06 & 0.23 & 052840-5316.2~ & 0.36797 & 13.48 & 0.59\\
053022-3234.8~? & 0.233120 & 11.79 & 0.22 & 053628-3837.0~ & 0.37137 & 12.78 & 0.45 & 053830-3554.4~: & 0.270632 & 13.31 & 0.82\\
053902-5925.9~ & 0.40997 & 13.30 & 0.56 & 054449-2659.5~ & 0.262611 & 12.22 & 0.21 & 054810-2001.4~?v & 0.225147 & 8.16 & 0.34\\
055122-6812.8~v & 0.321775 & 11.87 & 0.44 & 055239-0551.6~: & 0.37441 & 11.96 & 0.47 & 055251-8113.4~? & 0.256639 & 11.90 & 0.34\\
055322-5417.9~? & 0.245298 & 12.87 & 0.58 & 075836-8530.8~? & 0.224120 & 12.24 & 0.51 & 230724-7649.1~v & 0.304971 & 13.14 & 0.40\\
230815-8403.9~? & 0.277365 & 13.13 & 0.55 & 232521-6750.1~:v & 0.34966 & 12.95 & 0.88 & 233446-4854.3~? & 0.319542 & 11.07 & 0.44\\
234001-5001.3~ & 0.309888 & 13.90 & 0.75 & 234106-4208.8~?v & 0.38049 & 12.83 & 0.49 & 234429-4210.3~ & 0.34579 & 13.66 & 0.58\\
234439-0148.6~ & 0.276854 & 12.23 & 0.38 & 235139-7334.4~ & 0.326350 & 12.56 & 0.48 & 235408+0057.8~ & 0.306258 & 10.39 & 0.43\\
235622-5329.4~?: & 0.42253 & 12.73 & 0.49 & 235721-6414.6~?v & 0.302975 & 9.48 & 0.31 &  &  &  & \\
\multicolumn{12}{|c|}{\em  Stars classified as RRAB.}\\
000248-2456.7~v & 0.49337 & 9.90 & 1.24 & 000301-7041.5~ & 0.55381 & 13.27 & 0.91 & 000316-1125.1~: & 0.36892 & 13.64 & 0.89\\
000348-1128.6~? & 0.74089 & 12.45 & 0.84 & 000405-1659.8~v & 0.60606 & 11.94 & 0.66 & 000410-4108.2~ & 0.52551 & 13.68 & 1.21\\
000452-5254.4~? & 0.58939 & 13.58 & 0.69 & 000602-3654.3~v & 0.43911 & 13.38 & 0.63 & 000621-3517.2~v & 0.57767 & 13.03 & 1.20\\
001141-0144.9~v & 0.52974 & 11.83 & 1.05 & 001323-4255.2~ & 0.74407 & 13.62 & 0.80 & 001528-6238.8~ & 0.51605 & 12.72 & 1.15\\
001543-5853.1~v & 0.62538 & 12.88 & 1.02 & 001611-3927.4~ & 0.53158 & 13.56 & 1.28 & 002418-7616.9~v & 0.56874 & 13.17 & 1.12\\
002443-6949.7~? & 0.72792 & 10.98 & 0.38 & 002649-4757.6~?: & 1.04301 & 11.69 & 0.22 & 002831-8054.7~?v & 0.59395 & 13.57 & 0.85\\
002843-4400.4~ & 0.59993 & 12.64 & 0.81 & 003338-1529.2~v & 0.57373 & 11.25 & 0.59 & 003358-1331.4~ & 0.43725 & 12.91 & 1.32\\
003543-2141.0~ & 0.56150 & 13.60 & 0.91 & 003640-5036.6~? & 0.56771 & 14.11 & 1.33 & 003706-4317.7~ & 0.62756 & 13.57 & 0.70\\
003752-3952.0~v & 0.58401 & 13.35 & 1.17 & 003857-5252.2~ & 0.53054 & 13.69 & 1.21 & 004025-2134.6~v & 0.56537 & 13.03 & 0.94\\
004115-6844.0~ & 0.67805 & 13.44 & 0.75 & 004336-6843.5~ & 0.68360 & 13.11 & 0.93 & 004453-2442.7~ & 0.57050 & 13.12 & 0.91\\
004458-7352.9~? & 0.73288 & 12.55 & 0.54 & 004548-4355.2~ & 0.58981 & 13.21 & 0.93 & 004554-6404.0~ & 0.47387 & 13.86 & 1.16\\
005001-6238.1~v & 0.41453 & 11.56 & 1.22 & 005031-4340.8~ & 0.56705 & 13.80 & 1.30 & 005453-6642.5~v & 0.60254 & 12.46 & 1.02\\
005559-2623.0~v & 0.52064 & 13.11 & 1.09 & 005705-6855.2~ & 0.59317 & 13.91 & 1.13 & 005712-1935.9~ & 0.48053 & 12.53 & 1.06\\
005810-6323.8~v & 0.64226 & 10.90 & 1.11 & 005829-1508.8~ & 0.51716 & 13.07 & 1.14 & 010028-2812.3~v & 0.46370 & 12.57 & 1.27\\
010040-1557.4~?v & 0.58629 & 11.16 & 1.00 & 010308-2530.3~? & 0.57506 & 13.62 & 1.06 & 010658-6423.0~? & 0.53067 & 13.97 & 1.07\\
010726-3218.6~ & 0.55011 & 11.97 & 1.06 & 010949-4418.9~v & 0.48436 & 12.77 & 1.04 & 010958-4207.7~?v & 0.61554 & 12.69 & 0.48\\
011022-7351.9~v & 0.53329 & 14.04 & 1.08 & 011200-5002.8~ & 0.68120 & 13.37 & 1.09 & 011506-5920.5~v & 0.45677 & 12.71 & 1.21\\
011515-6255.5~ & 0.52187 & 13.98 & 1.00 & 011815-3912.8~v & 0.51094 & 10.48 & 1.23 & 011825-1724.9~ & 0.62088 & 13.17 & 0.85\\
012254-2617.6~v & 0.45168 & 13.02 & 1.29 & 012500-3438.9~v & 0.42637 & 13.53 & 1.22 & 012612-1738.6~ & 0.46005 & 13.51 & 1.33\\
012848-1127.2~? & 0.51669 & 12.83 & 1.02 & 013055-5935.2~ & 0.53789 & 12.57 & 0.98 & 013113-7829.1~v & 0.55576 & 12.45 & 0.78\\
013140-4957.3~ & 0.46042 & 12.23 & 1.05 & 013524-3507.7~v & 0.63702 & 11.65 & 0.87 & 013744-0954.2~ & 0.56674 & 13.56 & 1.04\\
013922-3304.4~v & 0.62201 & 12.67 & 0.51 & 014231-8000.6~v & 0.34837 & 12.09 & 1.12 & 014237-3001.6~ & 0.51455 & 13.09 & 1.26\\
014543-4526.0~v & 0.55941 & 14.13 & 0.73 & 014721-7321.0~v & 0.48758 & 13.61 & 1.02 & 014946-6614.8~ & 0.56747 & 13.36 & 0.95\\
015136-4939.3~ & 0.79292 & 13.14 & 0.75 & 015649-4421.1~v & 0.58266 & 13.72 & 1.27 & 015920-4025.7~? & 0.52871 & 13.35 & 0.62\\
020345-1729.1~ & 0.36206 & 11.34 & 0.18 & 020752-2651.9~v & 0.49545 & 9.74 & 1.28 & 020925-5210.1~v & 0.62557 & 13.80 & 1.11\\
021214-4422.3~ & 0.68463 & 13.44 & 0.90 & 021356-4624.4~ & 0.39641 & 13.59 & 1.24 & 021515-1048.0~v & 0.62335 & 10.62 & 0.73\\
021931-7333.9~v & 0.67535 & 12.70 & 0.92 & 022256-6725.1~ & 0.61340 & 14.13 & 1.10 & 022536-5519.4~ & 0.38710 & 14.07 & 0.72\\
022832-0821.5~v & 0.51059 & 11.41 & 0.81 & 023020-5908.1~ & 0.57290 & 11.90 & 0.63 & 023106-4406.0~: & 0.56779 & 13.35 & 0.95\\
023303-7102.0~v & 0.38691 & 13.95 & 1.18 & 023627-3736.0~ & 0.57995 & 12.99 & 0.84 & 024457-4351.1~ & 0.52851 & 13.66 & 1.22\\
024506-5007.0~? & 0.55207 & 12.82 & 0.92 & 024629-1357.0~? & 0.56154 & 13.57 & 0.97 & 024814-5435.2~ & 0.58196 & 12.38 & 0.73\\
024819-4523.0~ & 0.58725 & 13.93 & 1.50 & 024956-0125.2~v & 0.54814 & 12.27 & 1.05 & 025010-2615.9~: & 0.37043 & 12.40 & 0.71\\
025021-6415.7~?v & 0.57249 & 12.89 & 1.00 & 025118-2003.5~ & 0.48820 & 13.51 & 1.15 & 025136-7210.3~ & 0.58834 & 13.38 & 0.97\\
025336-6844.9~ & 0.47921 & 13.80 & 0.63 & 025430-2824.1~ & 0.60304 & 12.27 & 1.14 & 025546-5556.2~ & 0.51810 & 12.63 & 1.10\\
025733-5946.3~ & 0.68802 & 13.76 & 1.14 & 030015-4944.5~ & 0.51948 & 14.19 & 1.07 & 030109-3807.7~v & 0.80377 & 12.14 & 0.60\\
030313-5131.8~ & 0.56291 & 13.76 & 1.02 & 030534-3116.1~ & 0.49645 & 12.77 & 1.16 & 030811-3845.4~ & 0.55011 & 13.77 & 0.85\\
030850-5711.5~v & 0.73877 & 13.23 & 0.84 & 031100-3520.8~ & 0.60840 & 13.38 & 1.07 & 031113-2629.0~v & 0.59732 & 11.51 & 1.04\\
031119-2853.6~ & 0.55356 & 12.87 & 0.82 & 031141-5804.0~v & 0.46961 & 13.68 & 1.36 & 031158-0348.6~ & 0.45589 & 13.42 & 1.27\\
031245-5655.1~ & 0.65281 & 13.67 & 0.87 & 031346-1412.7~v & 0.64859 & 12.49 & 0.97 & 031541-4519.8~ & 0.53071 & 13.88 & 1.26\\
031826-4936.0~v & 0.64367 & 12.03 & 1.19 & 031948-3331.0~ & 0.63438 & 12.99 & 0.96 & 032145-5645.6~ & 0.54559 & 12.89 & 0.85\\
032229-8122.1~v & 0.45624 & 13.93 & 1.21 & 032234-3804.6~ & 0.72941 & 13.20 & 0.80 & 032347-4801.3~ & 0.56467 & 13.21 & 1.20\\
032429-4857.0~ & 0.66219 & 12.48 & 0.68 & 032438-2334.7~ & 0.62960 & 12.33 & 0.65 & 032520-6503.3~?:v & 0.49205 & 11.33 & 0.88\\
032601-4126.0~? & 0.53125 & 13.25 & 0.53 & 032749-3854.6~ & 0.60889 & 13.33 & 0.91 & 033022-3603.2~v & 0.60535 & 10.89 & 0.64\\
033103-4523.8~ & 0.64962 & 13.54 & 0.50 & 033625-4937.3~ & 0.54860 & 13.49 & 1.16 & 034307-1926.5~ & 0.61509 & 13.26 & 1.00\\
034920-2310.6~ & 0.63866 & 13.29 & 1.09 & 035618-2759.7~? & 0.62489 & 13.67 & 1.05 & 040011-1949.6~ & 0.60224 & 12.36 & 0.54\\
040054-4923.8~: & 0.71603 & 13.77 & 0.60 & 040142-7839.6~v & 0.61872 & 13.35 & 1.37 & 040312-1951.2~ & 0.60883 & 11.69 & 1.02\\
040327-8148.3~v & 0.63158 & 13.21 & 1.17 & 040446-2643.0~ & 0.60275 & 12.46 & 0.71 & 040500-4457.1~ & 0.56685 & 12.87 & 0.48\\
040615-3050.0~ & 0.44432 & 12.05 & 0.16 & 041021-2619.1~ & 0.50879 & 13.94 & 1.48 & 041117-1350.9~ & 0.55429 & 12.60 & 0.92\\
041148-3042.6~? & 0.68909 & 13.39 & 0.41 & 041321-3833.0~ & 0.57650 & 12.43 & 0.76 & 041605-1434.0~v & 0.48215 & 13.68 & 1.25\\
041749-4800.4~ & 0.57608 & 12.00 & 0.92 & 041801-0342.8~? & 0.48719 & 10.95 & 0.15 & 042014-4236.2~ & 0.47105 & 13.39 & 1.30\\
042055-2302.9~? & 0.58069 & 13.10 & 0.91 & 042116-3518.2~v & 0.66278 & 12.53 & 1.19 & 042135-6640.2~ & 0.45884 & 13.16 & 1.34\\
042245-8633.9~v & 0.52348 & 13.61 & 0.75 & 042602-5116.7~ & 0.48942 & 13.31 & 0.68 & 042640-3618.6~v & 0.38653 & 12.58 & 0.84\\
042716-4828.0~ & 0.61306 & 11.63 & 0.54 & 042805-1710.6~v & 0.65691 & 13.11 & 1.19 & 043641-0440.2~? & 0.63054 & 12.65 & 0.38\\
043659-2244.2~? & 0.57406 & 13.68 & 0.72 & 043801-0845.3~v & 0.48162 & 13.13 & 1.25 & 044030-7326.3~v & 0.53712 & 12.91 & 1.02\\
044040-0911.4~v & 0.66000 & 13.24 & 0.69 & 044200-2528.8~ & 0.58750 & 14.25 & 0.82 & 044203-5133.6~v & 0.73207 & 11.89 & 0.76\\
044525-2455.6~ & 0.50502 & 13.09 & 1.17 & 044541-3458.7~ & 0.65857 & 13.13 & 0.57 & 044620-6825.5~? & 0.51627 & 13.54 & 0.95\\
044753-1120.8~ & 0.63998 & 12.85 & 0.50 & 044801-2531.4~ & 0.54876 & 12.58 & 0.96 & 044901-4523.0~? & 0.58090 & 13.46 & 0.80\\
044944-1544.5~v & 0.58725 & 9.36 & 0.87 & 045007-5039.4~v & 0.44036 & 11.02 & 1.15 & 045018-4311.0~ & 0.47275 & 12.70 & 0.30\\
045129-7137.7~ & 0.68879 & 12.99 & 0.73 & 045314-3749.2~v & 0.41979 & 11.62 & 1.26 & 045337-1926.0~v & 0.56989 & 11.13 & 0.88\\
045426-6626.2~? & 0.60885 & 12.31 & 0.58 & 045618-2113.0~v & 0.58148 & 10.13 & 1.19 & 045647-3512.8~ & 0.58091 & 13.66 & 0.96\\
045833-7020.8~?v & 0.97341 & 12.74 & 1.07 & 050124-1849.5~ & 0.59882 & 12.79 & 0.68 & 050138-3908.0~v & 0.57088 & 12.46 & 1.01\\
050206-6716.9~ & 0.52530 & 13.56 & 1.03 & 050332-6543.0~v & 0.38661 & 13.68 & 1.06 & 050708-6853.3~v & 0.49152 & 12.97 & 1.04\\
\hline
}
\clearpage

\addtocounter{table}{-1}
\MakeTableSep{|l|r|r|r||l|r|r|r||l|r|r|r|}{10cm}{Continued}{
\hline
\multicolumn{1}{|c|}{ID} & \multicolumn{1}{c|}{$P$} & \multicolumn{1}{c|}{$V$} & \multicolumn{1}{c|}{$\Delta~V$} & \multicolumn{1}{|c|}{ID} & \multicolumn{1}{c|}{$P$} & \multicolumn{1}{c|}{$V$} & \multicolumn{1}{c|}{$\Delta~V$} & \multicolumn{1}{|c|}{ID} & \multicolumn{1}{c|}{$P$} & \multicolumn{1}{c|}{$V$} & \multicolumn{1}{c|}{$\Delta~V$}\\
\hline
\multicolumn{12}{|c|}{\em  Stars classified as RRAB.}\\
050747-3351.9~v & 0.48734 & 12.14 & 0.94 & 050838-5602.9~v & 0.51609 & 12.25 & 0.93 & 050956-7138.7~v & 0.40981 & 13.18 & 0.91\\
051001-4123.1~ & 0.58679 & 13.66 & 1.09 & 051101-3851.6~ & 0.55951 & 14.09 & 1.60 & 051118-0131.5~ & 0.69491 & 11.60 & 0.47\\
051209-1053.8~ & 0.56876 & 13.29 & 1.11 & 051320-5835.1~ & 0.45873 & 13.48 & 0.81 & 051423-5353.9~ & 0.54793 & 13.38 & 1.09\\
051508-4137.7~v & 0.47882 & 10.75 & 0.90 & 051550-7036.7~v & 0.55337 & 13.48 & 1.29 & 051623-7527.4~ & 0.83396 & 13.20 & 0.70\\
051651-2728.4~v & 0.53658 & 12.43 & 0.92 & 051703-0515.8~? & 0.36965 & 13.62 & 0.57 & 052122-6221.4~ & 0.64967 & 12.18 & 0.43\\
052402-2247.4~ & 0.64987 & 13.83 & 0.64 & 052406-6925.2~v & 0.55311 & 13.78 & 1.36 & 052415-1406.0~v & 0.56010 & 12.07 & 0.79\\
052722-5533.8~ & 0.54634 & 13.77 & 1.17 & 052727-0351.6~ & 0.58740 & 13.36 & 1.04 & 052945-6417.2~v & 0.48284 & 13.13 & 1.15\\
053104-6434.3~? & 0.49124 & 9.29 & 0.14 & 053212-1305.6~ & 0.53644 & 13.25 & 1.21 & 053436-7539.8~v & 0.36430 & 13.54 & 0.91\\
053552-0508.2~:v & 0.297630 & 12.40 & 0.38 & 053750-2247.5~? & 0.64660 & 12.58 & 0.29 & 053951-4925.7~ & 0.61025 & 12.82 & 0.45\\
054230-1622.9~ & 0.53890 & 12.01 & 0.99 & 054238-2557.3~ & 0.51690 & 13.37 & 1.10 & 054343-2036.2~ & 0.342338 & 14.28 & 1.15\\
054557-1441.5~ & 0.59940 & 12.97 & 1.02 & 054838-1929.5~ & 0.55879 & 13.25 & 1.00 & 054843-1627.0~?: & 0.37677 & 12.96 & 0.61\\
055109-7219.4~ & 0.55888 & 13.67 & 1.20 & 055133-6041.6~ & 0.44104 & 13.07 & 0.97 & 055620-1848.1~ & 0.73075 & 12.87 & 0.63\\
055651-2740.0~v & 0.46876 & 11.91 & 1.24 & 055740-2422.6~ & 0.53818 & 14.00 & 1.08 & 060031-2136.3~ & 0.58404 & 12.76 & 0.78\\
092907-8829.7~v & 0.38355 & 12.65 & 0.67 & 142904-8838.7~ & 0.64655 & 11.90 & 0.98 & 182858-8832.5~v & 0.57302 & 13.32 & 0.99\\
194301-8746.9~ & 0.58171 & 13.80 & 1.20 & 214719-8739.1~v & 0.45803 & 12.53 & 0.96 & 215336-8246.8~v & 0.62187 & 11.50 & 1.01\\
223645-7807.2~v & 0.63902 & 12.48 & 0.69 & 231215-7434.7~v & 0.39402 & 12.28 & 1.18 & 232933-7232.6~v & 0.55005 & 12.29 & 0.98\\
233743-7950.0~v & 0.41243 & 13.69 & 1.37 & 233951-3530.4~ & 0.67750 & 12.67 & 0.40 & 234023-3818.9~v & 0.58585 & 12.80 & 0.91\\
234033-5336.9~ & 0.64335 & 13.22 & 0.74 & 234507-5000.0~? & 0.48061 & 12.93 & 0.27 & 234631-8130.0~v & 0.58339 & 14.01 & 1.12\\
234854-4141.3~ & 0.73826 & 13.97 & 0.81 & 234948-7411.6~v & 0.55894 & 13.92 & 0.87 & 235033-7840.9~v & 0.56772 & 12.83 & 0.74\\
235118-7416.2~v & 0.56025 & 13.78 & 1.25 & 235401-0740.7~ & 0.52891 & 12.09 & 0.87 & 235442-3328.8~v & 0.61676 & 12.57 & 0.83\\
235522-2618.1~v & 0.72775 & 12.29 & 0.75 & 235702-5826.0~: & 0.68802 & 13.72 & 0.94 & 235807-3345.2~ & 0.65024 & 13.68 & 1.16\\
235811-2000.7~? & 0.53128 & 13.60 & 0.90 &  &  &  &  &  &  &  & \\
\multicolumn{12}{|c|}{\em  Stars classified as DCEP-FU.}\\
000614-3249.0~:v & 150 & 7.62 & 1.69 & 001255-0341.6~ & 74.4 & 11.40 & 0.19 & 001507-0320.0~? & 8.834 & 11.52 & 0.27\\
001902-4853.7~ & 14.41 & 12.24 & 0.29 & 001912-7509.3~? & 9.190 & 11.32 & 0.16 & 002606-1542.3~? & 20.56 & 9.56 & 0.16\\
002737-8321.4~ & 9.721 & 12.33 & 0.21 & 003005-3319.1~ & 24.75 & 11.30 & 0.17 & 003041-4416.4~ & 1.9007 & 13.02 & 1.02\\
003256-7349.4~v & 15.85 & 14.31 & 1.02 & 003307-3201.3~? & 27.60 & 9.84 & 0.18 & 003510-5020.1~?: & 8.354 & 9.85 & 0.13\\
003918-7202.0~v & 18.86 & 13.51 & 0.97 & 004045-7343.0~v & 28.45 & 13.75 & 1.15 & 004156-7332.3~v & 16.79 & 13.71 & 0.65\\
004224-7747.6~ & 2.5716 & 10.29 & 0.09 & 004349-7336.8~v & 32.08 & 13.45 & 1.25 & 004506-1854.2~v & 106.1 & 12.53 & 0.97\\
004655-7242.9~v & 65.7 & 12.09 & 1.05 & 005028-7245.1~v & 84.6 & 11.69 & 0.71 & 005345-7217.3~v & 73.7 & 12.00 & 0.71\\
005557-7159.0~v & 42.6 & 12.92 & 0.85 & 005632-0809.1~ & 31.06 & 11.98 & 0.28 & 005754-7224.8~v & 33.12 & 13.11 & 1.28\\
005905-7203.8~v & 16.24 & 13.90 & 0.89 & 005919-5954.7~ & 5.868 & 13.00 & 0.53 & 010012-3818.6~ & 20.43 & 10.57 & 0.21\\
010145-1208.1~ & 26.89 & 10.09 & 0.18 & 010150-7205.8~?v & 50.2 & 12.84 & 0.66 & 010318-7233.1~v & 17.24 & 13.69 & 0.76\\
010322-7054.3~:v & 215 & 12.17 & 0.89 & 010416-7245.3~v & 208 & 11.80 & 1.04 & 010424-7200.6~v & 29.09 & 13.50 & 1.24\\
010647-7316.2~ & 15.94 & 13.61 & 0.91 & 010718-7313.4~v & 32.86 & 13.36 & 0.80 & 011241-2910.8~v & 7.935 & 8.89 & 0.10\\
011355-5308.3~? & 14.36 & 10.06 & 0.18 & 011400-1222.0~? & 8.569 & 11.84 & 0.20 & 011423-7157.7~ & 11.27 & 14.10 & 1.13\\
011428-7239.9~ & 41.9 & 12.59 & 0.96 & 011658-7343.1~v & 25.42 & 13.74 & 0.92 & 011753-7337.4~v & 22.65 & 13.31 & 0.74\\
011948-6933.4~ & 22.92 & 9.58 & 0.22 & 012000-8133.4~:v & 108.0 & 12.60 & 1.19 & 012008-7254.2~?v & 20.06 & 13.23 & 0.80\\
012831-7347.8~v & 28.94 & 13.03 & 1.29 & 013552-7353.3~v & 15.16 & 13.72 & 0.92 & 014032-7430.4~v & 33.36 & 12.70 & 1.23\\
014049-6729.7~? & 1.1228 & 13.85 & 1.47 & 014818-7303.5~v & 17.67 & 13.85 & 0.79 & 015147-7334.1~? & 6.985 & 10.82 & 0.14\\
015825-2422.7~?v & 91.4 & 10.66 & 0.37 & 015930-3129.3~v & 128 & 10.70 & 1.28 & 015939-1734.8~ & 2.5005 & 11.13 & 0.11\\
020731-2053.2~ & 42.4 & 10.64 & 0.22 & 020851-0721.5~ & 8.932 & 10.24 & 0.19 & 020909-7626.2~ & 53.2 & 12.06 & 0.55\\
021313-8042.6~? & 80.7 & 10.92 & 0.14 & 021710-6325.0~ & 31.47 & 10.04 & 0.27 & 022049-7305.1~v & 21.38 & 13.47 & 0.94\\
023010-6506.2~v & 237 & 10.86 & 1.37 & 023425-0243.8~ & 40.5 & 11.10 & 0.23 & 024704-4018.1~? & 21.20 & 11.84 & 0.21\\
025112-4753.2~ & 13.18 & 11.74 & 0.43 & 030439-8114.0~ & 17.90 & 9.99 & 0.11 & 031115-5701.5~? & 10.608 & 11.33 & 0.23\\
031332-3644.5~? & 38.2 & 11.39 & 0.19 & 031339-1026.5~v & 2.2132 & 11.09 & 0.66 & 031347-1701.9~ & 11.71 & 11.34 & 0.21\\
032642-5901.1~ & 28.03 & 10.31 & 0.16 & 032738-8050.6~ & 20.30 & 10.85 & 0.14 & 032740-5809.8~ & 4.075 & 10.99 & 0.34\\
032919-0701.9~ & 15.48 & 13.27 & 0.56 & 033149-6331.9~? & 2.7320 & 10.89 & 0.14 & 033410-4143.8~? & 15.35 & 11.46 & 0.26\\
033936-2922.0~ & 22.33 & 12.77 & 0.26 & 034025-7054.6~? & 18.83 & 10.69 & 0.29 & 034037-6601.8~ & 22.54 & 10.86 & 0.12\\
034103-3029.6~ & 32.09 & 11.28 & 0.17 & 034324-6126.0~: & 13.36 & 12.94 & 0.26 & 034325-4525.0~ & 18.51 & 10.28 & 0.17\\
034951-2355.8~:v & 125 & 11.89 & 0.74 & 035830-5016.4~? & 8.325 & 9.91 & 0.11 & 035937-3953.3~ & 9.922 & 9.59 & 0.24\\
040404-3801.0~ & 6.307 & 10.67 & 0.15 & 040730-3526.1~? & 28.89 & 10.71 & 0.17 & 040958-5520.2~ & 9.209 & 13.86 & 0.67\\
041425-2049.5~ & 11.39 & 9.39 & 0.11 & 041443-1852.2~? & 60.3 & 9.42 & 0.10 & 041847-3003.8~ & 8.888 & 11.39 & 0.13\\
042538-2702.7~? & 11.96 & 11.42 & 0.20 & 043030-2806.5~? & 33.81 & 11.60 & 0.18 & 043332-7835.7~ & 38.6 & 11.59 & 0.22\\
043939-0501.9~ & 1.7829 & 11.19 & 0.47 & 044049-2150.9~? & 25.33 & 11.14 & 0.12 & 044312-3748.9~? & 45.7 & 11.41 & 0.14\\
044324-6913.7~v & 8.864 & 14.01 & 0.90 & 044518-6916.6~ & 21.17 & 13.37 & 0.39 & 044529-4656.4~ & 16.38 & 12.91 & 0.30\\
044537-7015.0~v & 28.34 & 13.10 & 1.22 & 044707-6917.7~v & 22.52 & 13.62 & 1.12 & 044908-4400.4~? & 31.28 & 11.31 & 0.17\\
044955-6945.3~v & 27.04 & 11.80 & 0.30 & 045004-6815.6~v & 22.33 & 13.03 & 0.87 & 045110-5920.7~? & 21.82 & 11.83 & 0.16\\
045124-3817.4~? & 22.73 & 9.67 & 0.14 & 045205-0941.8~ & 4.778 & 11.19 & 0.22 & 045247-6820.9~ & 20.11 & 14.15 & 0.85\\
045319-3131.9~ & 1.2215 & 13.07 & 0.59 & 045423-7054.1~ & 34.5 & 12.47 & 1.05 & 045506-6728.6~v & 29.85 & 13.32 & 0.66\\
045541-6625.7~v & 98.2 & 11.79 & 0.53 & 045627-6922.9~v & 30.32 & 12.55 & 0.38 & 045702-6759.7~v & 45.1 & 13.05 & 0.65\\
045713-6723.0~v & 22.70 & 13.17 & 0.76 & 045745-6542.5~v & 24.21 & 12.64 & 0.76 & 045751-6750.3~v & 13.98 & 13.59 & 0.74\\
045751-6957.4~v & 23.34 & 13.06 & 1.17 & 045805-6927.2~v & 36.8 & 12.92 & 1.01 & 045811-6957.0~v & 39.4 & 12.57 & 1.06\\
045836-7006.6~v & 17.28 & 13.46 & 1.39 & 045847-7003.7~v & 15.89 & 13.77 & 0.83 & 045848-6719.0~v & 15.22 & 13.21 & 0.63\\
045941-6927.4~v & 31.80 & 12.96 & 1.12 & 045949-4328.0~ & 16.72 & 11.59 & 0.29 & 050008-6827.0~v & 133 & 11.68 & 1.37\\
050055-6638.2~v & 27.89 & 12.64 & 1.05 & 050154-6854.3~v & 20.78 & 13.10 & 0.75 & 050252-7142.1~? & 12.25 & 14.28 & 1.35\\
050308-6613.8~v & 23.96 & 12.81 & 1.22 & 050308-6913.4~v & 17.50 & 13.13 & 0.56 & 050326-6909.0~ & 21.26 & 13.10 & 0.80\\
050346-6852.6~v & 22.34 & 13.41 & 0.75 & 050349-6856.1~v & 25.63 & 13.18 & 0.74 & 050609-7115.4~ & 42.2 & 12.34 & 1.20\\
050615-6640.8~ & 36.5 & 12.21 & 1.34 & 050620-8641.8~? & 7.980 & 11.30 & 0.21 & 050648-7002.2~v & 47.4 & 12.45 & 1.03\\
050716-6853.0~v & 52.2 & 12.07 & 0.40 & 050720-7027.2~v & 26.36 & 12.83 & 1.21 & 050737-0140.1~: & 6.416 & 12.34 & 0.26\\
050803-7157.6~ & 15.63 & 13.69 & 1.30 & 050819-6846.7~v & 30.40 & 12.74 & 0.73 & 050904-7021.9~v & 13.64 & 14.02 & 0.83\\
050915-3813.2~?v & 50.6 & 9.35 & 0.26 & 050920-7027.4~v & 37.5 & 12.43 & 1.13 & 051354-6703.8~ & 48.4 & 12.25 & 1.29\\
051514-6549.1~v & 17.57 & 13.70 & 1.42 & 051833-6712.9~ & 27.96 & 13.66 & 0.82 & 051908-3740.6~? & 17.98 & 10.56 & 0.21\\
052018-6756.8~v & 28.27 & 13.77 & 1.27 & 052032-3036.7~? & 54.4 & 10.72 & 0.15 & 052112-6903.1~v & 23.61 & 13.16 & 0.89\\
052429-7217.1~ & 21.29 & 13.52 & 0.89 & 052502-0052.4~: & 70.6 & 10.71 & 0.22 & 052508-6738.7~v & 48.1 & 11.88 & 1.09\\
052545-7402.7~ & 20.72 & 11.93 & 0.21 & 052548-6721.2~v & 13.77 & 13.57 & 1.26 & 052606-6710.9~? & 1.6510 & 12.29 & 0.37\\
052655-6958.9~v & 28.12 & 13.09 & 0.71 & 052700-7138.6~v & 16.32 & 13.60 & 1.29 & 052926-6614.4~v & 16.89 & 13.77 & 1.21\\
053003-6556.1~ & 26.10 & 13.10 & 1.21 & 053014-6926.3~v & 23.09 & 13.15 & 0.67 & 053122-7057.4~v & 52.5 & 12.46 & 1.01\\
053259-6643.5~ & 16.19 & 13.67 & 1.01 & 053322-0309.9~ & 65.0 & 11.35 & 0.23 & 053416-3626.0~v & 22.35 & 9.53 & 0.18\\
053504-0508.2~ & 5.671 & 10.59 & 0.24 & 053537-6832.1~v & 24.28 & 12.82 & 1.23 & 053543-2955.7~?v & 3.1054 & 9.80 & 0.14\\
053550-6642.1~?v & 30.34 & 12.52 & 1.34 & 053813-2124.7~? & 17.93 & 11.25 & 0.12 & 053820-2021.6~: & 89.9 & 12.14 & 0.67\\
053954-6750.2~v & 24.12 & 13.49 & 0.93 & 054041-1058.2~: & 116 & 12.93 & 0.91 & 054118-7211.9~v & 12.73 & 13.82 & 1.39\\
054420-5523.5~: & 17.95 & 11.45 & 0.41 & 055137-1432.2~ & 1.08259 & 12.68 & 1.00 & 055223-2036.0~? & 25.54 & 9.95 & 0.09\\
060530-5513.2~ & 151 & 10.25 & 1.09 & 072738-8622.6~ & 60.6 & 11.93 & 0.38 & 100112-8844.6~ & 43.5 & 10.70 & 0.27\\
\hline
}
\clearpage

\addtocounter{table}{-1}
\MakeTableSep{|l|r|r|r||l|r|r|r||l|r|r|r|}{10cm}{Continued}{
\hline
\multicolumn{1}{|c|}{ID} & \multicolumn{1}{c|}{$P$} & \multicolumn{1}{c|}{$V$} & \multicolumn{1}{c|}{$\Delta~V$} & \multicolumn{1}{|c|}{ID} & \multicolumn{1}{c|}{$P$} & \multicolumn{1}{c|}{$V$} & \multicolumn{1}{c|}{$\Delta~V$} & \multicolumn{1}{|c|}{ID} & \multicolumn{1}{c|}{$P$} & \multicolumn{1}{c|}{$V$} & \multicolumn{1}{c|}{$\Delta~V$}\\
\hline
\multicolumn{12}{|c|}{\em  Stars classified as DCEP-FU.}\\
222914-8149.7~? & 62.3 & 11.72 & 0.14 & 224542-8654.8~ & 6.068 & 12.41 & 0.26 & 231359-7435.6~? & 18.81 & 10.79 & 0.11\\
231839-6752.5~ & 2.5526 & 13.81 & 1.05 & 233049-8346.7~? & 21.41 & 10.60 & 0.16 & 235119-5250.6~ & 43.3 & 12.50 & 0.30\\
235511-4958.4~ & 12.45 & 10.59 & 0.19 & 235638-2738.7~? & 26.77 & 12.95 & 0.33 & 235930-8636.3~? & 14.96 & 12.27 & 0.22\\
\multicolumn{12}{|c|}{\em  Stars classified as DCEP-FO.}\\
000735-1146.5~? & 0.87971 & 11.14 & 0.17 & 002743-4126.3~? & 0.48520 & 10.58 & 0.12 & 003414-2911.3~? & 5.649 & 12.41 & 0.20\\
003418-4152.3~? & 0.51340 & 12.75 & 0.26 & 005218-6826.4~ & 3.1286 & 12.08 & 0.18 & 005929-1206.9~? & 2.5932 & 12.04 & 0.14\\
010207-1234.1~? & 3.768 & 13.67 & 0.47 & 010241-4321.5~? & 0.88988 & 10.51 & 0.11 & 011329-0738.1~ & 0.57969 & 11.94 & 0.29\\
012110-3729.5~: & 0.63955 & 10.45 & 0.13 & 012943-4900.0~? & 0.57251 & 13.54 & 1.14 & 014028-1606.5~? & 1.5604 & 10.60 & 0.12\\
014508-1914.2~? & 2.5232 & 12.08 & 0.19 & 015523-3846.3~? & 0.89193 & 11.08 & 0.15 & 015656-4149.0~ & 2.5599 & 12.44 & 0.25\\
020017-1620.7~v & 0.82318 & 9.38 & 0.44 & 021947-1025.7~v & 0.53187 & 10.79 & 0.25 & 022023-3221.6~: & 0.62177 & 12.66 & 0.27\\
022821-3557.1~ & 0.55974 & 14.12 & 0.86 & 023314-3112.8~? & 0.94820 & 10.18 & 0.18 & 024033-4034.9~: & 0.49512 & 9.84 & 0.12\\
024755-1036.0~? & 0.75161 & 11.70 & 0.17 & 025357-3707.2~ & 1.5735 & 12.71 & 0.24 & 025436-1315.5~ & 0.69253 & 13.06 & 0.79\\
031152-1121.3~v & 0.71388 & 9.68 & 0.59 & 031243-6524.0~v & 0.54902 & 14.00 & 0.89 & 031305-8341.8~? & 0.71138 & 13.92 & 0.61\\
034338-2504.2~? & 0.55994 & 13.23 & 0.67 & 034612-2725.1~? & 0.61744 & 13.80 & 0.65 & 034731-4202.3~? & 0.55162 & 14.13 & 1.15\\
035525-3137.9~? & 0.73738 & 11.52 & 0.17 & 040037-6014.0~? & 3.2981 & 11.85 & 0.39 & 041015-7926.9~?v & 1.0912 & 13.42 & 0.64\\
041743-1754.4~? & 4.410 & 11.39 & 0.13 & 041928-1225.3~?: & 0.91161 & 12.02 & 0.20 & 042351-7654.7~ & 0.65192 & 12.84 & 0.62\\
043142-0322.1~ & 0.96551 & 12.03 & 0.22 & 043340-4437.7~ & 1.6478 & 11.37 & 0.11 & 043516-4112.4~ & 0.64100 & 12.68 & 0.29\\
043811-0726.3~? & 2.1092 & 10.52 & 0.13 & 044131-5216.6~ & 0.54869 & 13.72 & 0.87 & 044315-4106.3~: & 0.46628 & 10.39 & 0.14\\
044546-1748.7~ & 0.91961 & 12.84 & 0.92 & 045037-1651.7~ & 0.51627 & 11.62 & 0.14 & 045305-4844.6~? & 4.539 & 10.74 & 0.10\\
050421-2129.6~ & 5.378 & 11.63 & 0.18 & 050601-6906.3~v & 0.95220 & 13.79 & 1.15 & 050710-4923.1~? & 0.61643 & 13.77 & 0.79\\
050803-7903.8~ & 0.58334 & 13.75 & 1.20 & 051222-2111.6~ & 0.46053 & 13.41 & 0.58 & 051311-3031.8~? & 5.615 & 11.33 & 0.23\\
051605-6006.9~ & 1.07032 & 12.09 & 0.24 & 051819-2522.4~ & 1.2299 & 11.72 & 0.18 & 052652-2504.8~? & 0.48578 & 9.87 & 0.13\\
053236-0523.0~?v & 1.3269 & 11.98 & 0.39 & 053407-3245.6~ & 0.58504 & 14.13 & 0.86 & 053559-2437.1~? & 0.54686 & 13.66 & 1.04\\
053600-3116.1~? & 4.140 & 12.67 & 0.27 & 054738-0707.6~? & 0.63001 & 11.30 & 0.23 & 054806-2959.3~: & 0.51163 & 13.48 & 1.09\\
060138-2153.1~? & 0.52202 & 12.32 & 0.19 & 060537-2453.4~? & 0.87835 & 13.89 & 0.57 & 061712-7328.9~ & 0.54226 & 11.24 & 0.73\\
073113-8432.3~v & 0.55807 & 12.47 & 1.05 & 230450-7311.6~?: & 0.54612 & 14.29 & 0.93 & 231538-7848.9~v & 0.62762 & 14.00 & 0.78\\
232152-6942.2~ & 1.8732 & 9.96 & 0.12 & 232705-6540.6~: & 0.52201 & 14.00 & 1.35 & 234624-5642.0~: & 0.49061 & 13.91 & 1.18\\
\multicolumn{12}{|c|}{\em  Stars classified as MIRA.}\\
000208-1440.5~v & 352 & 8.35 & 5.27 & 000736-2529.5~v & 321 & 13.29 & 2.06 & 000837-3913.2~v & 304 & 9.26 & 5.41\\
000842-8611.3~v & 340 & 11.35 & 2.35 & 001522-3202.7~v & 425 & 6.72 & 5.42 & 001654-3013.8~v & 171 & 11.44 & 3.21\\
002231-1832.7~ & 197 & 12.37 & 2.47 & 002308-6140.3~v & 246 & 8.83 & 5.20 & 002404-0919.7~v & 320 & 8.46 & 5.40\\
002912-3754.5~?v & 214 & 9.06 & 4.49 & 003026-4624.6~v & 259 & 10.55 & 3.78 & 003419-4300.1~v & 144 & 10.81 & 2.05\\
003503-5012.3~v & 227 & 9.54 & 5.00 & 003939-8002.1~v & 302 & 11.94 & 2.37 & 004930-3454.8~v & 257 & 10.11 & 3.82\\
005300-6953.2~v & 221 & 9.10 & 4.87 & 005712-7459.9~v & 242 & 8.47 & 5.13 & 010529-3141.6~v & 329 & 9.76 & 2.58\\
010645-0128.9~v & 189 & 9.37 & 4.04 & 011028-7650.9~v & 218 & 9.98 & 4.43 & 011136-3006.5~ & 333 & 10.30 & 4.66\\
011402-5646.7~ & 278 & 11.15 & 3.81 & 011953-5555.1~v & 352 & 9.16 & 5.90 & 013334-7512.4~v & 311 & 9.57 & 3.32\\
013407-1014.1~v & 176 & 11.00 & 4.40 & 013426-1858.5~v & 258 & 10.22 & 2.05 & 015329-4900.4~v & 198 & 10.33 & 4.70\\
020534-5708.6~v & 295 & 9.56 & 2.61 & 021600-2031.2~ & 368 & 8.68 & 5.08 & 021921-0258.8~v & 418 & 4.89 & 4.27\\
022148-6305.1~v & 309 & 9.63 & 3.48 & 022516-5934.3~v & 336 & 9.48 & 4.95 & 022602-0010.7~v & 166 & 8.19 & 5.24\\
022735-6747.2~ & 140 & 12.15 & 1.99 & 022915-2605.9~ & 458 & 8.73 & 2.49 & 023124-4127.1~v & 279 & 10.94 & 2.23\\
023158-1930.9~ & 313 & 12.87 & 1.31 & 023344-1308.9~v & 219 & 8.14 & 5.64 & 023552-6235.0~v & 203 & 8.79 & 5.29\\
023723-2658.7~v & 523 & 10.09 & 4.71 & 023835-5408.1~v & 147 & 11.64 & 3.18 & 024422-2912.4~v & 265 & 8.67 & 1.99\\
024745-5903.1~v & 287 & 8.61 & 2.10 & 025353-4953.4~v & 382 & 7.22 & 4.32 & 025816-1557.5~ & 231 & 12.35 & 2.26\\
030052-5038.5~v & 218 & 8.24 & 4.98 & 031118-2341.7~v & 310 & 9.61 & 3.00 & 031153-1152.5~v & 164 & 10.92 & 3.90\\
031926-0103.9~v & 172 & 8.88 & 3.92 & 032844-5558.8~v & 155 & 11.52 & 2.90 & 033229-7626.9~v & 431 & 10.23 & 4.06\\
033412-1609.8~v & 360 & 10.22 & 3.48 & 034431-2513.6~v & 309 & 9.15 & 5.79 & 034748-6230.9~v & 268 & 9.81 & 4.65\\
035003-5721.2~v & 256 & 8.86 & 2.39 & 035029-2457.4~v & 275 & 9.18 & 3.73 & 035155-7358.1~ & 118 & 11.57 & 2.91\\
035247-4549.8~v & 344 & 7.51 & 6.04 & 035343-5456.1~ & 215 & 11.85 & 3.41 & 035514-2401.9~v & 239 & 7.77 & 4.68\\
040412-4550.3~ & 277 & 11.12 & 3.54 & 040551-5450.0~ & 162 & 10.01 & 2.55 & 041131-2508.0~v & 400 & 8.84 & 5.12\\
041741-6035.7~v & 265 & 11.56 & 2.76 & 041755-1830.4~ & 311 & 10.47 & 2.19 & 042020-5420.2~ & 161 & 9.79 & 4.20\\
042850-5430.1~ & 186 & 10.14 & 4.67 & 043333-6301.8~v & 277 & 7.06 & 6.65 & 043648-7018.6~v & 273 & 12.06 & 3.21\\
043917-7421.7~?v & 54.7 & 12.48 & 2.16 & 044030-3814.2~v & 387 & 8.98 & 5.10 & 044148-0757.3~ & 287 & 10.40 & 2.75\\
044509-2351.3~v & 196 & 11.13 & 3.89 & 044550-5947.2~v & 166 & 8.51 & 4.50 & 044730-5541.1~v & 236 & 10.50 & 3.83\\
045059-2859.9~ & 145 & 12.45 & 2.35 & 045524-0655.9~v & 275 & 10.46 & 3.59 & 045701-8416.1~v & 446 & 10.74 & 4.08\\
045936-1448.4~v & 397 & 7.52 & 1.90 & 050135-6805.9~v & 360 & 11.58 & 3.21 & 050417-2228.5~ & 192 & 12.48 & 2.42\\
050451-2154.3~v & 345 & 8.04 & 3.75 & 051009-6419.1~v & 365 & 10.47 & 3.74 & 051057-4830.4~v & 398 & 8.24 & 5.01\\
051102-0513.8~v & 126 & 11.54 & 2.83 & 051506-4655.1~v & 198 & 8.40 & 5.17 & 051639-4053.0~v & 488 & 10.40 & 2.33\\
051820-1621.1~v & 271 & 9.03 & 5.46 & 051917-3342.5~v & 233 & 8.04 & 3.86 & 051926-1350.9~v & 206 & 10.84 & 4.10\\
052145-1555.6~v & 199 & 10.55 & 2.67 & 052241-1208.5~ & 184 & 12.15 & 2.29 & 052428-1039.4~v & 451 & 9.78 & 2.11\\
052544-1423.5~v & 173 & 11.09 & 3.65 & 052606-8623.3~v & 436 & 8.19 & 5.29 & 052613-0016.9~v & 193 & 9.24 & 1.90\\
052613-2850.4~ & 170 & 10.74 & 4.32 & 052901-0441.6~v & 420 & 8.42 & 4.85 & 053221-1037.4~v & 145 & 10.13 & 2.33\\
053246-2800.8~v & 260 & 11.62 & 2.32 & 053410-5519.5~v & 298 & 10.56 & 4.31 & 053600-4404.6~?v & 190 & 11.18 & 3.16\\
053656-1304.6~ & 246 & 11.75 & 2.95 & 053717-5538.4~v & 315 & 9.87 & 4.16 & 054053-7320.9~ & 195 & 12.61 & 2.31\\
054134-0407.9~v & 136 & 10.42 & 2.68 & 054233-2341.7~v & 412 & 8.85 & 5.61 & 054906-2241.3~ & 191 & 11.71 & 2.11\\
054957-5252.1~ & 162 & 11.16 & 2.09 & 055033-2911.9~v & 345 & 8.06 & 4.18 & 055329-5226.0~ & 102 & 12.07 & 2.33\\
055412-2241.8~v & 466 & 9.91 & 2.24 & 055938-2555.9~v & 145 & 12.65 & 1.75 & 060316-1003.6~v & 264 & 12.02 & 1.63\\
060913-2743.1~ & 347 & 12.73 & 1.45 & 060932-6007.3~v & 152 & 9.87 & 2.95 & 060946-1306.9~v & 142 & 11.34 & 2.26\\
113924-8709.4~ & 186 & 11.81 & 3.28 & 174436-8629.2~v & 332 & 10.75 & 2.65 & 180844-8647.9~v & 248 & 8.13 & 5.52\\
225223-7918.3~v & 138 & 12.09 & 2.32 & 230109-8703.2~v & 175 & 9.93 & 3.87 & 232340-6622.0~v & 149 & 10.99 & 3.58\\
234349-1517.1~v & 343 & 8.00 & 2.95 & 234405+0022.9~ & 252 & 10.04 & 2.19 & 234917-4306.7~ & 265 & 10.67 & 3.71\\
235627-4947.2~v & 262 & 8.69 & 4.58 & 235630-5412.1~ & 160 & 12.21 & 1.96 & 235726-6523.1~v & 295 & 9.80 & 5.22\\
235754-0857.5~ & 243 & 9.98 & 4.21 & 235844-3927.0~v & 511 & 9.67 & 5.91 & 235905-5324.1~v & 268 & 11.05 & 4.88\\
\multicolumn{12}{|c|}{\em  Stars classified as PULS.}\\
004145-7343.4~v & 125 & 11.68 & 0.57 & 010401-7213.9~:v & 16.63 & 13.45 & 0.88 & 012727-3645.8~ & 10.442 & 11.04 & 0.34\\
015034-1739.0~v & 222 & 9.86 & 1.20 & 031101-6527.8~: & 64.2 & 12.09 & 0.59 & 033247-6356.4~ & 134 & 12.03 & 1.42\\
035510-5330.5~v & 166 & 10.35 & 1.48 & 044548-4300.9~ & 66.77 & 13.38 & 1.61 & 045627-6441.7~v & 112 & 11.72 & 1.29\\
\multicolumn{12}{|c|}{\em  Stars classified as MISC.}\\
000118-3551.7~ & 25.67 & 9.84 & 0.35 & 000119-3505.9~ & 37.3 & 10.77 & 0.27 & 000120-5834.8~ & 53.6 & 9.53 & 0.33\\
000139-0345.4~v & 150 & 12.82 & 1.04 & 000142-4229.3~ & 32.44 & 10.71 & 0.29 & 000155-6707.7~ & 228 & 12.74 & 2.41\\
000157-5250.1~ & 30.95 & 10.78 & 0.26 & 000239-1926.7~v & 72.8 & 9.69 & 0.35 & 000309-1050.5~ & 68.2 & 12.01 & 0.89\\
000341-3906.9~ & 72.6 & 11.65 & 0.59 & 000407-0818.7~ & 64.9 & 12.27 & 0.57 & 000416-5815.9~ & 88.9 & 11.01 & 0.18\\
000457-5554.1~ & 34.49 & 10.50 & 0.25 & 000506+0043.8~ & 68.5 & 10.79 & 0.58 & 000600-4840.8~v & 41.6 & 9.79 & 0.25\\
000633-3849.0~ & 37.2 & 11.41 & 0.23 & 000636-3235.6~v & 71.3 & 8.97 & 0.56 & 000636-3746.5~v & 53.3 & 9.54 & 0.41\\
000659-0846.7~ & 43.1 & 9.59 & 0.20 & 000716-6338.3~ & 39.7 & 10.12 & 0.78 & 000724-5212.2~ & 42.3 & 10.73 & 0.29\\
\hline
}
\clearpage

\addtocounter{table}{-1}
\MakeTableSep{|l|r|r|r||l|r|r|r||l|r|r|r|}{10cm}{Continued}{
\hline
\multicolumn{1}{|c|}{ID} & \multicolumn{1}{c|}{$P$} & \multicolumn{1}{c|}{$V$} & \multicolumn{1}{c|}{$\Delta~V$} & \multicolumn{1}{|c|}{ID} & \multicolumn{1}{c|}{$P$} & \multicolumn{1}{c|}{$V$} & \multicolumn{1}{c|}{$\Delta~V$} & \multicolumn{1}{|c|}{ID} & \multicolumn{1}{c|}{$P$} & \multicolumn{1}{c|}{$V$} & \multicolumn{1}{c|}{$\Delta~V$}\\
\hline
\multicolumn{12}{|c|}{\em  Stars classified as MISC.}\\
000732-1954.5~ & 59.1 & 13.12 & 0.49 & 000739+0120.8~ & 1.9399 & 10.07 & 0.53 & 000819-4949.0~ & 59.5 & 11.01 & 0.85\\
000829-2558.5~v & 41.9 & 8.78 & 0.27 & 000842-1151.2~ & 35.64 & 11.27 & 0.19 & 001027-5635.0~ & 20.95 & 9.43 & 0.17\\
001152-4807.8~ & 33.20 & 9.16 & 0.21 & 001201-2434.0~ & 101.2 & 9.91 & 0.73 & 001202-5440.0~ & 106.8 & 12.17 & 0.86\\
001205-1305.9~ & 256 & 11.61 & 0.27 & 001215-8312.0~ & 27.29 & 10.94 & 0.18 & 001217-3511.2~ & 64.4 & 11.12 & 0.33\\
001230-7405.1~v & 112 & 9.93 & 0.52 & 001237-1639.5~ & 100.6 & 11.58 & 0.62 & 001246-1101.3~ & 197 & 9.45 & 1.52\\
001256-2451.5~ & 4.004 & 11.30 & 0.14 & 001330-7010.3~ & 28.11 & 10.47 & 0.23 & 001347-0748.6~ & 120 & 11.06 & 1.00\\
001512-6013.5~v & 103.3 & 9.37 & 1.48 & 001513-6851.0~v & 34.86 & 8.77 & 0.16 & 001521-4637.9~ & 48.2 & 11.11 & 0.37\\
001549-7625.8~ & 46.2 & 9.71 & 0.21 & 001714-0536.7~ & 377 & 10.88 & 0.46 & 001724-7559.4~ & 41.6 & 11.70 & 0.26\\
001854-7401.9~v & 43.9 & 11.80 & 0.20 & 002020-5709.8~v & 53.0 & 8.82 & 0.39 & 002023-2323.0~v & 97.7 & 10.38 & 0.49\\
002108-4802.6~ & 350 & 10.25 & 0.45 & 002109-0221.5~ & 26.46 & 9.50 & 0.39 & 002209-7707.1~ & 58.6 & 10.43 & 0.59\\
002219-8351.1~ & 189 & 12.86 & 0.92 & 002251-5920.0~ & 71.1 & 10.69 & 0.67 & 002258-7207.0~v & 257 & 11.43 & 0.51\\
002331-7222.6~v & 64.9 & 11.87 & 0.39 & 002334-1751.5~ & 29.12 & 13.23 & 0.33 & 002340-2049.7~ & 44.1 & 10.21 & 0.51\\
002411-1649.7~ & 59.7 & 9.39 & 0.57 & 002417-7558.4~ & 513 & 11.83 & 1.04 & 002430-3244.2~ & 40.3 & 10.76 & 0.26\\
002431-0954.0~ & 103.9 & 10.67 & 0.43 & 002458-7041.8~ & 32.79 & 9.75 & 0.14 & 002504-7209.5~v & 44.7 & 11.44 & 0.49\\
002602-3451.8~ & 54.9 & 9.83 & 0.27 & 002656-7933.0~v & 103.2 & 12.01 & 1.41 & 002706-0636.3~v & 179 & 8.99 & 0.63\\
002708-8436.6~ & 10.545 & 10.24 & 0.09 & 002718-6819.8~ & 37.8 & 9.98 & 0.18 & 002808-8224.4~ & 45.5 & 9.96 & 0.21\\
002941-2034.8~ & 46.8 & 10.26 & 0.19 & 003030-5612.1~v & 127 & 11.44 & 1.07 & 003037-7550.1~ & 47.9 & 10.79 & 0.12\\
003114-1051.2~ & 32.01 & 9.01 & 0.13 & 003125-0938.5~ & 282 & 13.02 & 0.83 & 003222-1839.1~ & 64.1 & 11.72 & 0.48\\
003256-0553.1~ & 39.6 & 11.44 & 0.29 & 003314-1043.7~v & 96.8 & 9.47 & 0.13 & 003320-0623.6~ & 36.5 & 9.57 & 0.42\\
003336-4847.6~ & 71.5 & 12.97 & 0.53 & 003547-7352.7~v & 39.3 & 8.70 & 0.12 & 003606-5549.5~v & 46.8 & 10.88 & 0.35\\
003613-5231.2~ & 39.2 & 11.41 & 0.18 & 003615-1711.0~ & 44.2 & 10.56 & 0.26 & 003629-3020.0~v & 51.4 & 8.84 & 0.32\\
003651-4439.4~ & 48.1 & 9.65 & 0.20 & 003723-3357.4~ & 49.5 & 11.50 & 0.24 & 003753-8237.2~v & 107.7 & 9.22 & 1.02\\
003859-1927.7~ & 42.0 & 11.58 & 0.28 & 003925-6307.2~ & 50.2 & 9.74 & 0.31 & 004108-2720.8~ & 0.329590 & 10.38 & 0.20\\
004201-0938.9~v & 163 & 8.71 & 0.30 & 004237-6740.4~ & 78.8 & 11.38 & 0.63 & 004409-0214.8~ & 73.0 & 8.98 & 0.40\\
004430-3339.2~v & 31.76 & 9.00 & 0.25 & 004451-7615.3~ & 79.5 & 12.41 & 0.42 & 004506-7305.5~ & 324 & 12.51 & 0.63\\
004619-7240.6~ & 421 & 12.86 & 0.85 & 004658-4758.8~v & 69.1 & 9.22 & 0.50 & 004753-0119.0~v & 98.2 & 11.59 & 1.29\\
004842-0631.9~ & 25.67 & 10.17 & 0.34 & 004854-7322.7~ & 213 & 12.69 & 0.55 & 004901-5605.8~v & 510 & 10.41 & 0.18\\
004902-7259.6~ & 300 & 12.40 & 0.59 & 004927-7318.3~v & 384 & 11.59 & 0.23 & 004929-5545.9~v & 38.7 & 9.62 & 0.33\\
004942-1141.0~ & 59.2 & 10.34 & 1.07 & 004957-7337.7~ & 141 & 12.75 & 0.29 & 005002-4843.8~ & 200 & 9.91 & 1.08\\
005009-0223.4~ & 447 & 11.90 & 0.75 & 005106-7243.4~v & 420 & 11.16 & 0.22 & 005124-5937.8~v & 75.5 & 9.66 & 0.38\\
005125-7238.8~v & 425 & 12.49 & 0.63 & 005133-0814.9~ & 43.1 & 9.67 & 0.75 & 005151-5550.4~ & 64.8 & 12.08 & 0.49\\
005153-7211.6~ & 341 & 12.79 & 0.57 & 005311-7304.1~v & 326 & 11.95 & 0.21 & 005336-7301.6~ & 211 & 12.69 & 0.32\\
005338-6115.9~ & 10.828 & 9.48 & 0.12 & 005348-7202.2~ & 1484 & 12.42 & 0.38 & 005354-7142.8~ & 350 & 12.33 & 0.86\\
005444-7005.6~ & 91.4 & 12.41 & 0.71 & 005455-8002.4~v & 88.9 & 10.53 & 0.97 & 005507-7300.7~ & 309 & 12.43 & 0.35\\
005536-7236.4~v & 396 & 12.74 & 0.37 & 005553-7318.4~v & 79.3 & 11.67 & 0.57 & 005606-5810.0~ & 49.0 & 11.37 & 0.29\\
005611-1329.7~ & 39.5 & 12.45 & 0.21 & 005627-7328.4~ & 4.011 & 12.80 & 0.27 & 005827-2422.2~ & 50.6 & 10.27 & 0.29\\
005931-7215.8~v & 407 & 12.54 & 0.33 & 005935-7204.1~v & 321 & 12.53 & 0.48 & 010014-4831.1~ & 0.39889 & 12.97 & 0.60\\
010042-7210.6~v & 351 & 12.46 & 0.61 & 010044-3228.4~ & 248 & 11.25 & 0.40 & 010054-7251.7~v & 394 & 11.64 & 0.66\\
010056-7137.9~v & 400 & 11.68 & 1.62 & 010216-1050.3~ & 56.4 & 9.99 & 0.24 & 010319-6421.8~ & 73.3 & 9.31 & 0.40\\
010328-7252.2~ & 513 & 11.71 & 0.48 & 010329-5645.6~ & 145 & 11.60 & 0.45 & 010431-7032.4~v & 47.2 & 8.80 & 0.40\\
010439-7201.4~ & 412 & 12.06 & 0.65 & 010520-8152.6~v & 48.7 & 12.10 & 0.54 & 010534-1131.9~ & 48.6 & 12.21 & 0.39\\
010628-2055.6~ & 58.4 & 10.73 & 0.34 & 010646-4904.0~ & 27.30 & 10.78 & 0.15 & 010648-3615.1~ & 35.53 & 10.18 & 0.14\\
010648-7216.3~ & 379 & 11.55 & 0.59 & 010718-7228.1~v & 318 & 10.08 & 0.32 & 010753-7210.7~ & 256 & 12.13 & 0.25\\
010753-8631.2~ & 82.3 & 11.57 & 0.24 & 010807-0750.7~ & 48.3 & 11.31 & 0.33 & 010810-5114.0~ & 40.1 & 10.69 & 0.12\\
010839-1703.8~v & 56.6 & 8.70 & 0.43 & 010852-5530.4~ & 59.7 & 10.74 & 0.27 & 010906-7238.5~ & 85.1 & 9.43 & 0.51\\
010931-5249.7~ & 34.18 & 10.64 & 0.26 & 010939-7320.0~v & 545 & 11.56 & 1.10 & 011003-7236.9~ & 377 & 12.86 & 1.67\\
011022-5433.7~ & 24.43 & 9.71 & 0.16 & 011101-7136.5~ & 69.2 & 11.45 & 0.34 & 011109-4241.4~ & 63.9 & 9.35 & 0.67\\
011205-5145.1~ & 41.5 & 8.88 & 0.19 & 011313-1655.2~ & 34.76 & 10.05 & 0.14 & 011319-4004.4~ & 53.0 & 11.09 & 0.98\\
011323-6734.9~ & 28.34 & 11.11 & 0.26 & 011339-2617.2~v & 50.7 & 10.68 & 0.37 & 011415-0210.7~v & 325 & 8.60 & 0.59\\
011431-6722.1~ & 39.3 & 10.35 & 0.21 & 011437-7348.9~ & 129 & 13.00 & 0.56 & 011651-6101.1~ & 211 & 11.53 & 1.24\\
011653-2350.7~v & 67.8 & 9.62 & 0.64 & 011717-5851.1~ & 46.5 & 11.49 & 0.26 & 011725-7343.6~ & 38.9 & 11.04 & 0.32\\
011924-2314.2~ & 34.61 & 10.81 & 0.29 & 012010-5145.4~ & 38.8 & 9.19 & 0.21 & 012017-0944.9~v & 21.75 & 9.12 & 0.16\\
012037-0824.9~v & 151 & 9.23 & 0.56 & 012057-6938.4~ & 73.4 & 11.89 & 0.53 & 012236-0456.2~v & 96.6 & 9.87 & 0.30\\
012247-1609.0~ & 67.7 & 13.49 & 0.78 & 012318-6231.4~ & 48.9 & 9.83 & 0.33 & 012325-1515.3~ & 17.52 & 10.15 & 0.15\\
012326-4359.4~ & 99.0 & 12.40 & 0.67 & 012340-7823.2~v & 83.9 & 12.66 & 0.74 & 012401-3736.0~v & 46.6 & 9.07 & 0.27\\
012410-7223.5~v & 341 & 12.50 & 0.50 & 012416-1646.8~ & 51.3 & 10.48 & 0.49 & 012438-5733.8~ & 33.67 & 10.30 & 0.17\\
012450-4245.8~v & 19.37 & 8.94 & 0.11 & 012514-4555.6~v & 29.18 & 8.68 & 0.35 & 012646-6224.5~ & 52.9 & 10.60 & 0.36\\
012702-1837.8~ & 72.8 & 10.48 & 0.21 & 012806-3142.2~ & 61.4 & 10.21 & 0.60 & 012810-4524.6~v & 107.1 & 11.71 & 0.99\\
012850-2729.1~ & 31.04 & 11.85 & 0.17 & 012918-7302.0~ & 186 & 11.47 & 0.22 & 013032-7318.7~v & 391 & 12.40 & 0.76\\
013139-7056.0~ & 70.5 & 9.36 & 0.15 & 013200-1649.4~ & 98.3 & 11.38 & 0.32 & 013404-4314.5~v & 53.2 & 9.88 & 0.43\\
013407-2246.7~ & 21.24 & 9.04 & 0.15 & 013420-6431.1~v & 40.4 & 9.00 & 0.21 & 013503-8255.7~ & 49.2 & 11.97 & 0.44\\
013547-3837.1~ & 25.29 & 8.64 & 0.16 & 013548-1122.5~v & 218 & 9.58 & 1.50 & 013608-3511.2~ & 127 & 11.67 & 0.70\\
013632-5019.8~ & 0.174195 & 14.46 & 0.98 & 013732-3406.0~ & 68.1 & 11.76 & 0.48 & 013935-0754.4~ & 98.0 & 11.18 & 0.68\\
014000-1206.2~ & 102.4 & 11.38 & 0.50 & 014030-6116.8~ & 38.7 & 10.12 & 0.13 & 014124-6906.1~ & 77.4 & 10.36 & 0.83\\
014138-5934.9~ & 39.8 & 10.70 & 0.37 & 014224-4640.9~v & 319 & 9.51 & 0.59 & 014318-3618.9~ & 83.6 & 11.73 & 0.29\\
014348-6713.9~ & 60.3 & 11.50 & 0.32 & 014352-3409.0~ & 26.98 & 9.72 & 0.19 & 014420-2044.5~ & 51.0 & 11.47 & 0.56\\
014423-3453.5~ & 61.8 & 10.03 & 0.50 & 014438-7722.2~v & 46.5 & 11.35 & 0.32 & 014458-8011.1~v & 56.5 & 8.99 & 0.42\\
014532-4155.7~v & 89.1 & 10.24 & 0.46 & 014553-2404.0~? & 53.2 & 10.93 & 0.23 & 014648-6448.8~ & 71.2 & 10.95 & 0.33\\
014726-2658.1~ & 26.36 & 11.56 & 0.15 & 014820-8300.1~ & 138 & 10.34 & 0.23 & 014851-6439.4~ & 63.3 & 9.88 & 0.33\\
014930-6917.8~ & 42.9 & 11.11 & 0.25 & 014932-5421.4~ & 52.3 & 10.21 & 0.25 & 014944-3537.9~ & 69.8 & 10.00 & 0.35\\
014945-7134.3~ & 99.0 & 10.21 & 0.41 & 014951-4258.2~v & 46.9 & 9.87 & 0.21 & 014955-0451.6~v & 58.1 & 8.56 & 0.61\\
015019-4716.2~ & 44.0 & 10.41 & 0.33 & 015142-3648.3~v & 46.7 & 9.34 & 0.23 & 015221-1505.3~ & 45.2 & 9.72 & 0.37\\
015226-2646.0~ & 63.6 & 9.73 & 0.08 & 015254-7209.2~ & 51.9 & 10.11 & 0.48 & 015318-4614.4~ & 123 & 12.29 & 0.58\\
015417-7622.8~v & 62.4 & 9.91 & 0.32 & 015446-1809.1~ & 337 & 10.90 & 0.73 & 015452-6616.0~ & 58.3 & 11.43 & 0.49\\
015516-0219.4~ & 71.6 & 12.10 & 0.41 & 015518-3548.2~v & 63.9 & 9.19 & 0.79 & 015604-2526.3~ & 30.54 & 9.65 & 0.11\\
015756-7830.5~ & 31.15 & 9.26 & 0.28 & 015757-2628.9~v & 79.5 & 9.45 & 0.57 & 015827-3648.9~v & 100.5 & 9.48 & 0.35\\
020104-7805.5~v & 53.7 & 10.71 & 0.19 & 020213-1619.9~ & 39.2 & 10.09 & 0.17 & 020217-3057.5~ & 386 & 11.05 & 0.70\\
020219-2751.6~ & 107.0 & 11.35 & 0.37 & 020440-2018.0~ & 43.6 & 11.79 & 0.46 & 020453-1344.1~ & 34.54 & 11.20 & 0.23\\
020527-7448.6~ & 44.3 & 12.68 & 0.59 & 020531-4834.4~ & 79.2 & 9.98 & 0.41 & 020606-1012.7~v & 80.9 & 9.25 & 0.33\\
020638-4035.8~ & 55.2 & 11.90 & 0.32 & 020753-5645.6~ & 70.7 & 10.85 & 0.12 & 020801-3619.1~ & 66.2 & 11.57 & 0.56\\
020956-5909.3~ & 0.317042 & 13.85 & 0.67 & 021008-5430.6~v & 5.898 & 10.15 & 0.19 & 021011-5457.6~ & 35.26 & 11.57 & 0.23\\
021029-1835.8~ & 72.8 & 11.42 & 0.24 & 021106-5814.3~v & 49.6 & 12.07 & 0.52 & 021140-2925.5~? & 28.14 & 9.89 & 0.21\\
021148-2341.7~ & 79.6 & 10.00 & 0.37 & 021149-7129.1~v & 85.9 & 9.01 & 0.78 & 021253-3224.8~ & 0.267860 & 14.61 & 0.84\\
021255-6823.0~ & 56.2 & 11.66 & 0.47 & 021327-2338.3~ & 39.4 & 11.16 & 0.31 & 021411-8143.8~v & 129 & 12.06 & 0.74\\
\hline
}
\clearpage

\addtocounter{table}{-1}
\MakeTableSep{|l|r|r|r||l|r|r|r||l|r|r|r|}{10cm}{Continued}{
\hline
\multicolumn{1}{|c|}{ID} & \multicolumn{1}{c|}{$P$} & \multicolumn{1}{c|}{$V$} & \multicolumn{1}{c|}{$\Delta~V$} & \multicolumn{1}{|c|}{ID} & \multicolumn{1}{c|}{$P$} & \multicolumn{1}{c|}{$V$} & \multicolumn{1}{c|}{$\Delta~V$} & \multicolumn{1}{|c|}{ID} & \multicolumn{1}{c|}{$P$} & \multicolumn{1}{c|}{$V$} & \multicolumn{1}{c|}{$\Delta~V$}\\
\hline
\multicolumn{12}{|c|}{\em  Stars classified as MISC.}\\
021446-5837.7~ & 41.9 & 9.41 & 0.25 & 021511-1324.4~ & 45.4 & 10.41 & 0.41 & 021712-7925.4~v & 44.8 & 12.02 & 0.53\\
021741-5639.9~ & 81.9 & 11.13 & 0.45 & 021829-6821.1~ & 68.7 & 10.87 & 0.34 & 022145-3759.6~ & 167 & 10.83 & 0.88\\
022243-1012.2~v & 60.3 & 8.77 & 0.40 & 022332-4034.3~ & 68.2 & 12.22 & 1.10 & 022458-5548.8~ & 47.8 & 10.99 & 0.28\\
022508-1324.0~v & 73.8 & 10.92 & 0.18 & 022552-2955.7~ & 44.0 & 13.02 & 0.49 & 022600-1631.2~ & 45.9 & 10.72 & 0.35\\
022607-5814.5~ & 390 & 10.97 & 1.08 & 022729-1844.6~ & 1.4186 & 11.84 & 0.18 & 022748-5736.8~ & 69.6 & 11.30 & 0.24\\
022806+0005.2~ & 141 & 11.48 & 0.56 & 022808-5542.4~ & 36.7 & 11.07 & 0.15 & 022855-1553.8~ & 51.5 & 9.41 & 0.33\\
022856-3630.3~ & 32.29 & 10.97 & 0.35 & 022902-1637.2~ & 40.3 & 10.38 & 0.56 & 023143-1430.3~ & 68.2 & 10.67 & 0.69\\
023145-6222.6~ & 43.5 & 10.74 & 0.27 & 023216-5502.8~ & 113 & 12.88 & 0.50 & 023221-1359.7~ & 33.86 & 10.34 & 0.20\\
023237-1643.6~ & 65.2 & 9.96 & 0.91 & 023243-4550.7~ & 35.9 & 11.10 & 0.23 & 023334-4010.5~ & 157 & 9.79 & 0.14\\
023440-0247.7~v & 21.02 & 8.64 & 0.15 & 023442-2053.8~ & 126 & 11.17 & 0.49 & 023507-5935.9~ & 125 & 12.93 & 1.55\\
023527-6011.0~? & 52.2 & 10.04 & 0.15 & 023535-6118.0~ & 39.7 & 11.22 & 0.42 & 023553-1900.1~ & 60.1 & 11.04 & 0.48\\
023716-0253.2~ & 278 & 11.01 & 0.99 & 023747-4537.2~v & 43.9 & 8.66 & 0.32 & 023916-2927.3~ & 52.0 & 11.63 & 0.54\\
024003-0859.7~ & 29.06 & 10.44 & 0.24 & 024042-5025.7~ & 27.70 & 9.50 & 0.12 & 024101-6806.0~ & 22.80 & 10.22 & 0.09\\
024129-3731.5~ & 81.7 & 12.42 & 0.44 & 024156+0031.6~ & 273 & 9.97 & 0.57 & 024210-3848.4~ & 33.00 & 10.61 & 0.38\\
024229-8131.6~ & 28.40 & 11.16 & 0.16 & 024230-8804.4~ & 47.6 & 11.61 & 0.26 & 024234-3345.9~ & 51.7 & 11.54 & 0.36\\
024235-6246.0~ & 35.1 & 11.22 & 0.14 & 024258-2607.1~v & 90.5 & 8.68 & 0.37 & 024412-2001.8~ & 26.82 & 10.00 & 0.17\\
024415-5418.1~v & 64.6 & 8.84 & 0.38 & 024452-1347.6~ & 50.0 & 10.59 & 0.24 & 024528-1359.5~ & 75.8 & 8.89 & 0.32\\
024542-3121.5~ & 59.8 & 10.28 & 0.31 & 024642-4131.5~v & 68.8 & 9.60 & 0.63 & 024749-3838.1~ & 61.7 & 9.93 & 0.32\\
024820-6150.1~ & 67.0 & 11.77 & 0.26 & 024924-8218.7~ & 48.9 & 12.85 & 0.47 & 024943-6948.7~v & 59.5 & 8.65 & 0.34\\
024947-3512.5~ & 35.8 & 10.27 & 0.41 & 025006-8609.5~ & 138 & 12.61 & 0.67 & 025314-3746.3~v & 148 & 10.08 & 1.53\\
025351-7952.0~ & 61.0 & 11.06 & 0.14 & 025402-7009.2~ & 85.6 & 9.53 & 0.55 & 025416-3910.7~ & 39.1 & 10.68 & 0.65\\
025429-7556.7~ & 311 & 10.37 & 0.11 & 025434-2725.4~v & 26.63 & 11.51 & 0.24 & 025453-5157.2~v & 27.63 & 8.79 & 0.13\\
025502-4411.0~ & 60.5 & 10.88 & 0.35 & 025619-6611.3~ & 43.7 & 10.45 & 0.20 & 025620-2507.1~v & 79.7 & 11.02 & 0.52\\
025721-1615.7~ & 68.5 & 12.49 & 0.43 & 025728-4931.4~ & 32.82 & 9.41 & 0.30 & 025858-1320.7~v & 51.3 & 8.63 & 0.43\\
030027-1335.3~v & 48.9 & 10.78 & 0.33 & 030035-8803.1~ & 252 & 11.82 & 1.86 & 030036-2701.8~ & 33.84 & 11.97 & 0.17\\
030129-0850.4~ & 64.6 & 11.31 & 0.20 & 030203-8002.1~v & 424 & 10.46 & 1.47 & 030227-7652.0~ & 62.9 & 12.56 & 0.43\\
030311-2660.0~ & 66.5 & 10.60 & 1.12 & 030313-8538.3~v & 50.3 & 8.86 & 0.23 & 030349-1413.5~v & 33.46 & 9.55 & 0.23\\
030409-7318.8~ & 61.6 & 12.33 & 0.67 & 030456-5820.0~ & 0.97723 & 10.98 & 0.35 & 030521-5218.0~ & 73.1 & 10.35 & 0.58\\
030540-3827.8~ & 98.6 & 11.22 & 0.16 & 030615-3942.5~ & 64.0 & 11.23 & 0.55 & 030638+0109.5~ & 37.3 & 10.07 & 0.59\\
030754-2709.0~ & 30.33 & 10.53 & 0.22 & 030838-2626.8~v & 42.0 & 8.65 & 0.25 & 030842-7547.4~v & 85.7 & 11.35 & 0.98\\
030943-6442.5~v & 61.1 & 11.39 & 0.75 & 030954-2731.4~ & 117 & 11.53 & 0.42 & 031002-4047.7~ & 61.9 & 12.48 & 0.46\\
031108-8453.6~ & 57.0 & 10.62 & 0.14 & 031222-7259.1~ & 81.3 & 9.25 & 0.29 & 031333-4633.7~ & 57.5 & 10.64 & 0.19\\
031336-0239.5~ & 46.0 & 10.16 & 0.25 & 031454-6657.1~v & 59.0 & 11.10 & 0.32 & 031500-1008.0~ & 17.90 & 12.16 & 0.22\\
031544-4532.4~v & 20.45 & 9.04 & 0.16 & 031547-5750.0~ & 95.7 & 12.07 & 0.39 & 031555-7636.3~ & 56.3 & 11.55 & 0.23\\
031618-2727.3~ & 50.9 & 10.97 & 0.21 & 031625-1935.2~ & 71.3 & 10.69 & 0.22 & 031853-3129.2~ & 160 & 11.45 & 0.58\\
031939-6918.2~v & 119 & 11.32 & 0.97 & 032009-7136.2~v & 44.2 & 10.22 & 0.92 & 032016-0006.5~ & 335 & 9.65 & 0.85\\
032020-7551.2~ & 51.0 & 10.46 & 0.38 & 032029-0342.0~ & 68.8 & 11.58 & 0.38 & 032110-1713.9~v & 163 & 10.90 & 1.19\\
032124-1950.6~ & 47.0 & 12.25 & 0.44 & 032130-2127.4~v & 83.6 & 8.94 & 0.93 & 032229-3936.3~ & 19.03 & 10.16 & 0.12\\
032242-3041.6~ & 41.6 & 11.24 & 0.19 & 032255-7923.7~ & 39.3 & 10.79 & 0.29 & 032258-7047.3~ & 309 & 10.15 & 0.31\\
032323-8503.8~ & 449 & 12.37 & 1.03 & 032336-1953.0~v & 80.1 & 10.39 & 0.69 & 032353-2520.4~ & 160 & 11.87 & 0.31\\
032433-5409.2~ & 85.5 & 10.62 & 0.40 & 032503-7856.9~ & 47.7 & 10.62 & 0.38 & 032509-1221.3~v & 173 & 9.33 & 0.74\\
032512-7839.9~v & 50.4 & 8.50 & 0.18 & 032535-8139.2~? & 27.09 & 10.57 & 0.18 & 032659-5333.8~v & 24.57 & 9.10 & 0.18\\
032705-0416.0~ & 76.4 & 11.46 & 0.75 & 032719-0730.7~ & 75.5 & 9.95 & 0.21 & 032737-8233.8~ & 119 & 11.86 & 0.51\\
032822-1524.2~ & 77.8 & 11.76 & 0.43 & 032846-6041.4~ & 93.6 & 10.58 & 0.49 & 032935-2824.2~v & 90.2 & 9.49 & 1.16\\
032958-0655.6~ & 52.4 & 10.91 & 0.29 & 033015-4138.9~ & 27.01 & 9.32 & 0.21 & 033029-7150.4~ & 53.5 & 10.85 & 0.31\\
033042-0156.4~ & 58.8 & 9.85 & 0.33 & 033104-1524.8~v & 70.0 & 11.68 & 0.98 & 033104-6829.5~ & 29.75 & 10.92 & 0.27\\
033124-7527.4~v & 146 & 12.77 & 1.38 & 033227-2539.5~v & 77.5 & 7.93 & 0.50 & 033313-5324.6~ & 22.39 & 9.27 & 0.11\\
033316-0606.7~ & 411 & 11.49 & 0.39 & 033324-6000.9~ & 39.8 & 11.34 & 0.25 & 033405-4718.7~ & 105.6 & 12.46 & 0.92\\
033425-2537.7~ & 106.6 & 11.19 & 0.94 & 033454-6459.9~ & 51.1 & 11.16 & 0.24 & 033516-5105.8~ & 64.5 & 11.74 & 0.36\\
033553-6911.6~v & 32.29 & 8.55 & 0.12 & 033620-6129.4~v & 18.82 & 9.28 & 0.11 & 033630-3219.6~ & 40.9 & 9.97 & 0.25\\
033718-8516.7~ & 51.9 & 10.08 & 0.23 & 033719-5225.5~ & 53.0 & 10.41 & 0.24 & 033739-5148.0~ & 25.71 & 9.20 & 0.15\\
033745-5523.8~v & 255 & 8.41 & 0.56 & 033926-5745.6~v & 40.1 & 11.67 & 0.69 & 033953-1337.8~ & 151 & 10.88 & 0.58\\
034116-1045.1~v & 191 & 9.03 & 0.90 & 034136-3750.4~ & 151 & 11.07 & 0.18 & 034207-1347.2~ & 66.6 & 12.80 & 0.26\\
034234-1450.7~ & 54.7 & 9.19 & 0.42 & 034252-4909.6~ & 26.93 & 11.63 & 0.23 & 034303-3527.1~ & 43.2 & 12.22 & 0.28\\
034319-2929.1~ & 45.2 & 9.71 & 0.30 & 034418-3739.0~ & 31.35 & 10.59 & 0.12 & 034428-7640.0~ & 284 & 11.76 & 0.18\\
034459-1526.8~ & 0.33948 & 13.67 & 0.61 & 034520-4723.3~v & 206 & 11.58 & 1.64 & 034546-1157.9~ & 49.1 & 11.71 & 0.15\\
034552-5054.7~ & 95.5 & 12.47 & 0.23 & 034629-5557.3~ & 246 & 13.40 & 1.27 & 034639-4931.3~ & 42.2 & 10.09 & 0.22\\
034702-6646.2~ & 50.0 & 10.04 & 0.30 & 034740-6641.6~v & 79.3 & 8.15 & 0.98 & 034746-1952.1~ & 54.5 & 11.23 & 0.28\\
034750-6232.2~v & 75.0 & 11.43 & 0.36 & 034900-6952.7~v & 201 & 8.78 & 0.90 & 034910-2155.2~ & 114 & 11.48 & 0.31\\
035027-8608.5~ & 41.7 & 9.18 & 0.34 & 035116-0015.9~v & 171 & 11.01 & 0.52 & 035127-0122.3~v & 127 & 8.45 & 0.45\\
035157-3032.9~ & 117 & 11.91 & 0.42 & 035204-3539.0~ & 102.0 & 10.41 & 0.18 & 035214-4904.3~ & 80.0 & 9.02 & 0.36\\
035216-0703.2~ & 96.8 & 12.50 & 0.51 & 035230-0843.2~ & 37.1 & 9.64 & 0.18 & 035257-7402.6~v & 274 & 11.62 & 0.85\\
035302-4410.3~ & 38.0 & 10.88 & 0.24 & 035342-1116.1~ & 341 & 12.51 & 1.36 & 035411-3654.1~ & 169 & 11.25 & 0.43\\
035442-2914.4~v & 42.5 & 9.99 & 0.22 & 035443-8802.8~ & 70.7 & 12.39 & 0.32 & 035504-4710.2~ & 127 & 9.04 & 0.26\\
035620-3228.9~v & 184 & 10.80 & 1.63 & 035620-3948.6~ & 35.5 & 10.79 & 0.21 & 035803-7205.1~ & 99.5 & 9.38 & 0.26\\
035822-5021.5~? & 44.6 & 11.00 & 0.16 & 035900-0118.4~ & 80.7 & 10.92 & 0.26 & 035924-4554.8~ & 69.8 & 11.02 & 0.44\\
035928-4222.4~ & 39.7 & 11.25 & 0.18 & 035939-1224.8~ & 42.5 & 9.67 & 0.11 & 035947-5824.3~ & 163 & 13.20 & 0.55\\
035957-5823.8~ & 43.3 & 11.19 & 0.20 & 040016-5242.7~ & 31.55 & 9.97 & 0.15 & 040019-0623.1~v & 22.83 & 8.41 & 0.24\\
040024-2741.4~ & 21.70 & 9.20 & 0.10 & 040031-1559.5~ & 45.2 & 11.14 & 0.25 & 040053-5515.8~ & 146 & 9.66 & 0.85\\
040101-3726.8~ & 91.0 & 11.93 & 1.16 & 040125-5214.7~ & 62.2 & 12.35 & 0.56 & 040210-6422.7~ & 60.3 & 11.49 & 0.24\\
040215-3954.1~ & 16.58 & 12.20 & 0.21 & 040224-7754.1~ & 29.93 & 10.91 & 0.16 & 040251-7532.8~ & 35.9 & 10.16 & 0.18\\
040258-7950.7~v & 133 & 9.31 & 0.62 & 040326-3922.7~v & 48.4 & 9.01 & 0.25 & 040329-1121.8~? & 64.2 & 11.30 & 0.35\\
040340-1329.2~ & 36.2 & 9.05 & 0.18 & 040416-3039.9~ & 52.5 & 12.02 & 0.25 & 040418-2036.7~ & 39.2 & 8.93 & 0.05\\
040419-1543.6~v & 118 & 8.57 & 0.56 & 040436-3016.8~ & 200 & 11.85 & 0.92 & 040521-4541.4~ & 55.0 & 12.58 & 0.46\\
040537-2518.4~ & 52.0 & 11.18 & 0.33 & 040558-1838.9~v & 145 & 10.35 & 0.26 & 040602-0915.7~ & 50.4 & 12.38 & 0.70\\
040650-0740.4~v & 147 & 10.11 & 1.10 & 040705-1329.0~ & 80.9 & 12.69 & 0.50 & 040717-2800.5~ & 44.5 & 9.84 & 0.21\\
040719-6026.8~v & 65.5 & 9.21 & 0.53 & 040724-3929.9~v & 55.9 & 8.80 & 0.28 & 040738-1701.6~ & 48.2 & 9.27 & 0.27\\
040752-7313.6~ & 178 & 11.57 & 0.20 & 040814-5353.7~ & 33.80 & 10.38 & 0.28 & 040821-5752.5~ & 41.6 & 11.00 & 0.28\\
040826-7036.5~ & 116 & 10.49 & 0.55 & 040831-0439.8~ & 91.0 & 12.78 & 0.73 & 040850-2525.8~ & 40.0 & 11.54 & 0.43\\
040856-0806.0~v & 53.2 & 8.90 & 0.24 & 040907-0914.2~v & 208 & 9.96 & 1.26 & 040920-1326.4~ & 80.1 & 11.26 & 0.73\\
040936-8151.3~v & 9.685 & 9.54 & 0.38 & 040941-6631.5~ & 68.1 & 10.43 & 0.23 & 041010-0851.5~ & 28.72 & 10.32 & 0.13\\
041027-3435.2~ & 37.8 & 11.13 & 0.21 & 041038-8647.1~ & 45.7 & 9.80 & 0.23 & 041058-3153.8~ & 50.9 & 10.38 & 0.56\\
041104-2003.1~v & 24.43 & 9.03 & 0.10 & 041210-5815.4~ & 67.0 & 11.82 & 0.29 & 041210-7636.2~ & 89.3 & 10.99 & 0.56\\
041228-6508.8~v & 123 & 12.65 & 1.41 & 041239-0747.8~ & 145 & 13.34 & 0.88 & 041337-5035.2~v & 51.2 & 8.82 & 0.45\\
\hline
}
\clearpage

\addtocounter{table}{-1}
\MakeTableSep{|l|r|r|r||l|r|r|r||l|r|r|r|}{10cm}{Continued}{
\hline
\multicolumn{1}{|c|}{ID} & \multicolumn{1}{c|}{$P$} & \multicolumn{1}{c|}{$V$} & \multicolumn{1}{c|}{$\Delta~V$} & \multicolumn{1}{|c|}{ID} & \multicolumn{1}{c|}{$P$} & \multicolumn{1}{c|}{$V$} & \multicolumn{1}{c|}{$\Delta~V$} & \multicolumn{1}{|c|}{ID} & \multicolumn{1}{c|}{$P$} & \multicolumn{1}{c|}{$V$} & \multicolumn{1}{c|}{$\Delta~V$}\\
\hline
\multicolumn{12}{|c|}{\em  Stars classified as MISC.}\\
041359-3132.7~ & 8.317 & 10.63 & 0.16 & 041404-4413.4~ & 28.13 & 10.70 & 0.17 & 041505-5300.6~v & 74.8 & 8.80 & 0.49\\
041521-0630.3~ & 41.5 & 10.39 & 0.21 & 041540-5325.8~ & 131 & 11.03 & 0.39 & 041641-0912.0~ & 76.8 & 11.42 & 0.89\\
041657-6247.7~ & 131 & 11.07 & 0.53 & 041704-1213.3~ & 69.8 & 12.44 & 1.23 & 041707-2501.0~v & 138 & 10.31 & 1.04\\
041712-8344.8~ & 116 & 12.39 & 0.33 & 041737-0923.3~ & 31.44 & 12.43 & 0.30 & 041746-1006.2~ & 33.21 & 9.74 & 0.31\\
041954-1548.5~ & 85.1 & 11.52 & 0.33 & 041956-4303.8~ & 62.9 & 12.20 & 0.47 & 041957-7046.7~ & 175 & 13.31 & 0.52\\
041958-1843.3~v & 78.6 & 11.16 & 0.57 & 042019-1024.5~ & 252 & 12.62 & 1.89 & 042033-1554.9~ & 56.8 & 11.48 & 0.25\\
042110-0310.3~ & 78.6 & 10.96 & 0.59 & 042120-0250.0~ & 119 & 13.34 & 0.71 & 042131-8500.7~ & 30.46 & 9.82 & 0.18\\
042142-6339.5~ & 72.6 & 9.84 & 0.35 & 042144-1403.8~v & 271 & 11.82 & 0.26 & 042203-2241.1~ & 39.4 & 10.02 & 0.37\\
042212-6559.2~ & 57.6 & 11.65 & 0.42 & 042221-4623.3~ & 76.1 & 11.01 & 0.43 & 042235-0530.1~v & 127 & 9.96 & 0.42\\
042241-5032.2~ & 59.0 & 11.34 & 0.34 & 042338-0547.6~ & 32.87 & 12.62 & 0.32 & 042355-7802.4~ & 50.3 & 12.81 & 0.62\\
042401-6707.1~v & 99.9 & 12.28 & 0.97 & 042447-3745.7~ & 51.3 & 10.55 & 0.39 & 042515-5128.3~ & 0.200940 & 9.36 & 0.13\\
042518-3612.5~ & 50.0 & 9.90 & 0.28 & 042525-1136.9~ & 68.6 & 11.14 & 0.48 & 042554-3448.7~ & 22.47 & 10.05 & 0.12\\
042559-3618.3~ & 110 & 12.16 & 1.13 & 042627-0915.7~ & 116 & 10.61 & 0.54 & 042659-7054.0~ & 35.3 & 11.55 & 0.27\\
042706-0022.9~ & 32.6 & 10.62 & 0.22 & 042724-4635.6~ & 91.4 & 11.80 & 0.53 & 042740-6854.0~ & 50.6 & 13.81 & 0.51\\
042742-3213.3~ & 68.2 & 10.83 & 0.94 & 042756-2912.8~ & 49.8 & 9.09 & 0.24 & 042822-0148.6~ & 151 & 12.41 & 0.40\\
042917-8448.3~v & 50.8 & 8.53 & 0.25 & 042941-4033.3~ & 51.0 & 11.97 & 0.37 & 043006-1305.4~ & 63.4 & 11.13 & 0.46\\
043007-0803.2~v & 197 & 9.84 & 0.20 & 043032-1720.0~ & 54.3 & 9.87 & 0.21 & 043051-2817.5~ & 22.77 & 11.51 & 0.13\\
043053-8229.5~ & 30.90 & 10.30 & 0.26 & 043056-0414.6~ & 69.7 & 11.86 & 0.53 & 043116-0201.4~ & 71.5 & 10.25 & 0.38\\
043140-6358.7~ & 85.2 & 10.50 & 0.40 & 043211-6750.6~ & 31.23 & 10.67 & 0.15 & 043215-1029.7~ & 60.1 & 9.83 & 0.14\\
043257-7807.7~v & 79.2 & 9.09 & 0.48 & 043321-4605.9~ & 156 & 13.53 & 0.61 & 043329-1233.5~ & 66.4 & 10.86 & 0.15\\
043331-3334.8~ & 95.2 & 12.02 & 0.56 & 043358-5147.1~ & 35.5 & 11.96 & 0.24 & 043406-0706.9~ & 0.251433 & 13.52 & 0.53\\
043435-1054.5~ & 127 & 13.29 & 0.95 & 043554-8615.3~ & 140 & 12.76 & 0.49 & 043601-3030.9~ & 52.1 & 11.29 & 0.41\\
043603-3953.7~ & 46.1 & 10.75 & 0.28 & 043607-2731.3~ & 64.3 & 11.94 & 0.69 & 043608-2003.2~ & 106.0 & 9.71 & 0.54\\
043613-0516.2~ & 175 & 9.39 & 0.67 & 043634-2734.7~v & 158 & 9.48 & 0.42 & 043658-2403.0~ & 54.8 & 11.44 & 0.50\\
043701-2945.7~ & 81.0 & 11.14 & 0.28 & 043707-1238.3~ & 50.5 & 11.11 & 0.20 & 043712-1841.1~ & 121 & 11.27 & 0.83\\
043739-1740.9~ & 98.2 & 11.45 & 0.49 & 043812-6218.2~ & 96.5 & 10.27 & 0.62 & 043827-1710.9~ & 429 & 12.59 & 0.64\\
043923-3027.4~v & 46.9 & 8.93 & 0.28 & 043924-5212.0~ & 80.2 & 9.59 & 0.19 & 043926-3211.0~ & 58.6 & 11.29 & 0.30\\
043933-0348.7~ & 43.3 & 9.55 & 0.32 & 043935-8408.1~ & 71.4 & 10.92 & 0.55 & 043938-5636.8~ & 35.5 & 9.82 & 0.23\\
043945-2946.0~ & 79.0 & 11.30 & 0.43 & 044003-3924.3~ & 96.7 & 10.99 & 0.39 & 044003-5223.4~v & 30.58 & 9.16 & 0.13\\
044006-6344.0~ & 301 & 12.97 & 1.08 & 044016-2954.8~ & 43.9 & 11.75 & 0.27 & 044016-5336.6~ & 43.0 & 11.73 & 0.27\\
044033-1412.0~v & 80.0 & 10.42 & 0.30 & 044037-5236.0~ & 0.169289 & 13.86 & 0.55 & 044128-8235.2~ & 41.5 & 10.22 & 0.32\\
044139-0153.3~ & 194 & 11.22 & 0.49 & 044142-3314.9~v & 98.8 & 9.67 & 0.70 & 044244-6301.7~ & 89.1 & 12.13 & 0.38\\
044412-5009.1~ & 32.60 & 11.35 & 0.38 & 044421-1240.2~ & 72.8 & 10.06 & 0.48 & 044437-7043.0~v & 65.7 & 14.01 & 1.04\\
044458-0237.1~ & 90.2 & 11.08 & 0.93 & 044500-3722.0~ & 89.1 & 11.81 & 0.24 & 044502-5906.0~ & 84.9 & 13.11 & 0.87\\
044540-2212.7~ & 42.1 & 10.30 & 0.27 & 044557-3441.4~ & 39.2 & 11.12 & 0.15 & 044653-5847.6~ & 46.7 & 10.26 & 0.27\\
044704-1819.1~ & 33.42 & 10.45 & 0.18 & 044714-7504.6~ & 19.88 & 9.73 & 0.09 & 044723-7831.0~ & 75.4 & 11.90 & 0.35\\
044905-6747.2~ & 434 & 12.15 & 0.43 & 044909-1949.2~ & 192 & 11.22 & 1.67 & 044925-6649.8~v & 29.62 & 9.67 & 0.25\\
044926-2932.6~ & 188 & 11.64 & 0.68 & 045039-7633.3~ & 77.9 & 13.41 & 0.44 & 045109-5041.5~ & 41.7 & 10.08 & 0.21\\
045116-2407.3~ & 133 & 13.23 & 1.10 & 045119-0549.6~ & 64.2 & 9.41 & 0.43 & 045121-6929.2~ & 290 & 12.46 & 0.47\\
045133-0335.5~ & 58.1 & 9.72 & 0.44 & 045158-6654.9~ & 46.8 & 11.29 & 0.37 & 045213-4100.3~ & 29.10 & 9.06 & 0.17\\
045245-3914.0~ & 42.2 & 13.82 & 0.66 & 045252-0001.9~: & 0.288233 & 11.46 & 0.23 & 045254-6655.9~ & 393 & 11.22 & 0.23\\
045255-1711.4~ & 29.58 & 10.68 & 0.22 & 045319-6305.7~ & 51.6 & 11.85 & 0.35 & 045328-0534.9~v & 83.9 & 10.73 & 0.69\\
045331-6917.9~ & 135 & 12.78 & 0.60 & 045409-6832.5~ & 394 & 12.93 & 0.85 & 045414-6912.6~ & 330 & 10.04 & 0.16\\
045426-2810.0~ & 161 & 10.46 & 0.19 & 045434-4414.9~ & 48.5 & 10.57 & 0.20 & 045437-6920.4~ & 512 & 10.97 & 0.42\\
045439-1355.0~ & 48.7 & 11.06 & 0.44 & 045440-6904.6~v & 509 & 12.50 & 0.76 & 045441-0728.0~ & 89.6 & 11.79 & 0.41\\
045448-3614.3~ & 42.1 & 9.64 & 0.17 & 045459-5305.6~ & 49.4 & 12.01 & 0.37 & 045503+0025.1~ & 65.6 & 10.47 & 0.61\\
045510-4956.7~v & 48.0 & 8.21 & 0.28 & 045516-6919.2~v & 512 & 12.77 & 0.60 & 045522-0300.0~ & 56.2 & 10.95 & 0.26\\
045532-6650.6~v & 402 & 11.83 & 0.56 & 045550-2742.2~v & 24.15 & 8.45 & 0.12 & 045616-0858.2~ & 103.6 & 11.94 & 0.47\\
045648-6939.9~ & 512 & 12.30 & 0.72 & 045652-0632.1~v & 63.6 & 11.45 & 0.33 & 045706-0147.3~v & 270 & 12.14 & 0.84\\
045727-6623.4~ & 142 & 12.76 & 0.56 & 045731-7009.0~v & 391 & 12.38 & 1.29 & 045736-8708.3~? & 58.0 & 12.40 & 0.55\\
045742-6517.0~ & 20.82 & 9.73 & 0.11 & 045820-1734.3~ & 70.9 & 11.40 & 0.54 & 045822-1641.8~ & 79.7 & 9.95 & 0.51\\
045832-0604.1~v & 127 & 11.51 & 1.91 & 045856-7848.5~ & 51.5 & 11.23 & 0.40 & 045906-2145.0~ & 14.81 & 11.69 & 0.25\\
045909-6540.1~ & 345 & 9.79 & 0.32 & 045930-1627.4~ & 40.1 & 10.69 & 0.19 & 045931-4120.2~ & 77.6 & 12.30 & 0.63\\
045933-3406.6~ & 97.9 & 11.29 & 0.50 & 045939-2656.7~ & 36.8 & 11.51 & 0.21 & 045940-2551.3~ & 156 & 11.77 & 0.47\\
045954-4750.4~ & 151 & 12.77 & 0.82 & 050017-3401.2~ & 84.6 & 10.98 & 0.38 & 050041-6628.5~ & 131 & 10.75 & 0.16\\
050046-1300.3~ & 56.7 & 11.78 & 0.39 & 050049-6550.9~ & 447 & 12.54 & 0.61 & 050106-0439.8~ & 53.3 & 10.54 & 0.57\\
050129-6842.8~ & 409 & 12.85 & 0.50 & 050133-0451.8~ & 33.97 & 11.28 & 0.21 & 050140-5648.3~ & 0.41592 & 8.79 & 0.18\\
050141-1826.1~ & 79.6 & 11.05 & 0.60 & 050147-6859.3~ & 24.35 & 9.23 & 0.14 & 050201-2648.7~ & 49.0 & 11.64 & 0.31\\
050227-6509.1~ & 39.0 & 11.19 & 0.19 & 050231-3854.1~ & 62.3 & 11.04 & 0.37 & 050232-3129.7~ & 39.8 & 10.47 & 0.30\\
050247-2944.1~ & 108.1 & 12.28 & 0.71 & 050300-5405.9~v & 42.7 & 8.80 & 0.23 & 050324-1630.6~ & 16.44 & 10.03 & 0.13\\
050332-2436.5~? & 27.24 & 11.37 & 0.23 & 050357-1859.1~? & 21.53 & 11.17 & 0.16 & 050405-7022.8~v & 511 & 12.37 & 0.44\\
050414-6716.2~v & 418 & 11.57 & 0.59 & 050415-6715.1~ & 42.6 & 10.96 & 0.20 & 050425-2211.1~ & 63.7 & 11.96 & 0.56\\
050445-2837.0~ & 98.7 & 9.86 & 0.13 & 050503-6918.5~v & 156 & 13.05 & 0.55 & 050507-8043.6~ & 98.7 & 10.77 & 0.38\\
050524-2832.5~ & 26.50 & 10.56 & 0.10 & 050527-1628.0~ & 48.8 & 9.46 & 0.16 & 050542-1409.6~ & 43.2 & 10.43 & 0.15\\
050547-5533.7~ & 53.5 & 11.45 & 0.37 & 050557-7035.4~ & 412 & 12.98 & 0.81 & 050559-7048.2~v & 293 & 12.72 & 0.80\\
050701-2220.2~ & 38.8 & 10.49 & 0.16 & 050706-3741.9~v & 69.1 & 8.68 & 0.19 & 050706-7032.7~v & 349 & 11.71 & 0.25\\
050806-6819.1~ & 411 & 13.85 & 1.01 & 050811-6617.8~ & 54.7 & 12.71 & 0.56 & 050816-5013.1~ & 63.4 & 9.83 & 0.32\\
050838-4947.6~ & 51.4 & 11.28 & 0.50 & 050847-3431.0~v & 195 & 8.21 & 1.00 & 050851-2425.5~v & 108.9 & 8.77 & 0.90\\
050855-2043.9~ & 66.7 & 13.17 & 0.57 & 050904-0320.1~ & 61.4 & 9.85 & 0.35 & 050907-1122.5~ & 75.4 & 12.19 & 1.11\\
050911-6936.2~ & 99.0 & 10.71 & 0.21 & 050920-7650.9~ & 62.8 & 10.23 & 0.40 & 050928-2650.4~ & 196 & 11.73 & 0.59\\
051013-6026.3~v & 43.4 & 8.72 & 0.28 & 051112-4534.6~v & 79.6 & 8.54 & 0.43 & 051118-6526.1~v & 391 & 12.07 & 0.34\\
051127-3301.0~ & 56.5 & 10.25 & 0.37 & 051131-6129.1~ & 60.7 & 11.16 & 0.23 & 051200-0820.3~ & 96.6 & 11.93 & 0.50\\
051217-8744.7~ & 68.4 & 10.93 & 0.62 & 051218-3438.3~? & 20.26 & 10.34 & 0.18 & 051229-0233.5~ & 168 & 11.94 & 0.38\\
051246-6719.7~v & 376 & 12.32 & 1.08 & 051333-6720.7~v & 607 & 11.94 & 0.37 & 051344-2003.3~ & 98.7 & 12.95 & 0.98\\
051345-4932.8~v & 253 & 8.71 & 1.28 & 051448-6911.5~ & 63.5 & 11.00 & 0.72 & 051449-6727.4~v & 420 & 12.33 & 0.63\\
051502-7734.9~ & 20.56 & 12.24 & 0.17 & 051505-0939.7~ & 76.7 & 10.89 & 0.26 & 051510-3453.6~ & 51.2 & 12.15 & 0.44\\
051516-6933.1~ & 273 & 12.15 & 0.61 & 051534-6532.6~ & 54.9 & 10.82 & 0.35 & 051535-5309.4~ & 80.1 & 12.12 & 0.81\\
051536-1550.2~ & 28.21 & 9.94 & 0.25 & 051547-6355.9~ & 71.7 & 9.59 & 0.88 & 051600-0948.6~v & 41.2 & 9.90 & 0.31\\
051600-3417.5~ & 78.3 & 11.32 & 0.32 & 051632-1358.1~ & 65.8 & 11.51 & 0.32 & 051643-5456.7~ & 38.5 & 11.65 & 0.30\\
051644-1758.7~ & 64.0 & 11.62 & 0.30 & 051650-2435.5~?v & 60.1 & 9.12 & 0.42 & 051652-1140.6~ & 106.7 & 12.35 & 0.62\\
051653-1711.7~ & 108.4 & 11.77 & 0.22 & 051700-2220.2~ & 35.6 & 10.72 & 0.17 & 051703+0016.2~ & 81.0 & 11.81 & 0.35\\
051726-0528.2~ & 45.9 & 8.84 & 0.25 & 051745-3938.5~? & 15.35 & 11.77 & 0.30 & 051757-6808.7~v & 410 & 13.33 & 0.66\\
051806-1406.8~ & 39.5 & 11.42 & 0.32 & 051807+0001.3~ & 350 & 10.14 & 0.88 & 051831-3106.2~ & 73.0 & 10.59 & 0.59\\
051831-6806.5~ & 362 & 12.69 & 0.78 & 051846-2212.8~v & 26.96 & 8.44 & 0.13 & 051852-0001.3~ & 149 & 11.51 & 0.21\\
\hline
}
\clearpage

\addtocounter{table}{-1}
\MakeTableSep{|l|r|r|r||l|r|r|r||l|r|r|r|}{10cm}{Continued}{
\hline
\multicolumn{1}{|c|}{ID} & \multicolumn{1}{c|}{$P$} & \multicolumn{1}{c|}{$V$} & \multicolumn{1}{c|}{$\Delta~V$} & \multicolumn{1}{|c|}{ID} & \multicolumn{1}{c|}{$P$} & \multicolumn{1}{c|}{$V$} & \multicolumn{1}{c|}{$\Delta~V$} & \multicolumn{1}{|c|}{ID} & \multicolumn{1}{c|}{$P$} & \multicolumn{1}{c|}{$V$} & \multicolumn{1}{c|}{$\Delta~V$}\\
\hline
\multicolumn{12}{|c|}{\em  Stars classified as MISC.}\\
051857-6756.2~v & 394 & 12.22 & 0.42 & 051917-6751.8~v & 43.6 & 9.71 & 0.22 & 051921-1957.0~ & 0.38423 & 14.13 & 0.56\\
051931-6841.2~ & 118 & 11.78 & 0.63 & 051945-1824.6~ & 52.9 & 12.59 & 0.58 & 051952-1752.4~ & 365 & 10.45 & 0.77\\
051954-6804.1~v & 376 & 12.22 & 0.40 & 052013-1811.3~ & 83.4 & 13.01 & 0.51 & 052024-6933.4~v & 438 & 12.03 & 0.86\\
052045-5628.4~ & 41.6 & 12.19 & 0.28 & 052056-6546.1~ & 436 & 11.43 & 0.31 & 052056-6639.7~ & 2.1021 & 13.99 & 1.26\\
052107-3833.2~ & 159 & 12.61 & 0.39 & 052133-0635.8~ & 110 & 11.19 & 1.43 & 052135-8624.8~ & 63.6 & 12.79 & 0.48\\
052147-7119.7~ & 335 & 11.46 & 0.26 & 052156-6549.8~v & 348 & 12.29 & 0.72 & 052158-5701.0~ & 137 & 11.46 & 0.65\\
052158-8448.8~? & 23.59 & 10.39 & 0.14 & 052200-2937.1~ & 54.0 & 11.78 & 0.48 & 052222-3030.7~ & 49.4 & 10.53 & 0.19\\
052228-4650.2~ & 106.2 & 11.16 & 0.44 & 052248-1910.6~ & 102.3 & 11.96 & 0.72 & 052249-6607.8~ & 236 & 12.20 & 0.39\\
052252-1613.7~ & 37.1 & 10.08 & 0.19 & 052259-2032.9~ & 59.6 & 9.46 & 0.27 & 052303-0630.8~ & 76.1 & 12.45 & 0.41\\
052306-0702.2~ & 68.7 & 10.26 & 0.64 & 052320-2428.1~ & 99.4 & 12.71 & 0.63 & 052321-0434.3~v & 428 & 9.60 & 1.07\\
052411-7001.1~ & 37.5 & 9.68 & 0.22 & 052422-1020.8~ & 54.2 & 8.94 & 0.26 & 052423-6246.5~ & 20.61 & 10.07 & 0.11\\
052429-0608.8~v & 115 & 9.26 & 1.01 & 052500-4308.8~ & 32.81 & 8.49 & 0.18 & 052510-1922.5~ & 54.0 & 10.27 & 0.37\\
052528-2010.5~ & 37.1 & 10.90 & 0.38 & 052542-1808.1~ & 64.0 & 12.38 & 0.73 & 052545-6904.9~ & 199 & 13.48 & 0.75\\
052554-4352.4~ & 29.35 & 10.25 & 0.11 & 052558-7011.1~v & 3.1595 & 11.88 & 0.23 & 052612-5336.6~ & 36.2 & 11.94 & 0.20\\
052616-6607.0~v & 265 & 13.06 & 1.27 & 052622-6621.5~ & 475 & 11.45 & 0.22 & 052624-6952.4~ & 311 & 12.12 & 0.29\\
052630-6416.0~ & 35.6 & 9.49 & 0.20 & 052632-1659.4~ & 63.1 & 10.42 & 0.43 & 052634-0019.5~ & 50.2 & 12.07 & 0.66\\
052643-6724.5~ & 360 & 11.54 & 0.37 & 052650-0913.7~v & 154 & 11.77 & 1.81 & 052701-4802.8~ & 62.8 & 12.79 & 0.49\\
052725-0035.2~ & 54.4 & 10.83 & 0.42 & 052730-6714.2~v & 149 & 12.68 & 0.48 & 052734-6653.5~v & 403 & 12.89 & 1.49\\
052745-5110.7~ & 49.9 & 9.92 & 0.19 & 052751-6910.8~v & 201 & 12.56 & 0.76 & 052802-6607.4~ & 433 & 12.94 & 0.57\\
052804-4535.9~ & 84.9 & 11.22 & 0.58 & 052805-5720.5~ & 176 & 10.65 & 1.23 & 052815-6658.9~v & 184 & 12.66 & 0.50\\
052816-5908.9~v & 15.54 & 8.72 & 0.09 & 052817-1222.4~ & 29.54 & 12.26 & 0.27 & 052820-6724.3~ & 159 & 10.52 & 0.19\\
052828-6913.0~ & 415 & 10.81 & 0.45 & 052829-6807.1~ & 391 & 12.28 & 0.60 & 052834-1651.8~ & 191 & 12.84 & 0.53\\
052836-6826.3~v & 422 & 12.21 & 0.66 & 052845-6858.1~ & 402 & 11.97 & 0.42 & 052901-4821.8~ & 52.4 & 8.73 & 0.31\\
052908-6912.3~ & 253 & 12.73 & 0.43 & 052911-0608.1~v & 30.05 & 10.48 & 0.87 & 052921-6847.5~v & 104.4 & 11.92 & 0.22\\
052923-2211.8~ & 36.8 & 10.97 & 0.27 & 052931-5402.8~ & 42.2 & 10.67 & 0.15 & 052942-6857.4~ & 507 & 11.25 & 0.80\\
052945-5339.5~ & 28.70 & 10.26 & 0.33 & 052946-2745.9~ & 62.5 & 11.66 & 0.65 & 052946-6905.8~ & 100.3 & 12.02 & 0.46\\
052953-6907.4~ & 452 & 11.44 & 0.78 & 052955-6718.6~v & 385 & 12.16 & 0.62 & 053005-5248.7~ & 70.9 & 11.86 & 0.59\\
053019-7302.1~ & 161 & 10.88 & 0.50 & 053021-6720.1~v & 511 & 12.54 & 0.53 & 053022-6919.7~ & 66.9 & 10.66 & 0.17\\
053031-1303.4~ & 74.7 & 12.41 & 0.35 & 053036-6859.4~v & 188 & 13.02 & 0.98 & 053041-6915.5~v & 512 & 12.08 & 0.94\\
053101-0323.4~ & 85.2 & 12.19 & 1.07 & 053111-6635.6~ & 356 & 11.88 & 0.33 & 053114-1028.8~ & 129 & 9.67 & 0.20\\
053118-7900.5~v & 11.10 & 8.82 & 0.30 & 053138-6630.1~ & 438 & 11.62 & 1.17 & 053146-1408.2~ & 58.5 & 10.65 & 0.32\\
053153-6640.7~ & 425 & 12.44 & 1.08 & 053218-4019.6~ & 51.9 & 10.31 & 0.18 & 053219-3157.1~ & 171 & 10.30 & 0.82\\
053230-2226.2~ & 37.4 & 9.97 & 0.24 & 053232-6549.6~ & 53.1 & 9.41 & 0.59 & 053244-5054.5~ & 47.5 & 9.89 & 0.30\\
053251-6310.2~ & 47.9 & 11.60 & 0.43 & 053255-2646.1~ & 57.4 & 11.84 & 0.46 & 053300-0607.0~v & 408 & 11.73 & 0.91\\
053300-6835.9~ & 440 & 11.70 & 0.28 & 053301-7157.7~v & 82.7 & 12.09 & 0.76 & 053308-6648.1~v & 428 & 12.30 & 0.94\\
053313-5846.2~ & 55.6 & 9.95 & 0.31 & 053314-6703.8~ & 166 & 12.79 & 0.41 & 053315-7422.1~v & 35.6 & 11.88 & 0.26\\
053325-6622.3~ & 156 & 12.39 & 0.36 & 053326-1216.9~ & 173 & 12.21 & 1.03 & 053328-6704.3~v & 163 & 12.18 & 0.40\\
053342-1621.6~v & 108 & 12.78 & 0.50 & 053352-6911.2~v & 362 & 13.34 & 0.85 & 053357-5826.6~ & 35.36 & 10.13 & 0.14\\
053409-4013.7~ & 50.9 & 12.16 & 0.81 & 053414-2343.0~ & 54.3 & 11.28 & 0.34 & 053421-5629.9~ & 357 & 13.07 & 0.98\\
053422-6225.4~ & 205 & 11.05 & 1.58 & 053450-2704.3~ & 43.3 & 11.74 & 0.45 & 053451-3828.8~ & 23.79 & 9.79 & 0.14\\
053453-7414.1~ & 60.6 & 10.29 & 0.46 & 053456-7539.0~ & 137 & 11.44 & 0.60 & 053458-2643.2~ & 22.77 & 11.83 & 0.22\\
053500-5156.5~ & 52.8 & 10.89 & 0.43 & 053514-1423.8~ & 0.35030 & 12.50 & 0.22 & 053514-6743.9~v & 342 & 12.96 & 1.22\\
053539-0825.2~ & 406 & 11.71 & 0.74 & 053541-6641.3~v & 392 & 12.36 & 0.49 & 053544-3049.6~v & 100.3 & 8.70 & 0.46\\
053601-6650.7~ & 70.0 & 12.40 & 0.47 & 053612-3319.7~ & 46.7 & 11.51 & 0.31 & 053620-2020.0~ & 41.7 & 11.24 & 0.20\\
053625-6941.5~ & 171 & 11.96 & 0.17 & 053630-5730.2~ & 21.90 & 9.87 & 0.13 & 053642-3603.4~ & 80.5 & 11.37 & 0.51\\
053657-5557.6~ & 127 & 9.94 & 0.35 & 053700-3027.2~ & 40.8 & 8.85 & 0.18 & 053700-7526.6~ & 120 & 11.86 & 0.62\\
053704-3453.1~ & 112 & 13.06 & 0.75 & 053713-0635.0~ & 50.3 & 9.94 & 1.27 & 053727-3344.3~ & 79.2 & 11.70 & 0.55\\
053735-7346.7~ & 160 & 11.92 & 0.61 & 053753-5519.6~ & 38.7 & 9.66 & 0.15 & 053756-2316.9~ & 187 & 11.92 & 0.46\\
053832-4547.8~ & 101.6 & 8.96 & 0.61 & 053853-1402.4~v & 445 & 10.42 & 0.29 & 053914-4641.8~ & 74.9 & 11.76 & 0.41\\
053917-8450.4~ & 116 & 13.28 & 0.60 & 053926-5125.4~ & 163 & 11.76 & 1.40 & 053929-7404.7~ & 116 & 12.12 & 0.61\\
053940-1456.6~ & 31.09 & 10.92 & 0.28 & 053943-0809.2~v & 409 & 11.76 & 1.52 & 053955-4504.2~ & 39.2 & 9.78 & 0.23\\
054004-1505.2~ & 42.6 & 11.58 & 0.28 & 054005-4105.2~ & 283 & 10.38 & 1.61 & 054023-2149.7~ & 101.5 & 11.89 & 0.51\\
054041-7203.1~ & 55.5 & 9.86 & 0.29 & 054049-3124.1~ & 18.35 & 8.68 & 0.15 & 054051-0530.5~v & 69.4 & 12.42 & 0.49\\
054059-2757.2~v & 61.8 & 8.38 & 0.59 & 054059-6918.6~v & 188 & 13.08 & 0.54 & 054108-3908.7~ & 31.81 & 9.91 & 0.23\\
054111-6938.0~v & 329 & 12.66 & 0.59 & 054112-1652.5~v & 38.9 & 8.65 & 0.14 & 054123-1743.2~ & 45.9 & 11.99 & 0.24\\
054128-2047.0~ & 80.1 & 10.85 & 0.51 & 054130-7002.0~ & 24.00 & 9.06 & 0.20 & 054149-1402.5~ & 49.5 & 9.26 & 0.20\\
054149-2103.1~ & 278 & 12.77 & 1.17 & 054210-1333.7~ & 49.9 & 11.74 & 0.28 & 054232-1914.0~ & 33.64 & 10.33 & 0.18\\
054240-5625.1~v & 106.9 & 10.19 & 1.05 & 054241-7515.0~v & 309 & 8.11 & 2.03 & 054253-5937.3~ & 70.7 & 10.74 & 0.46\\
054256-2218.8~ & 48.7 & 11.89 & 0.34 & 054257-1726.6~v & 460 & 10.97 & 0.79 & 054304-1336.0~ & 42.0 & 12.00 & 0.35\\
054317-8804.1~ & 426 & 12.46 & 0.33 & 054324-1444.0~ & 86.7 & 12.08 & 0.35 & 054328-6828.2~ & 30.84 & 10.22 & 0.22\\
054339-0504.0~v & 36.5 & 13.28 & 1.27 & 054346-6741.9~v & 103.6 & 9.70 & 0.54 & 054353-2331.7~ & 33.53 & 9.13 & 0.10\\
054416-5701.2~ & 161 & 11.48 & 0.32 & 054426-5801.5~ & 21.78 & 9.90 & 0.46 & 054444-7627.3~ & 127 & 11.90 & 0.65\\
054450-6729.6~ & 73.7 & 11.60 & 0.42 & 054508-2932.3~ & 32.76 & 9.06 & 0.26 & 054530-1903.5~ & 57.7 & 10.64 & 0.48\\
054549-2709.6~ & 46.4 & 11.65 & 0.32 & 054634-2027.9~ & 24.89 & 8.38 & 0.14 & 054648-1204.0~ & 77.7 & 12.58 & 0.57\\
054649-7551.1~ & 191 & 11.34 & 0.58 & 054720-2315.3~ & 46.3 & 9.84 & 0.26 & 054722-2233.6~ & 162 & 11.74 & 1.13\\
054728-2147.4~ & 195 & 12.06 & 1.25 & 054740-0727.8~ & 57.2 & 12.45 & 0.73 & 054745-0708.8~ & 379 & 12.86 & 0.58\\
054747-6022.1~ & 79.1 & 12.53 & 0.67 & 054754-1737.6~ & 44.0 & 10.57 & 0.27 & 054813-1809.6~ & 111.2 & 9.23 & 0.24\\
054840-2224.5~ & 38.4 & 10.19 & 0.18 & 054841-7003.2~ & 79.1 & 12.31 & 0.51 & 054907-2216.6~? & 24.26 & 9.52 & 0.21\\
054937-2606.3~ & 96.4 & 11.17 & 0.17 & 055019-1745.5~ & 36.6 & 9.06 & 0.25 & 055050-8543.0~ & 67.3 & 12.21 & 0.40\\
055102-7001.4~ & 47.4 & 11.96 & 0.43 & 055103-1930.9~ & 69.0 & 11.59 & 0.26 & 055119-1308.1~ & 162 & 12.43 & 0.67\\
055130-1246.9~ & 54.9 & 9.18 & 0.34 & 055141-1031.8~ & 57.0 & 12.02 & 0.35 & 055152-2916.9~ & 180 & 12.22 & 1.15\\
055202-0716.5~ & 74.9 & 12.87 & 0.51 & 055213-1852.0~ & 41.3 & 11.56 & 0.18 & 055218-1214.2~ & 138 & 11.47 & 0.48\\
055252-0447.9~ & 39.6 & 10.82 & 0.34 & 055315-2157.9~ & 49.1 & 11.78 & 0.37 & 055321-6056.9~ & 127 & 12.76 & 0.86\\
055326-1525.9~ & 64.0 & 12.10 & 0.63 & 055327-2307.5~ & 32.68 & 9.31 & 0.37 & 055335-6112.6~ & 69.3 & 9.62 & 0.20\\
055338-1327.9~ & 70.5 & 9.32 & 0.33 & 055341-2013.9~ & 55.3 & 12.50 & 0.58 & 055343-1024.0~ & 361 & 11.81 & 2.04\\
055354-5505.6~v & 47.4 & 8.66 & 0.19 & 055358-0625.3~v & 85.1 & 10.90 & 0.84 & 055406-2509.2~ & 42.1 & 11.63 & 0.40\\
055423-2550.1~ & 104.7 & 12.30 & 0.72 & 055430-0343.2~ & 279 & 12.02 & 0.86 & 055445-6947.2~ & 254 & 12.75 & 0.71\\
055451-1526.2~ & 78.6 & 11.24 & 0.86 & 055458-1829.3~ & 164 & 10.46 & 0.32 & 055508-5457.0~ & 27.21 & 9.71 & 0.19\\
055554-0758.1~ & 106 & 12.82 & 0.51 & 055559-1535.3~ & 30.22 & 11.75 & 0.25 & 055559-2123.0~ & 52.0 & 9.12 & 0.22\\
055616-0803.8~v & 96.3 & 10.89 & 0.48 & 055648-5925.4~v & 74.5 & 8.81 & 0.61 & 055717-1317.3~ & 52.0 & 11.95 & 0.48\\
055734-5538.4~ & 70.7 & 12.49 & 0.52 & 055742-1916.3~ & 77.5 & 11.64 & 0.53 & 055746-3036.5~ & 54.1 & 11.93 & 0.40\\
055815-0948.0~ & 50.7 & 11.89 & 0.36 & 055907-1052.6~v & 120 & 9.24 & 0.86 & 055915-2254.7~ & 24.11 & 12.10 & 0.20\\
055940-1627.9~? & 25.19 & 10.86 & 0.24 & 060048-5510.2~ & 71.5 & 10.87 & 0.40 & 060049-0739.1~ & 0.284760 & 11.47 & 0.30\\
060056-0929.7~ & 49.2 & 12.69 & 0.36 & 060109-2733.4~ & 20.52 & 8.93 & 0.09 & 060127-1417.3~ & 67.1 & 13.25 & 0.86\\
\hline
}
\clearpage

\addtocounter{table}{-1}
\MakeTable{|l|r|r|r||l|r|r|r||l|r|r|r|}{10cm}{Continued}{
\hline
\multicolumn{1}{|c|}{ID} & \multicolumn{1}{c|}{$P$} & \multicolumn{1}{c|}{$V$} & \multicolumn{1}{c|}{$\Delta~V$} & \multicolumn{1}{|c|}{ID} & \multicolumn{1}{c|}{$P$} & \multicolumn{1}{c|}{$V$} & \multicolumn{1}{c|}{$\Delta~V$} & \multicolumn{1}{|c|}{ID} & \multicolumn{1}{c|}{$P$} & \multicolumn{1}{c|}{$V$} & \multicolumn{1}{c|}{$\Delta~V$}\\
\hline
\multicolumn{12}{|c|}{\em  Stars classified as MISC.}\\
060128-0354.1~ & 27.96 & 10.85 & 0.20 & 060130-2527.9~ & 90.9 & 12.49 & 1.17 & 060139-1558.1~ & 55.7 & 9.98 & 0.46\\
060143-6931.5~ & 61.4 & 11.43 & 0.52 & 060150-7035.6~v & 193 & 10.03 & 1.36 & 060151-2106.3~v & 424 & 10.12 & 0.61\\
060151-2127.7~ & 64.4 & 11.41 & 0.36 & 060201-0422.0~ & 48.5 & 9.19 & 0.21 & 060205-7040.5~v & 31.21 & 8.52 & 0.16\\
060246-1943.7~ & 73.5 & 10.02 & 0.64 & 060249-2100.1~ & 21.84 & 12.22 & 0.20 & 060306-5754.7~ & 57.7 & 11.17 & 0.44\\
060346-6942.6~v & 122 & 8.95 & 0.45 & 060349-1217.8~ & 67.3 & 10.59 & 0.25 & 060400-0357.8~v & 54.5 & 10.87 & 0.35\\
060421-1709.7~ & 45.9 & 11.90 & 0.22 & 060457-2212.5~ & 108.7 & 12.60 & 0.86 & 060520-2505.6~ & 128 & 9.81 & 0.49\\
060524-5729.6~ & 127 & 10.76 & 0.34 & 060600-6040.7~ & 33.96 & 9.73 & 0.25 & 060612-0830.3~v & 85.2 & 10.92 & 0.31\\
060617-0705.9~ & 336 & 12.54 & 1.93 & 060628-0927.3~ & 31.21 & 10.30 & 0.17 & 060639-2706.1~ & 48.7 & 9.91 & 0.19\\
060643-1245.5~ & 121 & 12.47 & 0.85 & 060645-1239.0~ & 127 & 11.24 & 0.74 & 060649-6044.6~ & 121 & 11.75 & 0.87\\
060652-0812.0~ & 51.1 & 10.86 & 0.36 & 060701-0254.9~ & 26.13 & 9.35 & 0.78 & 060741-2052.4~ & 63.9 & 8.86 & 0.19\\
060810-1406.8~ & 49.0 & 10.87 & 0.24 & 060834-0042.8~v & 87.3 & 11.19 & 0.53 & 060848-1302.4~ & 150 & 12.67 & 0.30\\
060912-1428.8~ & 58.1 & 12.92 & 0.65 & 060920-2104.6~ & 38.9 & 10.73 & 0.27 & 060925-1306.1~ & 45.2 & 11.01 & 0.30\\
060942-8541.4~ & 59.6 & 11.95 & 0.34 & 060947-2228.8~v & 168 & 10.41 & 0.65 & 060949-1143.7~ & 84.7 & 9.94 & 0.39\\
060951-2231.8~v & 47.3 & 9.53 & 0.48 & 061046-6024.0~ & 62.8 & 10.90 & 0.64 & 061248-7736.7~ & 52.8 & 11.55 & 1.67\\
061446-7659.7~ & 27.22 & 11.28 & 0.23 & 062432-8343.0~ & 62.8 & 10.69 & 0.58 & 064333-8555.8~ & 316 & 13.61 & 0.93\\
064624-8538.4~ & 320 & 11.18 & 0.24 & 064629-8450.5~ & 203 & 12.92 & 1.31 & 064951-8243.6~ & 184 & 10.37 & 0.62\\
065208-8517.8~ & 146 & 11.56 & 0.37 & 074530-8449.7~ & 300 & 12.68 & 1.76 & 075342-8727.4~ & 128 & 11.14 & 0.56\\
080813-8556.6~ & 54.1 & 10.88 & 0.36 & 091527-8714.5~ & 77.4 & 13.37 & 0.80 & 092930-8721.6~ & 60.2 & 11.09 & 0.50\\
094432-8713.6~ & 39.2 & 10.11 & 0.19 & 095013-8725.3~ & 15.13 & 10.69 & 0.12 & 100512-8615.0~ & 48.4 & 11.98 & 0.37\\
102744-8617.9~ & 48.2 & 11.03 & 0.26 & 110752-8632.4~ & 157 & 12.87 & 0.87 & 111701-8835.6~ & 51.8 & 11.57 & 0.34\\
112039-8704.1~ & 166 & 11.70 & 0.31 & 114738-8622.9~ & 110.1 & 12.66 & 1.07 & 120449-8723.1~ & 127 & 11.54 & 0.40\\
122647-8702.0~? & 57.9 & 11.45 & 0.36 & 124002-8530.7~ & 71.5 & 11.96 & 0.42 & 125351-8609.9~ & 47.5 & 12.13 & 0.61\\
125353-8727.4~v & 54.5 & 10.03 & 0.36 & 130015-8520.2~ & 53.8 & 10.65 & 0.39 & 132030-8552.4~v & 253 & 8.81 & 0.48\\
140041-8519.4~ & 20.2 & 10.75 & 0.28 & 140634-8522.9~ & 26.8 & 9.76 & 0.17 & 155630-8547.5~v & 167 & 11.32 & 1.43\\
162726-8602.7~ & 63.9 & 11.21 & 0.64 & 164538-8611.8~ & 77.7 & 12.37 & 0.65 & 165107-8708.2~v & 59.0 & 9.09 & 0.32\\
170225-8703.3~ & 32.46 & 11.21 & 0.33 & 171030-8623.0~ & 222 & 9.97 & 1.85 & 171931-8638.5~v & 84.5 & 11.77 & 2.30\\
182927-8654.3~ & 93.5 & 12.83 & 0.42 & 192802-8813.4~ & 75.8 & 12.09 & 0.31 & 193325-8900.4~ & 53.9 & 10.83 & 0.40\\
202901-8605.2~ & 153 & 11.92 & 0.32 & 203912-8437.6~ & 79.4 & 11.75 & 0.70 & 204429-8713.6~ & 54.9 & 12.00 & 0.43\\
212153-8627.9~ & 29.10 & 8.68 & 0.21 & 215503-8606.7~ & 32.06 & 9.80 & 0.14 & 215650-8151.5~ & 158 & 12.10 & 0.53\\
220020-8654.1~ & 89.1 & 10.67 & 0.87 & 222132-8024.9~ & 91.1 & 12.10 & 0.63 & 222201-8440.0~v & 310 & 8.22 & 0.65\\
223301-8039.8~ & 48.8 & 9.95 & 0.32 & 223908-8011.3~ & 40.3 & 11.37 & 0.16 & 224737-7827.3~ & 302 & 13.51 & 1.85\\
224759-7834.0~ & 46.0 & 10.92 & 0.40 & 225242-8618.5~ & 29.50 & 10.43 & 0.16 & 225636-8359.0~ & 41.4 & 11.63 & 0.31\\
225937-7745.6~ & 394 & 11.48 & 0.36 & 230409-7856.5~ & 156 & 9.27 & 0.53 & 230856-8425.8~ & 80.6 & 9.71 & 0.53\\
231138-8357.9~ & 100.8 & 8.86 & 0.69 & 231447-7507.5~ & 74.5 & 10.67 & 0.89 & 231708-8548.1~ & 79.0 & 10.39 & 0.62\\
231818-8523.3~ & 255 & 11.52 & 0.36 & 231908-8702.2~ & 0.200480 & 11.04 & 0.14 & 232054-8434.2~ & 28.40 & 9.40 & 0.19\\
232310-6212.3~ & 42.3 & 11.21 & 0.33 & 232348-7128.0~ & 77.6 & 10.62 & 0.55 & 232511-7606.3~v & 40.8 & 8.33 & 0.26\\
232513-7703.0~v & 263 & 11.85 & 2.00 & 232641-6901.0~ & 130 & 11.03 & 0.80 & 232859-6002.0~ & 166 & 12.86 & 0.74\\
232903-6446.0~v & 79.7 & 9.30 & 0.59 & 232950-8213.9~ & 86.1 & 12.89 & 0.76 & 233003-6821.1~ & 55.8 & 11.30 & 0.44\\
233524-5503.1~v & 39.8 & 10.78 & 0.56 & 233614-8331.0~ & 86.6 & 12.13 & 0.69 & 233713-6044.2~ & 45.8 & 9.44 & 0.22\\
233849-6923.0~ & 56.2 & 11.61 & 0.35 & 233951-5549.9~ & 50.8 & 11.94 & 0.50 & 234303-0730.2~ & 61 & 8.77 & 0.24\\
234406-3405.2~ & 61.1 & 10.83 & 0.64 & 234423-5829.6~ & 38.5 & 10.03 & 0.21 & 234523-0208.8~ & 47.7 & 10.11 & 0.38\\
234547-6257.7~ & 42.4 & 10.83 & 0.22 & 234708-4548.9~ & 78.9 & 10.93 & 1.12 & 234945-4317.1~v & 57.6 & 10.01 & 0.29\\
235007-1417.8~ & 84.7 & 10.16 & 0.49 & 235010-8651.1~ & 68.7 & 11.19 & 0.63 & 235054-3517.8~ & 159 & 12.07 & 0.60\\
235115-0245.8~ & 127 & 12.22 & 1.14 & 235215-1551.3~v & 139 & 8.26 & 1.89 & 235258-7752.9~v & 48.2 & 11.01 & 0.28\\
235331-1000.3~ & 16.14 & 9.04 & 0.17 & 235440-3046.1~v & 46.8 & 9.98 & 0.24 & 235442-7700.5~ & 106.9 & 11.71 & 0.49\\
235514-4459.8~ & 49.1 & 11.96 & 0.49 & 235651-7928.1~ & 46.2 & 10.15 & 0.40 & 235705-5213.3~ & 50.0 & 10.10 & 0.37\\
235905-5634.6~v & 244 & 7.14 & 1.77 & 235935-2952.0~ & 218 & 10.52 & 0.11 & 235952-3452.2~ & 67.1 & 11.75 & 0.24\\
\hline
}

%% file: appendix_short
\begin{center}
\bf Appendix.\\
 ASAS Atlas of Variable Stars. 0$^{\rm h}$ - 6$^{\rm h}$
Quarter of the Southern Hemisphere.
{ \footnotesize Only first 16 light curves of each type are presented here. 
Complete set of thumbnails  can be obtained from\\
 http://www.astrouw.edu.pl/$\sim$gp/asas/appendix.ps.gz or\\
 http://archive.princeton.edu/$\sim$asas/appendix.ps.gz} 
\end{center}
\vspace*{26mm}
\noindent{\footnotesize Stars classified as EC.}\\[35mm]
\noindent{\footnotesize Stars classified as ESD.}\\[36mm]
\noindent{\footnotesize Stars classified as ED.}\\[35mm]
\noindent{\footnotesize Stars classified as DSCT.}\\
\FigurePs{2.8cm}{Stars classified as RRC.}{im15}{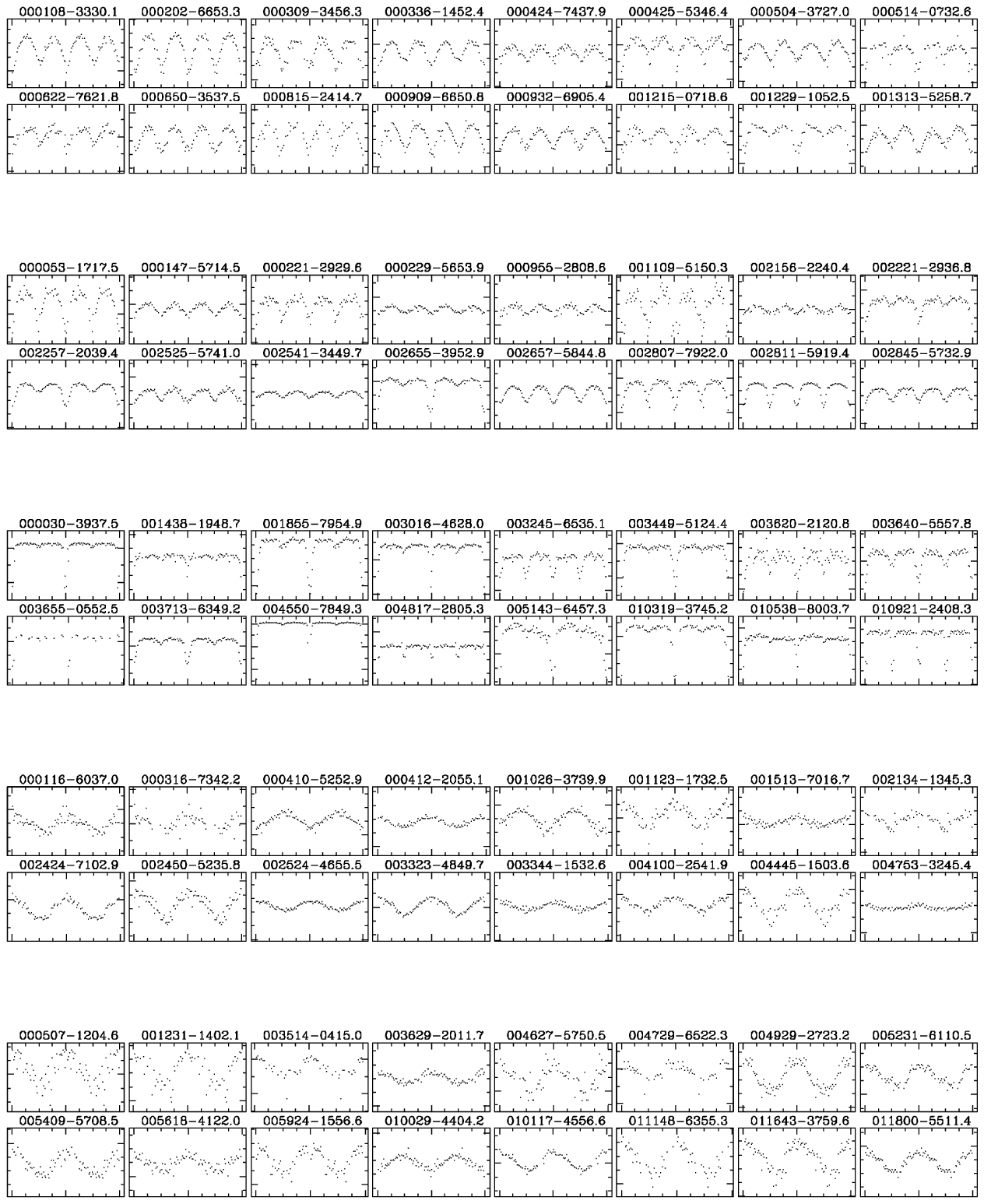}{120}{-40}{-35} \clearpage
\vspace*{26mm}
\noindent{\footnotesize Stars classified as RRAB.}\\[35mm]
\noindent{\footnotesize Stars classified as DCEP-FU.}\\[35mm]
\noindent{\footnotesize Stars classified as DCEP-FO.}\\[36mm]
\noindent{\footnotesize Stars classified as PULS.}\\[35mm]
\noindent{\footnotesize Stars classified as MIRA.}\\
\FigurePs{2.8cm}{Stars classified as MISC.}{im33}{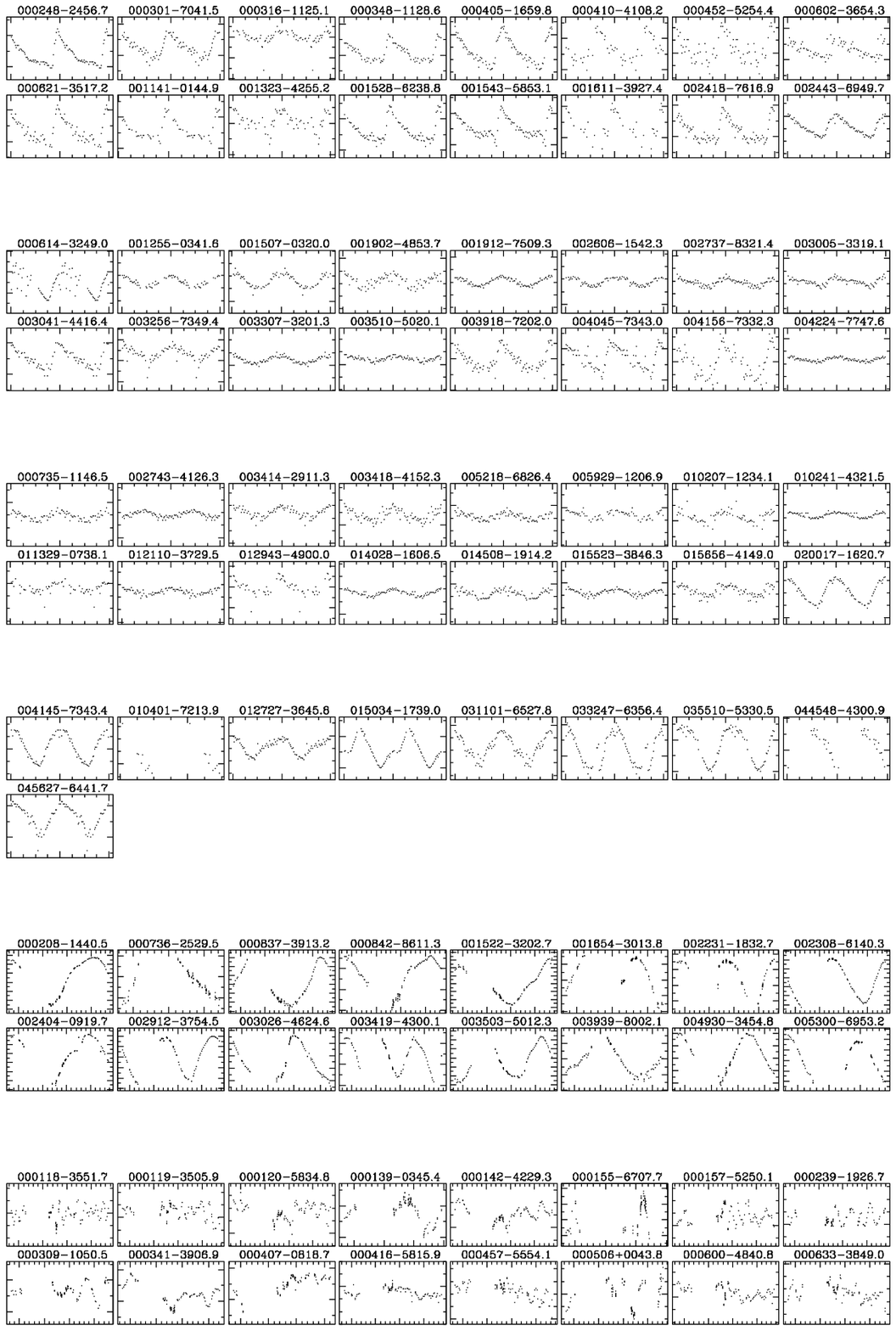}{120}{-40}{-75}